\PassOptionsToPackage{table}{xcolor}
\documentclass[sigconf,screen]{acmart}
\PassOptionsToPackage{pagebackref}{hyperref}
\usepackage{xspace}
\usepackage[normalem]{ulem}
\usepackage{lipsum}
\usepackage{amsmath}
\usepackage{tikz}
\usetikzlibrary{positioning, arrows.meta, patterns, fit, calc}
\usepackage{subcaption}
\usepackage{multirow}
\usepackage{adjustbox}
\usepackage[nameinlink]{cleveref}
\usepackage{backref}
\usepackage{geometry}
\usepackage{xcolor}
\usepackage{tabularx}
\usepackage{booktabs}

\iftrue
\newcommand{\todo}[1]{{\color{blue}\textbf{[todo]: #1}}}
\fi

\crefname{section}{Sec.}{Secs.}
\Crefname{section}{Sec.}{Secs.}
\crefname{figure}{Fig.}{Figs.}
\Crefname{figure}{Fig.}{Figs.}
\crefname{table}{Tab.}{Tabs.}
\Crefname{table}{Tab.}{Tabs.}
\crefname{equation}{Eq.}{Eqs.}
\Crefname{equation}{Eq.}{Eqs.}

\providecommand{\improv}[1]{{\color{green!60!black}\scriptsize #1}}
\providecommand{\degrad}[1]{{\color{red!70!black}\scriptsize #1}}

\title{Beyond Gestures: Estimating Full Hand Pose and Contact Forces from Wrist-Worn Pressure Sensor Array}

\author{Svetoslav Kolev, Lingni Ma, Michael Goesele, Renzo De Nardi, Jakob Engel,
  Richard Newcombe}
\affiliation{\vspace{2mm}\institution{Meta Reality Labs Research}\country{USA}}

\begin{document}

\begin{abstract}
Capturing hand motion and interaction forces is critical for interactive computing, VR, and high-fidelity tactile demonstrations for robot learning.
We introduce a wrist-worn pressure-sensing wristband that recovers continuous full-hand pose and distributed contact force on a single wearable. The system consists of flexible capacitive sensor arrays around the wrist, which require no electrical skin contact, and a recurrent network that maps the resulting pressure signal to hand state. Our key insight is that muscle contraction and tendon displacement produce pressure patterns, which correlate strongly with hand pose and interaction force. To validate this, we collect synchronized recordings of wrist pressure, optical motion-capture hand pose, and tactile-glove interaction force, covering isolated finger motion, fingertip-force stress tests, and natural hand--object manipulation. On isolated single-user motion the wristband attains $4.6^\circ$ mean finger-joint MAE, and across four users manipulating everyday objects it estimates per-finger contact force at $R^2{=}0.57$, which an external pose signal brings up to $0.75$. We see the wristband as one node in a constellation of everyday wearables --- e.g.\ paired with an egocentric camera --- adding the contact force that vision cannot observe and taking over when the hand is occluded.

\end{abstract}

%
%

\settopmatter{printacmref=false}
\renewcommand\footnotetextcopyrightpermission[1]{}
\pagestyle{plain}

\begin{teaserfigure}
  \centering
  \def\iw{0.195\textwidth}
  \begin{tikzpicture}[inner sep=1pt]
    \node(p01)                            {\includegraphics[width=\iw]{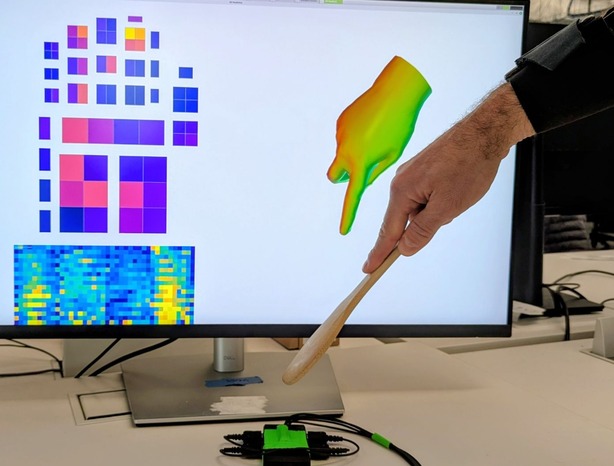}};
    \node(p02) at (p01.east) [anchor=west]{\includegraphics[width=\iw]{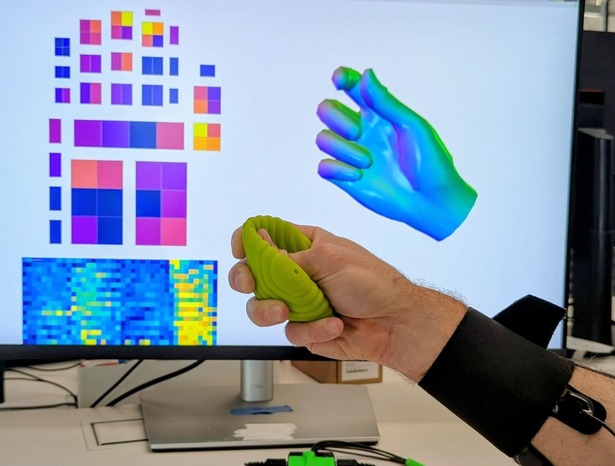}};
    \node(p03) at (p02.east) [anchor=west]{\includegraphics[width=\iw]{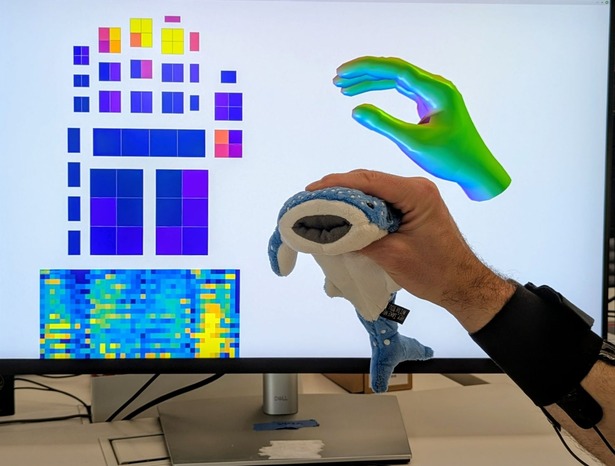}};
    \node(p04) at (p03.east) [anchor=west]{\includegraphics[width=\iw]{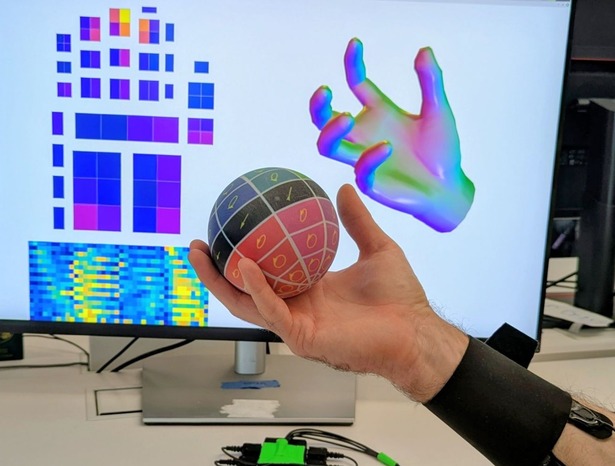}};
    \node(p05) at (p04.east) [anchor=west]{\includegraphics[width=\iw]{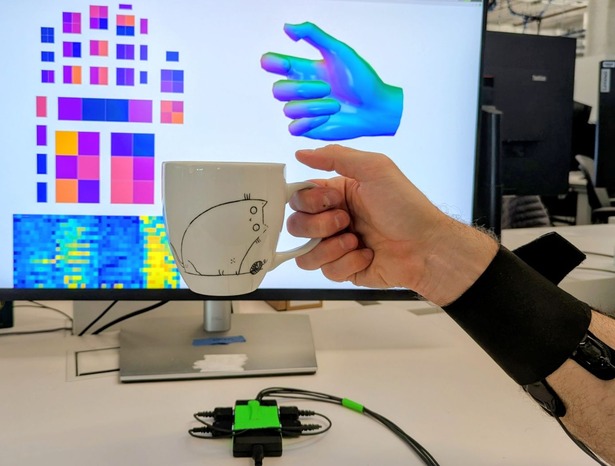}};
  \end{tikzpicture}
  \caption{\label{fig:teaser}Five frames from our end-to-end system running in real time. The user wears a wristband with four pressure-sensor arrays. The display behind the user shows, from bottom to top, the raw wristband pressure heatmap, the predicted tactile-glove pressure map, and the predicted 3D hand pose rendered from the 20 finger and 3 wrist angles from a per-user model.}
  \Description{Five frames from the end-to-end system running in real time, showing the user wearing the pressure-sensing wristband while manipulating different objects, with the predicted hand pose and pressure maps on a monitor behind them.}
\end{teaserfigure}

\maketitle
\raggedbottom

\section{Introduction}\label{sect:intro}

The future of everyday immersive computing relies on capturing the full expressive capability of the human hand --- both the finger motion and the forces exerted by them. Motion and force together enable a richer understanding of object interactions, better virtual manipulation in VR and gaming, and easier user input. Force in particular provides higher-fidelity human demonstrations for robot training. However, capturing this remains a major unsolved problem in everyday settings.

There are many ways to capture human hand state. Vision-based techniques are very effective at capturing finger motion when they can see the hand. However, occlusions are common, especially during object interaction. Even with dedicated egocentric cameras, the hands can move out of the field of view. Cameras and vision processing also require significant power. Moreover, vision-based systems are intrinsically blind to force, which is an internal phenomenon inside the body.

Gloves are another way to obtain hand state. Motion-capture gloves are accurate, but impractical for all-day use due to their fragility and obtrusiveness, which can severely limit hand dexterity. Gloves, however, do allow us to capture forces exerted by the hand. Pressure-sensing coverage is often limited only to the front of the fingers, missing important hand surfaces that participate in everyday object handling.

Wrist-worn devices offer a practical alternative for sensing hand state. The distal forearm (just proximal to the wrist joint), where people already wear watches and bands, is a convenient and information-rich sensing site. As forearm muscles and tendons actuate the hand, they produce characteristic pressure changes at the band--skin interface. This makes wrist-worn sensing unobtrusive, socially acceptable, and suitable for everyday use.

We do not envision the wristband acting alone. We see it as one node in a constellation of everyday wearables --- most naturally paired with an egocentric camera (e.g.\ Project Aria~\cite{aria23surreal}). When the hand is visible, vision supplies an accurate pose estimate; when it is occluded or out of frame, the wristband carries the estimate, and it supplies the force that vision cannot observe at all. This complementary role also shapes our evaluation: because a companion device can provide user-specific pose, and because a deployed system would calibrate to its wearer, we train and report \emph{per-user} models as our primary setting rather than treating the per-user scope as a limitation. Our ``pose-as-input'' force experiments make this constellation concrete --- an external pose signal, such as one from vision, is fed to the wristband to sharpen its force estimates.

In this paper, we present a wrist-worn pressure-sensing approach that measures these wrist pressure patterns and learns a mapping from them to hand pose and interaction force. Our contributions are summarized as follows:

\begin{itemize}
  \item The first continuous full-hand pose from a \emph{wrist-worn} pressure band; other full-pose pressure sensing has required a full-forearm sleeve~\cite{niu2026sleevepose}.
  \item The first demonstration of continuous force estimation from a wrist-worn pressure array, at both per-finger and per-taxel granularity --- a regime where prior wrist-pressure work has been limited to aggregate grip force or discrete gestures.
  \item A demonstration that conditioning force estimation on an externally provided hand pose substantially reduces force MAE for wrist-worn force myography (FMG) --- extending a benefit previously shown only for electromyography (EMG)-based force estimation~\cite{seo2024pimforce} to the pressure-sensing modality.
\end{itemize}

\section{Related Work}\label{sect:literature}

Hand-state sensing falls into three physically distinct categories that differ in \emph{what} they measure: electromyography senses the muscle \emph{activation} that drives the hand, whereas the other two observe its \emph{outcome} --- either at the source or indirectly through the forearm tissue deformation it produces. \textbf{Direct observation} --- vision, instrumented gloves, force plates --- senses finger configuration and contact force at the source, but is limited by line-of-sight and by obtrusive instrumentation. \textbf{Electromyography} measures muscle activity electrically and is the natural channel for intrinsic and isometric activity where no external displacement occurs; backed by a large public corpus~\cite{salter2024emg2pose,emg2qwerty,kaifosh2025generic}, it is a complementary signal source rather than a competitor to our approach. \textbf{Wrist morphology} --- the indirect category we target --- measures how forearm tissue deforms in response to hand action, spanning a depth spectrum from volumetric imaging (CT, MRI), through partial-penetration methods (ultrasound, NIRS, and electrical impedance tomography (EIT)), to pure surface mechanics (FMG).

For wearables, this spectrum effectively collapses. Ultrasound genuinely resolves internal muscle and tendon morphology and may offer the richest signal, but at the cost of bulk and acoustic coupling~\cite{bimbraw2023ultrasoundforce,engdahl2024sonomyography}. We hypothesize that, in tightly worn wrist systems for hand-state inference, the task-relevant signals from EIT and capacitive sensing are often dominated by pressure-dependent contact impedance and near-surface tissue deformation under the band --- mechanical phenomena that FMG measures directly, although each modality transduces them differently~\cite{yao2020wearableeit,lou2024advancing,truong2018capband,xiao2019review}.

We use FMG, which senses this surface deformation mechanically and without skin-electrode coupling. We use it for continuous full-hand pose and distributed contact force on a single wearable, and show how it complements other modalities that already cover the rest of hand state. \Cref{tab:comparison} summarizes representative systems across modalities; the subsections below discuss each layer in turn.

\begin{table}[t]
\centering
\caption{Representative wearable forearm/wrist systems for hand pose and interaction force estimation. Metrics and protocols differ across sources, so values are not directly comparable; per-row operating points and metric definitions are given in \cref{sec:supp_related}.}
\label{tab:comparison}
\small
\setlength{\tabcolsep}{3pt}
\begin{tabular}{@{}llrr cc@{}}
\toprule
\textbf{System} & \textbf{Year} & \textbf{Modality} & \textbf{\#\,Ch} & \textbf{Pose Err} & \textbf{Force Err} \\
\midrule
emg2pose~\cite{salter2024emg2pose}        & 2024 & sEMG          & 16  & 15.8\,mm & -- \\
NeuroPose~\cite{liu2021neuropose}         & 2021 & sEMG          & 8   & 6.2$^\circ$ & -- \\
Rahimi et al.~\cite{rahimi2024simulcontrol} & 2024 & HD-sEMG     & 320 & 11.0\,mm & 0.8\,N \\
Li et al.~\cite{li2024graphdriven}        & 2024 & HD-sEMG       & 192 & $R^2{=}0.80$ & $R^2{=}0.85$ \\
EtherPose~\cite{kim2022etherpose}         & 2022 & EF            & 2   & 11.6\,mm & -- \\
EITPose~\cite{kyu2024eitpose}             & 2024 & EIT           & 8   & 11.1\,mm & -- \\
Zadok et al.~\cite{zadok2022ultrasound}   & 2022 & Ultrasound    & --  & 4.9$^\circ$ & -- \\
Bimbraw et al.~\cite{bimbraw2022config}   & 2023 & Ultrasound    & --  & 7.4$^\circ$ & -- \\
Bimbraw \& Zhang~\cite{bimbraw2023ultrasoundforce} & 2023 & Ultrasound & -- & -- & 0.2\,N \\
EchoWrist~\cite{lee2024echowrist}         & 2024 & Acoustic      & 2   & 4.8\,mm & -- \\
SleevePose~\cite{niu2026sleevepose}       & 2026 & FMG           & 117 & 14.0\,mm & -- \\
Sakr et al.~\cite{sakr2019forcetorque}    & 2019 & FMG           & 60  & -- & $R^2{=}0.77$ \\
PiMForce~\cite{seo2024pimforce}           & 2024 & sEMG/cam    & 8   & -- & $R^2{=}0.89$ \\
Wrist2Finger~\cite{xiao2025wrist2finger}  & 2025 & sEMG/ring   & 1+1 & 5.7\,mm & 0.21\,N \\
\midrule
\textbf{Ours} & \textbf{2026} & \textbf{FMG} & \textbf{720} & \textbf{7.2\,mm} & \textbf{0.59\,N} \\
\bottomrule
\end{tabular}
\end{table}

\subsection{Wrist-Worn Sensing and Continuous Pose}
\label{sec:modality}

Surveys document a broad spectrum of wrist-worn modalities for hand inference~\cite{jiang2022survey, zheng2022reviewemgfmgeit}: sEMG, EIT, ultrasound, and active acoustic sensing~\cite{lee2024echowrist}. FMG instead senses mechanical deformation at the skin surface from muscle contraction and tendon displacement, using FSRs~\cite{dementyev2014wristflex}, barometric sensors~\cite{shull2019hand}, or capacitive arrays~\cite{truong2018capband}; it is low-power, low-cost, and robust to sweat~\cite{sherif2024fmgsurvey}. GestureWrist~\cite{rekimoto2001gesturewrist} established the wristband form factor and WristFlex~\cite{dementyev2014wristflex} showed FSR sensing is viable for always-on input, while denser barometric~\cite{zhu2018wrist, shull2019hand} and capacitive~\cite{truong2018capband, wang2023flexible} arrays push gesture-classification accuracy higher. The common thread, however, is \emph{discrete gestures}.

Continuous pose is harder. Environment-mounted cameras set the accuracy bar, with parametric models such as MANO~\cite{romero2017mano} as the backbone, but require instrumented spaces; wrist-worn cameras~\cite{kim2012digits, xi2026wristp2} need line-of-sight to the hand. Eyes-free pose has been shown from sEMG~\cite{salter2024emg2pose}, EIT/RF~\cite{kyu2024eitpose, kim2022etherpose}, and acoustics~\cite{lee2024echowrist}. Among pressure-based methods, only a few attempt continuous kinematics~\cite{kadkhodayan2016continuous, shull2019hand, kammarchedu2024realtime}, and these recover only partial hand representations (per-finger scalars or a few joints); full articulated pose from pressure has been shown only very recently, and from a full-forearm sleeve rather than a wristband~\cite{niu2026sleevepose}. We regress a complete hand pose from a wrist-worn capacitive pressure array; \cref{tab:comparison} places dense capacitive pressure favourably among eyes-free modalities at a comparable within-user operating point.

\subsection{Force and Combined Pose--Force Estimation}
\label{sec:force}

Per-finger force has so far required high-complexity sensing: HD-EMG motor-unit decomposition~\cite{rubin2022finger} or forearm ultrasound~\cite{bimbraw2023ultrasoundforce}. Vision can infer contact force from hand--object tracking~\cite{pham2018hand} but fails under the same occlusion that limits vision-based pose. FMG-based force work, by contrast, has targeted only aggregate grip or multi-axis force/torque from forearm bands~\cite{wininger2008pressure, sakr2019forcetorque}. To our knowledge, no prior work recovers per-finger --- let alone spatially resolved per-taxel --- interaction force from a wrist-worn pressure array.

A few systems estimate pose and force together. Multi-task HD-EMG models predict wrist angle or full joint positions together with grip force~\cite{li2024graphdriven, rahimi2024simulcontrol}, but rely on dense electrode grids (192--320 channels around the forearm) far from a practical wearable. PiMForce~\cite{seo2024pimforce} shows that conditioning force estimation on 3D hand posture markedly improves prediction, using an external camera for pose and sEMG for force --- we adopt the same pose-conditioning in our force model. Wrist2Finger~\cite{xiao2025wrist2finger} couples a ring IMU with watch sEMG, and WristP$^2$~\cite{xi2026wristp2} recovers pose and per-vertex pressure but needs continuous line-of-sight. Our system instead uses a single wrist-worn capacitive array for both full hand pose and per-taxel force, optionally conditioning force on an externally provided pose signal.

\section{Hand State from Pressure-sensing Wristband}
\label{sect:method}

\subsection{Motivation}
\label{sect:overview}

We investigate the pressure sensed around the distal forearm as a high-fidelity source for decoding human hand state --- both hand pose and the contact force distributed across the inner surface of the hand.

Our approach rests on a \textbf{biomechanical coupling} between forearm muscle--tendon units and hand--object interaction. Because the digits and wrist are actuated by extrinsic muscles in the forearm, any change in hand pose or interaction force requires a corresponding change in the internal loading of these tissues. As muscles contract and tendons displace, they alter the local stiffness and contour of the distal forearm, transmitting mechanical changes radially through the fascia to the skin. Surface pressure measured by the wristband therefore serves as a proxy for both the \textbf{kinematic} (pose) and \textbf{kinetic} (force) outcomes of the neuromuscular system. One limitation follows directly: intrinsic hand muscles are largely invisible above the wrist, bounding how observable purely intrinsic actuation --- including parts of thumb motion --- can be from any forearm-worn modality~\cite{salter2024emg2pose}; we detail the underlying anatomy in \cref{sec:supp_biomechanics}.

We treat hand state as three coupled but distinct components: kinematic pose, external contact force, and internal muscle tension under co-contraction or isometric loading without external contact. This paper targets the first two; the third lies outside what either motion capture or a tactile glove can directly report, and we return to it in the discussion.

Our choice of \emph{pressure} sensing is deliberate. We hypothesize that, for tightly worn wrist systems used for hand-state inference, task-relevant EIT and capacitive signals are often dominated by pressure-dependent contact effects and near-surface deformation. Pressure sensing measures this mechanical component directly rather than through an electrode--skin interface (\cref{sec:supp_why_pressure}); our matched-protocol EIT comparison in \cref{app:eit} tests this hypothesis empirically.

To study this coupling, our customized wristband embeds dense arrays of capacitive pressure sensors, and for supervision we synchronize its measurements with hand pose from OptiTrack Motion Capture (MoCap) and interaction force from PPS tactile gloves.

\subsection{Sensing Hardware}
\label{sect:hardware}
In this section, we describe our customized wristband for measuring pressure around the distal forearm, and the ground-truth system that provides hand-pose and interaction-force supervision.

\begin{figure}[t]
  \centering
  \def\iw{5.0em}
  \resizebox{\columnwidth}{!}{%
  \begin{tikzpicture}[inner sep=1pt]
    \node(p01)                             {\includegraphics[height=\iw]{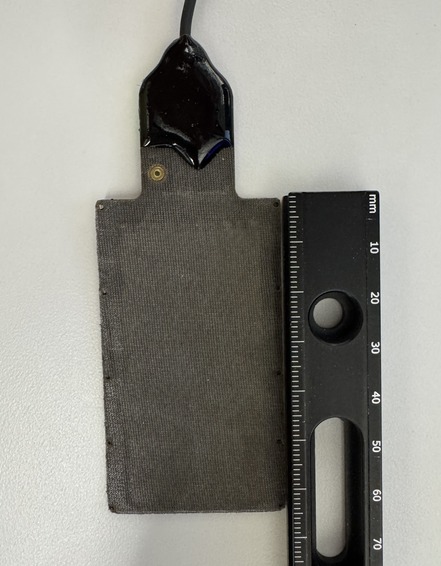}};
    \node(p02) at (p01.east) [anchor=west] {\includegraphics[height=\iw]{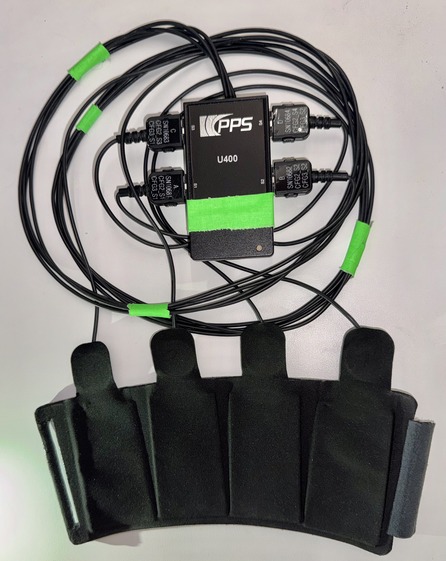}};
    \node(p03) at (p02.east) [anchor=west] {\includegraphics[height=\iw]{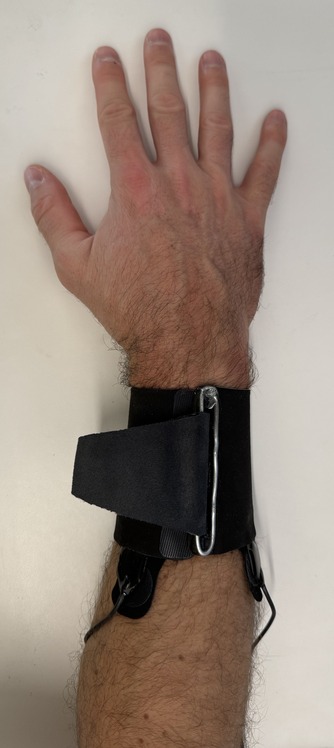}};
    \node(p04) at (p03.east) [anchor=west] {\includegraphics[height=\iw]{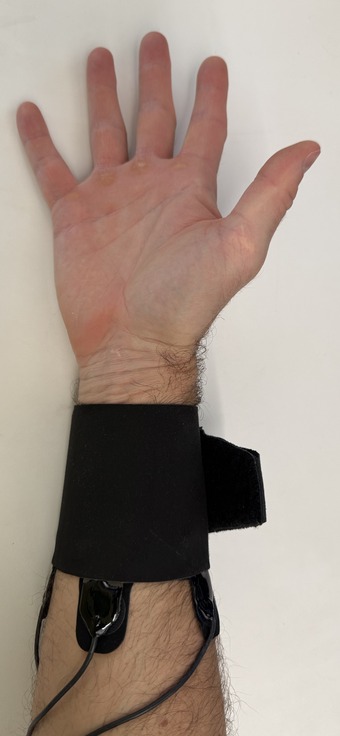}};
  \end{tikzpicture}%
  }
  \caption{Sensing hardware. From left to right: a single PPS \#6383 pressure-sensor array; our large customized pressure wristband with four embedded sensor arrays (visible as indentations under the fabric) and the ADC base station; and dorsal and palmar views of the pressure wristband worn on the distal forearm.}
  \label{fig:wristband}
\end{figure}

\subsubsection{Customized Pressure-sensing Wristband}\label{sect:wristband}
To measure pressure around the distal forearm, we adopt commercially available capacitive pressure sensor arrays from Pressure Profile Systems (PPS)~\cite{pps}. We use the PPS \#6863 array, which measures $39 \times 68\;\text{mm}^2$, with an active sensing area of $30 \times 56\;\text{mm}^2$. Each array contains a $10\times 18$ grid of sensing elements, with approximately $3\times 3\;\text{mm}^2$ area per taxel (\textit{cf.} the single array shown in \cref{fig:wristband}). Each sensor includes a tab that houses the amplifier electronics, and all tabs are routed by cable to a central hub for ADC conversion and USB communication with the host computer. The sustained streaming rate at the PPS hub depends on the number of attached arrays: the small 3-array wristband runs at approximately 50~Hz, and the large 4-array wristband at approximately 45~Hz. Each frame includes an internal device timestamp, and the hub emits periodic timing packets between sensor frames. Together, these signals allow us to align the wristband clock to host time.

\paragraph{Row-banding artefact.}
\label{para:row-banding}
The \#6863 array shows a mild row-to-row baseline difference --- visible as horizontal banding in the raw heatmap (\cref{fig:dataset-handpose})  --- which we attribute to per-row electronics. We do not correct for it and treat it as part of the sensor's nature, relying on our learning-based model to compensate for it.

To package the arrays into a wearable form factor, we design a thin fabric wristband with dedicated pockets to hold each sensor in place. The band has a tapered, conical geometry for better fit around the distal forearm and uses a rectangular-loop buckle with a hook-and-loop fastener. To accommodate different forearm sizes, we build a small wristband with 3 arrays (540 total taxels) and a large one with 4 arrays (720 total taxels). \Cref{fig:wristband} shows our wristband laid flat as well as worn by a participant in dorsal and palmar views.

\begin{figure}[t]
  \centering
  \def\iw{4.6em}
  \resizebox{\columnwidth}{!}{%
  \begin{tikzpicture}[inner sep=1pt]
    \node(p01)                            {\includegraphics[height=\iw]{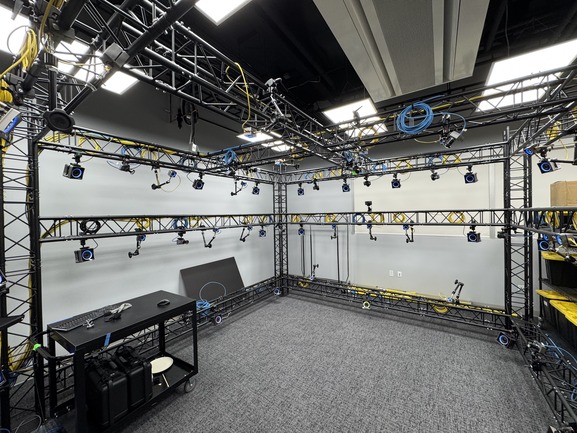}};
    \node(p02) at (p01.east) [anchor=west]{\includegraphics[height=\iw]{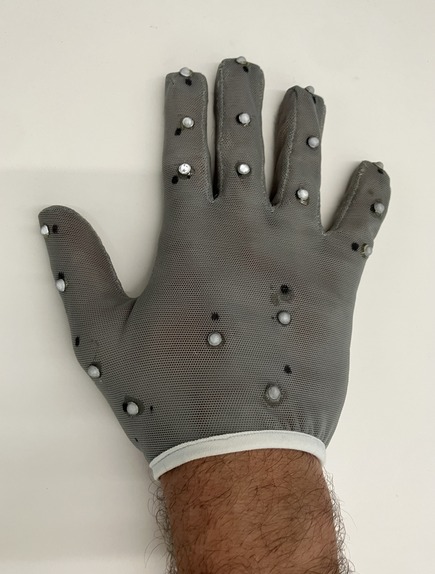}};
    \node(p03) at (p02.east) [anchor=west]{\includegraphics[height=\iw]{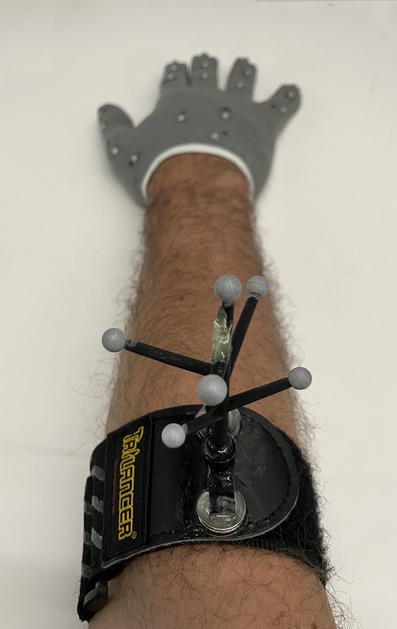}};
    \node(p04) at (p03.east) [anchor=west]{\includegraphics[height=\iw]{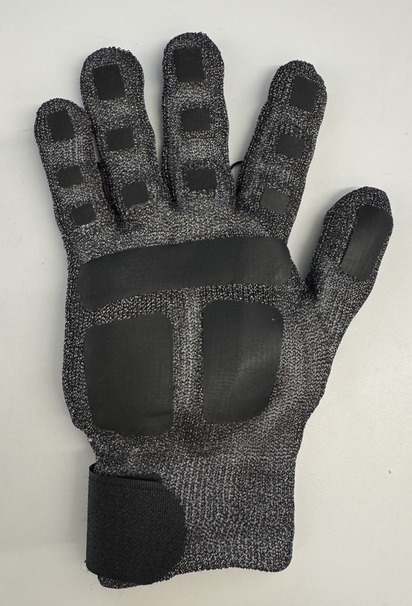}};
    \node(p05) at (p04.east) [anchor=west]{\includegraphics[height=\iw]{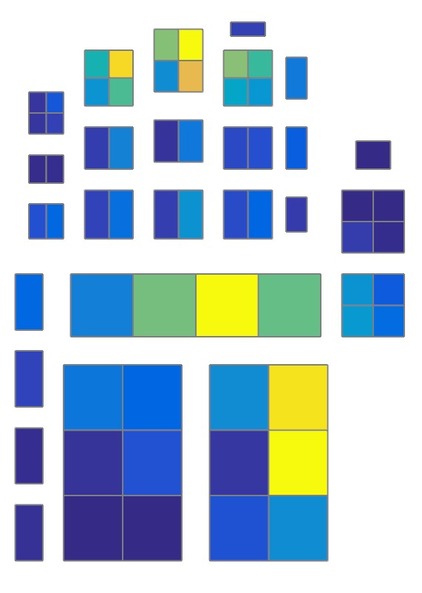}};
  \end{tikzpicture}%
  }
  \caption{Ground-truth collection setup. From left to right: the OptiTrack-based MoCap system to capture hand and wrist poses; the thin MoCap glove with 19 reflective markers attached for hand tracking; the rigid marker tree mounted on the proximal forearm to recover wrist rotation; the PPS tactile glove; and the 65-taxel layout of the PPS glove.}
  \label{fig:gt}
\end{figure}

\subsubsection{Ground-truth Hand Pose from MoCap}
\label{sect:mocap}
We use an OptiTrack MoCap system to obtain ground-truth hand poses, inside a $4\times 4\times 3$~meter capture volume instrumented with 50 infrared MoCap cameras. The intrinsic and extrinsic parameters are calibrated with the standard MoCap procedure. To capture hand motion, we place 19 reflective optical markers on a thin mocap glove (\cref{fig:gt}). The marker locations are then semi-automatically registered to the 3D hand model. During data collection, the markers are tracked by MoCap cameras and the hand pose is computed by fitting the participant's UmeTrack hand model using the algorithm of~\cite{mocaphand}.
The hand tracking algorithm also outputs the 6-DoF global transform of the hand with respect to the MoCap system. However, this is not the rotation we need for supervision. We seek the 3D rotation of the distal forearm with respect to the elbow. To obtain this rotation, we mount another rigid-body marker tree on the proximal forearm (\cref{fig:gt}). By combining the pose of the proximal forearm with the global root transformation of the hand, we obtain the needed rotation. We use the standard timecode synchronization protocol provided by OptiTrack to align MoCap data with the host PC clock, which bridges to the synchronization with the wristband recording. \Cref{fig:gt} also shows the MoCap lab.

\subsubsection{Ground-truth Interaction Force from Tactile Gloves}
\label{sect:glove}
We use PPS tactile gloves~\cite{pps} to measure interaction forces. Each glove provides 65 sensing taxels across the fingers and palm (see the layout in \cref{fig:gt}). Each finger has at least eight taxels: four on the fingertip pad and four distributed over the two intermediate finger pads. The thumb and index finger each include one additional distal taxel near the fingernail, and the index finger includes three lateral taxels on the side opposite the thumb. The remaining taxels are distributed across the palm and along the ulnar (pinky-side) edge of the hand. Each taxel reports pressure normal to the glove surface, converted to a force via the taxel area; we report it \emph{per-taxel}, \emph{per-finger} (the sum over a finger's taxels), or \emph{per-fingertip} (the sum over its fingertip-pad taxels). These forces are purely normal (no shear), but taxels across a finger face different directions, so a summed value is a total force magnitude, not a single-direction force.

A PPS tactile glove streams data at 100~Hz over Bluetooth. While it provides internal timestamps, we observe substantial latency (about 35~ms) and clock drift. The drift behaves like a random walk, accumulating up to about 1~second over a 10-minute recording. To compensate, we estimate a fixed latency by repeatedly pressing the glove on the wristband and aligning peak-force events across both streams, and we continuously track and correct the glove clock during recording to avoid misalignment.

\subsection{Hand States Learning Pipeline}
\label{sect:algorithm}

In this work, we aim to learn the mapping from a pressure sensor array around the distal forearm to hand state.
We formulate the learning as sequence-to-sequence prediction to leverage the temporal coherence of the measurements and predictions.

\subsubsection{Notations}
\label{sect:notations}
A sequence of pressure array measurements is denoted by $\boldsymbol{P}:=\{P_i\}$, where $i$ is the timestamp and ${P}_i\in\mathbb{R}^{w\times h}$ is the pressure measured with $w\times h$ spatial resolution.
Hand state includes kinematics and kinetics. The \textit{kinematics} are represented by a sequence of parameterized hand poses $\boldsymbol{q}:=\{q_i\}$. We adopt the UmeTrack hand format~\cite{umetrack22nimble}, which consists of 20 joints articulated by 20 degree-of-freedom (DoF) joint angles $\beta\in\mathbb{R}^{20}$, together with a 3D wrist rotation $\omega\in\mathbb{R}^3$ (flexion--extension, radial--ulnar deviation, and pronation--supination; these wrist motions are defined in \cref{sec:supp_biomechanics}). Together, they form the full kinematic state $q_i = (\beta_i, \omega_i) \in\mathbb{R}^{23}$. We ignore global translation and orientation --- neither correlates with the pressure signal --- and track only the wrist orientation relative to the elbow. Following UmeTrack, the hand pose animates the corresponding hand mesh via linear blend skinning (LBS).
The \textit{kinetics} are represented by a sequence of interaction force $\boldsymbol{\eta}:=\{\eta_i\}$ exerted by the hand. In theory, hand-object interaction produces a continuous spatial force distribution across the contact manifold of the hand. However, capturing such a continuous field is challenging and physically intractable. In practice, we discretize the hand's inner manifold into distinct sensing regions corresponding to the taxel layout of the tactile gloves used to produce the ground truth. In this work, we use $\eta_i\in\mathbb{R}^{65}$ (see \cref{sect:glove} for the taxel layout and the per-finger and per-fingertip aggregations).

\subsubsection{Problem Formulation}
\label{sect:problem}
Given wrist pressure measurements, we explore several learning variants to study the correlation between wrist pressure and hand state. First, we analyze the mapping from pressure array to hand pose and interaction force in isolation, \textit{i.e.,} learning $\widetilde{\boldsymbol{q}} = f_{\boldsymbol{\theta}}(\boldsymbol{P})$ and $\widetilde{\boldsymbol{\eta}} = f_{\boldsymbol{\theta}}(\boldsymbol{P})$, respectively.
The success of these variants leads us to further explore simultaneous estimation of hand pose and force, \textit{i.e.,} learning
$\widetilde{\boldsymbol{q}}, \widetilde{\boldsymbol{\eta}} = f_{\boldsymbol{\theta}}(\boldsymbol{P})$.
Since many vision-based hand tracking algorithms have already achieved impressive results~\cite{umetrack22nimble, zhang2020mediapipe},
whereas estimating interaction force is ill-posed for vision-based solutions and remains much less explored,
we also study whether conditioning force estimation on known hand pose improves performance. This formulates the learning as
$\widetilde{\boldsymbol{\eta}} = f_{\boldsymbol{\theta}}(\boldsymbol{P}, \boldsymbol{q})$.

\begin{figure*}[t]
  \centering
  \adjustbox{max width=\linewidth}{%
  \begin{tikzpicture}[
    node distance=0.4cm,
    block/.style={
      draw, rounded corners=8pt, minimum height=0.9cm, minimum width=1.2cm,
      align=center, font=\small, thick
    },
    hatchblock/.style={
      draw, rounded corners=8pt, minimum height=0.9cm, minimum width=1.2cm,
      align=center, font=\small, thick,
      pattern=north east lines, pattern color=teal!20
    },
    arrow/.style={-{Stealth[length=2mm]}, thick},
  ]

    \node[block] (input) {pressure\\array};
    \node[hatchblock, right=0.5cm of input] (linraw) {Linear Projection\\+ LayerNorm + GeLU};

    \node[block, below=0.4cm of input] (vel) {temporal\\derivative};
    \node[hatchblock, right=0.5cm of vel] (linder) {Linear Projection\\+ LayerNorm + Tanh};

    \node[inner sep=0pt, left=0.15cm of input, yshift=0.0cm] (sensorimg) {\includegraphics[height=1cm]{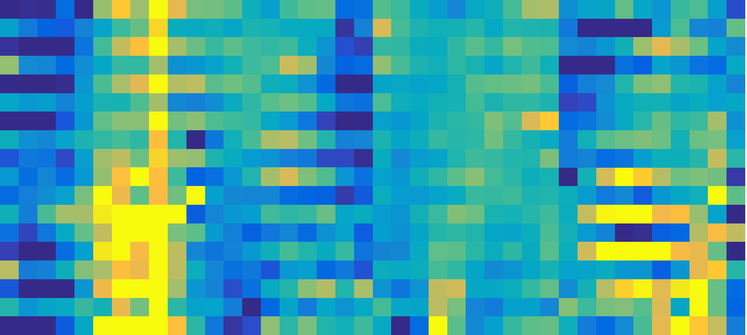}};

    \coordinate (midpoint) at ($(linraw.east)!0.5!(linder.east)$);
    \node[draw, circle, inner sep=1.5pt, thick] (concat)
      at ([xshift=0.5cm]midpoint) {$\oplus$};

    \node[hatchblock, right=0.4cm of concat] (dropout) {Dropout};
    \node[hatchblock, right=0.4cm of dropout] (gru) {GRU};
    \node[hatchblock, right=0.4cm of gru] (outproj) {Output\\Projection};

    \node[block, right=1.0cm of outproj] (output) {hand\\states};

    \node[draw, thick, rounded corners=10pt,
      inner xsep=0.3cm, inner ysep=0.1cm,
      fit=(linraw)(linder)(concat)(dropout)(gru)(outproj)] (bbox) {};
    \node[font=\normalsize, anchor=north east] at ([xshift=-0.15cm, yshift=-0.05cm]bbox.north east) {$f_{\boldsymbol{\theta}}$};

    \node[inner sep=0pt, right=0.0cm of output, yshift=0.15cm] (handimg)  {\includegraphics[height=2.0cm]{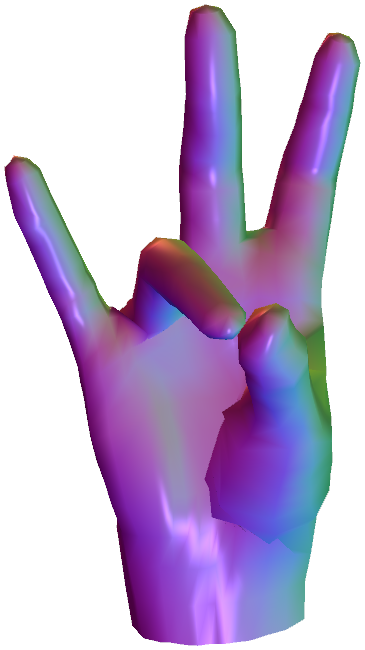}};
    \node[inner sep=0pt, right=1.2cm of output, yshift=0.15cm] (gloveimg) {\includegraphics[height=2.0cm]{figures/data-setup/projection-2d-hand.jpg}};

    \node[below=0.10cm of sensorimg, font=\normalsize] (plabel) {$\{\boldsymbol{P}_i\}$};
    \node[below=0.05cm of handimg,   font=\normalsize] {$\{\boldsymbol{\beta}_i, \boldsymbol{\omega}_i\}$};
    \node[below=0.05cm of gloveimg,  font=\normalsize] {$\{\boldsymbol{\eta}_i\}$};

    \draw[arrow] (input) -- (linraw);
    \draw[arrow] (input.south) -- (vel.north);
    \draw[arrow] (vel) -- (linder);

    \draw[arrow] (linraw.east) to[out=0, in=90] (concat.north);
    \draw[arrow] (linder.east) to[out=0, in=-90] (concat.south);

    \draw[arrow] (concat) -- (dropout);
    \draw[arrow] (dropout) -- (gru);
    \draw[arrow] (gru) -- (outproj);
    \draw[arrow] (outproj) -- (output);

  \end{tikzpicture}%
  }
  \caption{Learning pipeline. All variants are formulated as sequence-to-sequence prediction using the same GRU-based architecture. The input at each timestamp is the pressure array concatenated with its temporal derivative. The output can be hand poses (20 finger joint angles and 3 wrist angles), interaction forces, or both. We also study conditioning force estimation on known hand poses, in which case the hand pose is concatenated with the pressure array as input.}
  \label{fig:nn}
\end{figure*}

\subsubsection{Neural Network Architecture}
\label{sect:architecture}
To support all learning variants, we use a unified neural architecture for temporal modeling, illustrated in \cref{fig:nn}. Given an input sequence $\{\boldsymbol{P}_i\}$, we first compute its temporal derivative. The raw signal and its derivative are then passed through separate linear projection layers, each followed by LayerNorm and an activation: GeLU on the raw branch and Tanh on the derivative branch. Tanh is a natural fit for the derivative signal, since it is symmetric around zero and bounded, matching the symmetric, zero-mean distribution of finite differences. LayerNorm in both branches auto-tunes the per-channel scale of the inputs feeding into the GRU, which removes one source of per-recording variability and avoids manual feature normalisation. Although the pressure array has an explicit spatial layout of $w\times h$, we do not apply 2D convolutions; instead, each map is flattened into a 1D vector. In practice, this design works well in our setting and keeps ablations simple (e.g., dropping sensor rows or columns). Including temporal derivatives helped dynamic modeling in our earlier experiments, so we adopt it as the default.
The two processed branches are concatenated, passed through a Dropout layer, and fed into a recurrent neural network (RNN). In this work, we use gated recurrent units (GRUs)~\cite{gru14} as the sequence backbone. The GRU hidden states are linearly projected to the task-specific output dimension: 23 for hand pose ($q$), 65 for interaction force ($\eta$), and 88 for predicting pose and force together. For pose-conditioned force estimation, we concatenate hand pose with the pressure input and keep the rest of the pipeline unchanged.

To train the network, we adopt mean squared error (MSE) for supervised learning. For simultaneous estimation of hand pose and interaction force, the loss is weighted by modality, decomposing the kinematic state $q_i = (\beta_i, \omega_i)$ into separate finger and wrist terms:
\begin{equation}
  \mathcal{L} = \frac{1}{T} \sum_{i=1}^{T} \left(
    \lambda_{\beta} \left\lVert \widetilde{\boldsymbol{\beta}}_i - \boldsymbol{\beta}_i \right\rVert_2^2
    + \lambda_{\omega} \left\lVert \widetilde{\boldsymbol{\omega}}_i - \boldsymbol{\omega}_i \right\rVert_2^2
    + \lambda_{\eta} \left\lVert \widetilde{\boldsymbol{\eta}}_i - \boldsymbol{\eta}_i \right\rVert_2^2
  \right)
\end{equation}
where $\boldsymbol{\beta}_i$ and $\boldsymbol{\omega}_i$ are the finger and wrist joint angles defined above. Separating the two allows independent weighting, since wrist rotation produces a much stronger signal in the pressure array than finger articulation.

We train and run this architecture with a sliding-window scheme over the pressure sequence; the context length, supervision policy, and windowed averaging are detailed in \cref{sect:training-eval}. The model is lightweight enough to maintain real-time performance.

\subsection{Data Collection}
\label{sect:data}

We collect three datasets to support our experiments. A summary is provided in \cref{tab:dataset-summary}. Each recording is a separate donning session: participants put on the wristband at the start and remove it afterwards, so consecutive recordings differ in band placement and tension. When donning the wristband, we place the D-ring at the center of the dorsal distal forearm and set strap tension by visual and tactile inspection. Participants are asked to keep consistent wristband tension across sessions, and we maintain approximately the same duration per recording. Below we describe the two main collections; a third, auxiliary Fingertip-Force (FF) dataset is used as additional training data in several ablations and is detailed in the appendix (\cref{sec:supp_data}). After collection, all sensor streams are time-aligned and resampled onto a common 50~Hz timeline, with per-taxel calibration and baseline correction applied to the tactile-glove force. Representative samples, the full per-dataset recording protocols, and these processing steps are also provided in the appendix.

\begin{table}[!ht]
  \centering
  \small
  \caption{Summary of collected datasets: HP (hand pose), FF (fingertip force), and HOM (hand-object manipulation). For each, we mark the available hand-pose (hp.) and force ground truth, and list participants (ppl.), sessions, and total duration. The lower block breaks HOM down by participant; the single HP/FF participant is P4 of HOM.}
  \label{tab:dataset-summary}
  \begin{tabular}{l c c r r r}
    \toprule
    \textbf{dataset} & \textbf{hp.} & \textbf{force} & \textbf{\# ppl.} & \textbf{\# session} & \textbf{dur. (min)} \\
    \midrule
    \textbf{HP}  & \checkmark & \textbf{--} & \textbf{1 (P4)} & \textbf{20} & \textbf{415} \\
    \textbf{FF}  & \textbf{--} & \textbf{finger tip}  & \textbf{1 (P4)} & \textbf{18} & \textbf{348} \\
    \textbf{HOM} & \checkmark & \textbf{full hand} & \textbf{4} & \textbf{55} & \textbf{1160} \\
    \midrule
    HOM-P1 & \checkmark & full hand & 1  & 19 & 366 \\
    HOM-P2 & \checkmark & full hand & 1  & 11 & 285 \\
    HOM-P3 & \checkmark & full hand & 1  & 15 & 271 \\
    HOM-P4 & \checkmark & full hand & 1  & 10 & 238 \\
    \bottomrule
  \end{tabular}
\end{table}

\subsubsection{Data Collection Focused on Hand Poses (HP)}
\label{sect:data-hp}
The first collection is a single-participant dataset focused on hand poses. During data collection, the participant wears the pressure wristband along with the thin marker glove and forearm marker tree required for motion capture (\cref{sect:mocap}); no tactile glove is worn, since HP targets pose only. Their hand motion is recorded by MoCap. We emphasize diverse finger motion in combination with different wrist rotation, since we observe that wrist rotation produces a strong signal in wrist pressure. In total, we collect 20 recordings of total 415~min data, with each recording between 16--24 minutes in duration.

\subsubsection{Data Collection Focused on Hand-Object Manipulation (HOM)}
\label{sect:data-hom}
In the third collection, we focus on all available sensing modalities and on hand-object manipulation scenarios. Four participants contribute to this dataset. They wear the pressure wristband together with a PPS tactile glove instrumented with optical markers for MoCap hand tracking. Each recording session contains two activity blocks: \textit{pick-and-place} of everyday objects on table tops, and \textit{object manipulation}. We choose a wide set of everyday objects, e.g., cans, water bottles, toys, boxes, drill, hammer, etc. Object manipulation follows their natural functionality, e.g.,  opening and closing bottle caps, in-hand object inspection, simulated tool usage, etc. During data collection, participants are instructed to maintain firm, stable grasps and apply natural unscripted motion.

\section{Experiments and Results}
\label{sect:results}

We present results in two parts: hand pose estimation and force estimation, using the data collected in \cref{sect:data}. For hand pose, we report results on the HP and HOM datasets. Force results are reported on HOM; the auxiliary FF dataset (\cref{sec:supp_ff}) additionally serves as a reference baseline in the analysis below.

\textbf{Cross-user generalization.} All results in this section use \emph{per-user} models (\cref{sect:eval-protocol}) --- partly by design (a companion device supplies user-specific pose), and partly because our cohort is too small to test transfer: one of the four HOM users wears a different-sized band, leaving only a train-on-two, test-on-one split. Under it, pre-training on the other users helps neither directly nor as initialization for per-user adaptation. We read this as a limitation of a small, hardware-mixed cohort rather than a property of the modality, and leave a large-\emph{N}, hardware-matched pre-train-and-adapt study to future work.

\subsection{Evaluation Protocol}
\label{sect:eval-protocol}

\subsubsection{Metrics.}
\label{sect:metrics}

We evaluate each estimated scalar quantity---individual joint angle, wrist angle, or force channel---with three complementary metrics. \textbf{Mean absolute error (MAE)} reports the average magnitude of the prediction error, in degrees for joint angles, Newtons for force, and PSI for taxel pressure. The \textbf{coefficient of determination ($R^2$)} measures the fraction of ground-truth variance the predictions explain, on a scale where $1$ is perfect and $0$ matches simply predicting the mean (it can go negative for worse-than-mean predictions). Finally, the \textbf{ground-truth standard deviation ($\sigma$(GT))} reports the spread of the signal itself and serves as a reference scale for MAE: an MAE approaching $\sigma$(GT) indicates the model captures little beyond the marginal distribution. Formal definitions are given in the appendix (\cref{sec:supp_metrics}).

\textbf{Position error.} For hand pose, we additionally report position error (mm), computed via forward kinematics from predicted joint angles using the skeletal model from the motion capture system. We report both the fingertip position error and the mean per-joint position error (MPJPE, the mean Euclidean landmark distance over the hand skeleton); precise joint definitions are given in the appendix (\cref{sec:supp_metrics}). Both come in two variants: ``No Wrist'' uses the ground-truth wrist orientation as the root, isolating finger-articulation error from wrist-prediction error, while ``With Wrist'' uses the predicted wrist orientation as the root, capturing the full end-to-end error. These position metrics are only meaningful at finger granularity.

\textbf{Aggregation across DOFs.} Aggregating these metrics across DOFs requires care to avoid bias; we detail our conventions in the appendix (\cref{sec:supp_aggregation}).

\subsubsection{Training and Evaluation}
\label{sect:training-eval}

Unless otherwise stated, we train and evaluate \emph{per-user} models, as motivated above.

For most experiments, we use leave-one-recording-out cross-validation within each dataset: for each held-out recording, we train on all remaining recordings and report the average error over all held-out folds. Because every recording is a separate donning of the band (worn and removed between recordings, \cref{sect:data}), each held-out fold is evaluated on a different don--doff session than any used for training; all reported numbers are therefore cross-session, capturing the placement and strap-tension variability that arises from re-wearing the device. We use mean L2 loss per output task and assign relative weights for each task; the loss formulation is detailed in \cref{sect:architecture}.

Unless otherwise noted, training uses Adam (learning rate $3\times10^{-4}$, weight decay $10^{-2}$), dropout $0.3$, batch size 128, a context length of $N=100$ samples (2~s at 50~Hz), and 50k iterations. Supervision is applied only to the last $N/2$ outputs of each window, so the model has sufficient warm-up context. At inference we use a sliding window of stride 1 and average the overlapping per-timestamp predictions, and the reported results ensemble 20 checkpoints from the final 10k iterations. Further training details are given in the appendix (\cref{sec:supp_training}).

\subsubsection{Data augmentation}
\label{sect:data-augmentation}

Although we treat wristband taxels as independent channels, the raw signals remain sensitive to practical recording variability, especially placement and strap tension across sessions. To improve robustness, we apply five augmentations to the wristband input during training.

The five transforms each model a session-to-session nuisance: small vertical and horizontal shifts (proximal--distal placement and circumferential rotation), in-plane rotation (sensor misalignment), isotropic zoom (day-to-day wrist-contour variation), and global intensity scaling (strap-tension changes). No augmentation is applied at inference. The specific magnitudes and resampling details are given in the appendix (\cref{sec:supp_augmentation}).

We apply this augmentation in all results that follow. A detailed comparison of performance with and without augmentation, together with a per-target search over the augmentation ranges, is provided in the appendix (\cref{sec:supp_augmentation}).

\subsection{Hand Pose Estimation}
\label{sect:results-hand-pose}

For this problem we use two of our datasets: Hand-Pose (HP) and Hand-Object Manipulation (HOM). As mentioned, we only do per-user training/test, and the HP dataset has a single user, who also happens to be one of the users in the HOM dataset. Moreover, the two data distributions differ substantially (object manipulation and force application vs.\ free finger/wrist movement), so we train and evaluate per-dataset. We set the relative wrist-angle loss weight to 0.4, which balances wrist and finger accuracy (see the ablation of relative weights in \cref{sec:supp_loss_weights}).

\subsubsection{Evaluation on Hand Pose dataset}
\label{sect:results-hp}

\begin{table}[t]
  \centering
  \caption{Joint-angle and position errors for the combined wrist-and-finger model on the single-user Hand Pose (HP) dataset. See \cref{sect:metrics} for metric details; the three wrist DOFs (pronation, flexion, deviation) are defined in \cref{sec:supp_biomechanics}.}
  \label{tab:joint_angle_errors}
  \small
  \setlength{\tabcolsep}{3pt}
  \begin{tabular}{ll c cc !{\hskip 3pt\vrule\hskip 3pt} cc}
    \toprule
    \textbf{Category} & \textbf{DOF} & $\sigma$(GT)$^\circ$ &
      MAE$^\circ$ & $R^2$ &
      \multicolumn{2}{c}{\textbf{Tip\,/\,MPJPE (mm)}} \\
    \midrule
    \multirow{4}{*}{Wrist}
      & Pronation     & 21.62 & 6.39 & 0.851 & & \\
      & Flexion       & 19.07 & 4.04 & 0.919 & & \\
      & Deviation     & 9.05  & 3.83 & 0.684 & & \\
    \cmidrule(lr){2-7}
      & \textit{Mean} & \textit{19.48} & \textit{4.75} & \textit{0.863} & {\footnotesize No Wrist} & {\footnotesize With Wrist} \\
    \midrule
    \multirow{6}{*}{Finger}
      & Thumb         & 15.51 & 4.64 & 0.802 & 10.8\,/\,6.0 & 20.7\,/\,13.9 \\
      & Index         & 21.47 & 4.61 & 0.846 & 10.8\,/\,7.7 & 21.3\,/\,17.1 \\
      & Middle        & 23.41 & 4.58 & 0.886 & 11.3\,/\,8.1 & 20.6\,/\,16.3 \\
      & Ring          & 23.27 & 4.53 & 0.878 & 11.0\,/\,8.0 & 19.1\,/\,15.0 \\
      & Pinky         & 21.81 & 4.45 & 0.854 & 9.5\,/\,6.6  & 18.2\,/\,14.0 \\
    \cmidrule(lr){2-7}
      & \textit{Mean} & \textit{21.65} & \textit{4.56} & \textit{0.859} & \textit{10.7\,/\,7.2} & \textit{20.0\,/\,15.3} \\
    \bottomrule
  \end{tabular}
\end{table}

\Cref{tab:joint_angle_errors} reports MAE and $R^2$ for each wrist angle, the mean joint-angle error across the four joints in each finger, and fingertip position error in millimeters. The appendix expands these into a per-joint breakdown for every finger (\cref{tab:per_joint_errors}) and a comparison of model variants---finger and wrist estimated jointly, each alone, and finger conditioned on wrist (\cref{tab:kinematics_comparison}).

Notably, the thumb shows a slightly lower $R^2$ ($0.802$ vs.\ a $0.859$ finger average). A plausible explanation is that thumb motion relies more on intrinsic/thenar muscles in the hand, which are less observable at the distal forearm than extrinsic finger flexor and extensor activity.

\begin{figure}[t]
  \centering
  \includegraphics[width=\columnwidth]{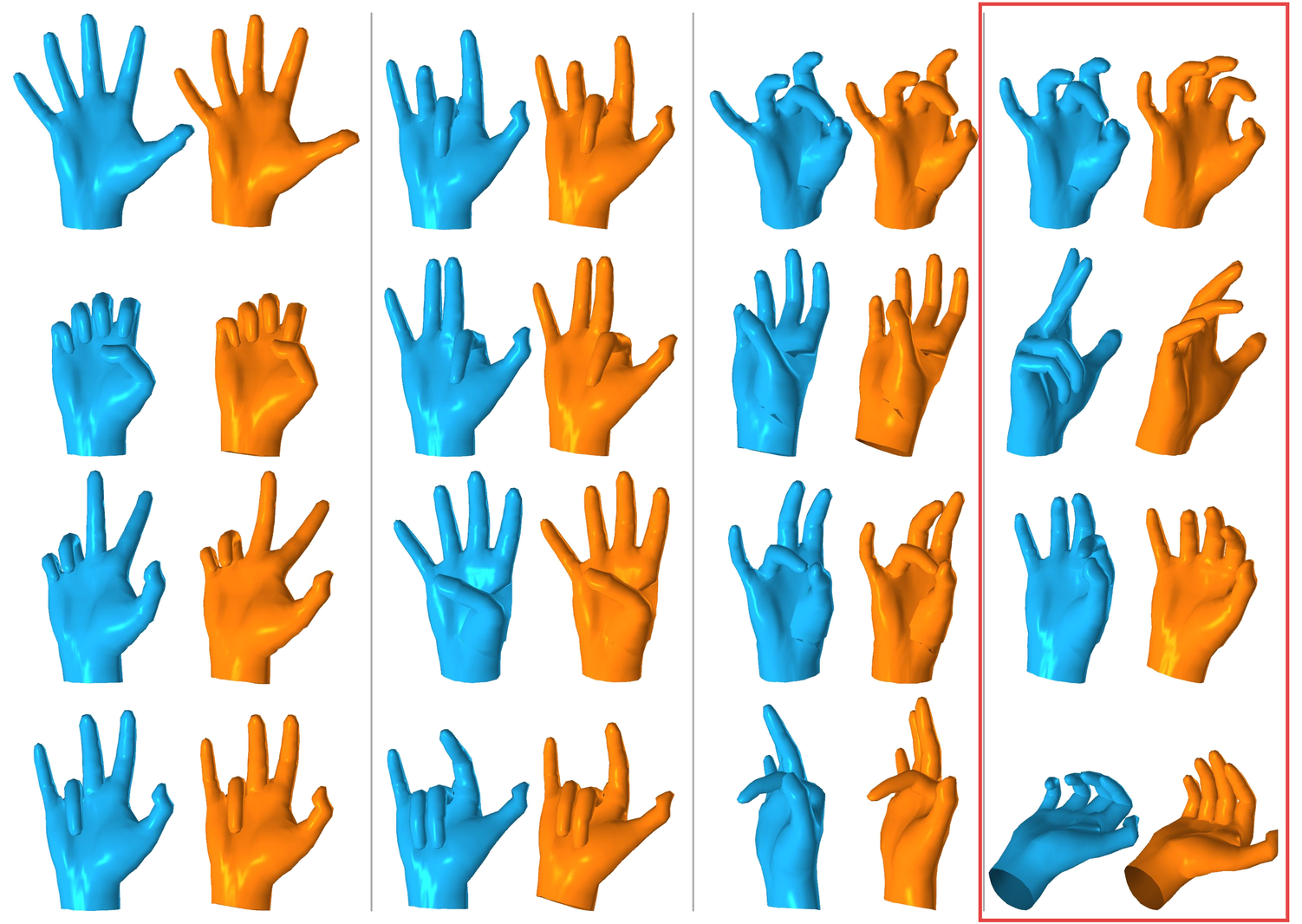}
  \caption{Sample visualization of hand poses (blue = ground truth, orange = prediction). The last column (highlighted) shows examples where the estimate diverges more noticeably from the ground truth. A larger set with per-vertex error heat maps is given in \cref{fig:hand-pose-selected-full}.}
  \label{fig:hand-pose-results}
\end{figure}

\Cref{fig:hand-pose-results} shows representative configurations (ground truth in blue, prediction in orange). Predictions are usually close, but occasional pronounced errors --- missed flexions or wrong-finger attributions --- tend to coincide with wrist rotation during the finger motion, since wrist rotation produces a much stronger pressure signal than finger articulation. These cases are worst where the wrist's range of motion is under-sampled in training: the model must then interpolate across wrist orientations, which it does only partially (\cref{sec:supp_wrist_analysis}).

We also evaluate pose on the HOM dataset, where the bulkier tactile glove and cluttered object interaction make the MoCap ground truth noticeably noisier (marker occlusion, dropped frames); the model nonetheless recovers hand pose together with force there. We report these per-user results in the appendix (\cref{sec:supp_hom_pose}): \cref{tab:pps_kinematics_summary} summarizes per-user HOM kinematics and \cref{tab:pps_kinematics} gives the full per-user, per-DOF breakdown.

\subsection{Force Estimation Results}
\label{sect:results-force}

\subsubsection{Evaluation on the Hand Object Manipulation Dataset}
\label{sect:results-hom-force}

This dataset contains 4 users with 10--20 recordings per user, evaluated per-user. \Cref{tab:pilot_force} summarizes force-estimation errors across targets (whole finger and per-taxel) and users.

\providecommand{\improv}[1]{{\color{green!60!black}\scriptsize #1}}
\providecommand{\degrad}[1]{{\color{red!70!black}\scriptsize #1}}

\newif\ifhomforcefull
\homforcefullfalse

\ifhomforcefull
\begin{table}[t]
  \centering
  \caption{Force prediction error across model configurations and users on the HOM dataset, for the per-finger, per-fingertip, and per-taxel granularities (with augmentation). We compare two configurations: \textit{Force+Pose} (estimating force and joint angles together) and \textit{Pose Input} (force estimation with joint angles as input). Percentages in the Pose Input column show MAE change relative to the in-row Force+Pose baseline. Finger and fingertip force are in Newtons; per-taxel force is in PSI.}
  \label{tab:pilot_force}
  \small
  \setlength{\tabcolsep}{4pt}
  \begin{tabular}{ll c cc !{\hskip 4pt\vrule\hskip 4pt} r@{\;}l c}
    \toprule
    & & &
      \multicolumn{2}{c}{Force+Pose} &
      \multicolumn{3}{c}{Pose Input} \\
    \cmidrule(lr){4-5} \cmidrule(lr){6-8}
    \textbf{Type} & \textbf{User} & $\sigma$(GT) &
      MAE & $R^2$ &
      \multicolumn{2}{c}{MAE} & $R^2$ \\
    \midrule
    \multirow{5}{*}{\shortstack[l]{Finger\\(N)}}
      & P1 & 2.86 & 0.83 & 0.42 & 0.57 & \improv{$-$31\%} & 0.70 \\
      & P2 & 2.16 & 0.52 & 0.67 & 0.45 & \improv{$-$13\%} & 0.76 \\
      & P3 & 1.83 & 0.60 & 0.64 & 0.44 & \improv{$-$27\%} & 0.79 \\
      & P4 & 1.62 & 0.31 & 0.77 & 0.28 & \improv{$-$10\%} & 0.81 \\
    \cmidrule(lr){2-8}
      & \textit{Mean} & \textit{2.12} & \textit{0.59} & \textit{0.57} & \textit{0.45} & \improv{$-$24\%} & \textit{0.75} \\
    \midrule
    \multirow{5}{*}{\shortstack[l]{Fingertip\\(N)}}
      & P1 & 2.22 & 0.64 & 0.42 & 0.46 & \improv{$-$28\%} & 0.67 \\
      & P2 & 1.85 & 0.39 & 0.67 & 0.34 & \improv{$-$13\%} & 0.74 \\
      & P3 & 1.35 & 0.44 & 0.63 & 0.34 & \improv{$-$23\%} & 0.76 \\
      & P4 & 1.41 & 0.28 & 0.73 & 0.25 & \improv{$-$11\%} & 0.78 \\
    \cmidrule(lr){2-8}
      & \textit{Mean} & \textit{1.71} & \textit{0.45} & \textit{0.57} & \textit{0.35} & \improv{$-$22\%} & \textit{0.72} \\
    \midrule
    \multirow{5}{*}{\shortstack[l]{Taxel\\(PSI)}}
      & P1 & 3.02 & 0.69 & 0.30 & 0.53 & \improv{$-$23\%} & 0.55 \\
      & P2 & 1.96 & 0.25 & 0.53 & 0.22 & \improv{$-$12\%} & 0.62 \\
      & P3 & 2.29 & 0.53 & 0.54 & 0.42 & \improv{$-$21\%} & 0.67 \\
      & P4 & 1.35 & 0.18 & 0.64 & 0.16 & \improv{$-$11\%} & 0.70 \\
    \cmidrule(lr){2-8}
      & \textit{Mean} & \textit{2.15} & \textit{0.43} & \textit{0.44} & \textit{0.35} & \improv{$-$19\%} & \textit{0.61} \\
    \bottomrule
  \end{tabular}
\end{table}

\else
\begin{table}[t]
  \centering
  \caption{Force prediction error across model configurations and users on the HOM dataset, for the per-finger and per-taxel granularities (with augmentation). We compare two configurations: \textit{Force+Pose} (estimating force and joint angles together) and \textit{Pose Input} (force estimation with joint angles as input). The final row (\textbf{FF}) reports the auxiliary Fingertip-Force dataset as a reference (single user P4; per-finger force in N, estimated from pressure without pose), so its Pose-Input entry is empty; see \cref{tab:force5} for the per-finger breakdown.}
  \label{tab:pilot_force}
  \small
  \setlength{\tabcolsep}{4pt}
  \begin{tabular}{ll c cc !{\hskip 4pt\vrule\hskip 4pt} r@{\;}l c}
    \toprule
    & & &
      \multicolumn{2}{c}{Force+Pose} &
      \multicolumn{3}{c}{Pose Input} \\
    \cmidrule(lr){4-5} \cmidrule(lr){6-8}
    \textbf{Type} & \textbf{User} & $\sigma$(GT) &
      MAE & $R^2$ &
      \multicolumn{2}{c}{MAE} & $R^2$ \\
    \midrule
    \multirow{5}{*}{\shortstack[l]{Finger\\(N)}}
      & P1 & 2.86 & 0.83 & 0.42 & 0.57 & \improv{$-$31\%} & 0.70 \\
      & P2 & 2.16 & 0.52 & 0.67 & 0.45 & \improv{$-$13\%} & 0.76 \\
      & P3 & 1.83 & 0.60 & 0.64 & 0.44 & \improv{$-$27\%} & 0.79 \\
      & P4 & 1.62 & 0.31 & 0.77 & 0.28 & \improv{$-$10\%} & 0.81 \\
    \cmidrule(lr){2-8}
      & \textit{Mean} & \textit{2.12} & \textit{0.59} & \textit{0.57} & \textit{0.45} & \improv{$-$24\%} & \textit{0.75} \\
    \midrule
    \multirow{5}{*}{\shortstack[l]{Taxel\\(PSI)}}
      & P1 & 3.02 & 0.69 & 0.30 & 0.53 & \improv{$-$23\%} & 0.55 \\
      & P2 & 1.96 & 0.25 & 0.53 & 0.22 & \improv{$-$12\%} & 0.62 \\
      & P3 & 2.29 & 0.53 & 0.54 & 0.42 & \improv{$-$21\%} & 0.67 \\
      & P4 & 1.35 & 0.18 & 0.64 & 0.16 & \improv{$-$11\%} & 0.70 \\
    \cmidrule(lr){2-8}
      & \textit{Mean} & \textit{2.15} & \textit{0.43} & \textit{0.44} & \textit{0.35} & \improv{$-$19\%} & \textit{0.61} \\
    \midrule
    FF\,(N) & P4 & 1.28 & 0.33 & 0.70 & \multicolumn{3}{c}{---} \\
    \bottomrule
  \end{tabular}
\end{table}
\fi

\Cref{tab:pilot_force} compares two configurations: \emph{Force+Pose} (jointly predicting force and pose) and \emph{Pose Input} (force estimation given pose as an input). Whole-finger force is reported in Newtons and full tactile-sensor regression in PSI. The relative force-loss weight $\lambda_F$ is selected per granularity from the sweep in \cref{sec:supp_loss_weights} ($\lambda_F{=}0.4$ for per-finger, $\lambda_F{=}1.2$ for per-taxel/glove). A per-user, per-finger breakdown of these force errors is given in the appendix (\cref{tab:pilot_force_per_finger}).

The table shows that externally provided kinematic cues improve force decoding. This is practically relevant because hand pose can often be obtained from vision pipelines, while force is harder to recover from vision alone. In contrast, predicting joint angles and force together --- or conditioning force on the model's own pressure-derived pose estimate --- does not significantly improve force estimation, whereas using an externally provided pose as model input yields a clear benefit. This suggests that explicit kinematic inputs help disambiguate finger-motion effects from true force-production effects.

\begin{figure}[t]
  \centering
  \begin{tikzpicture}[inner sep=0pt]
    \node (img) {\includegraphics[width=\columnwidth]{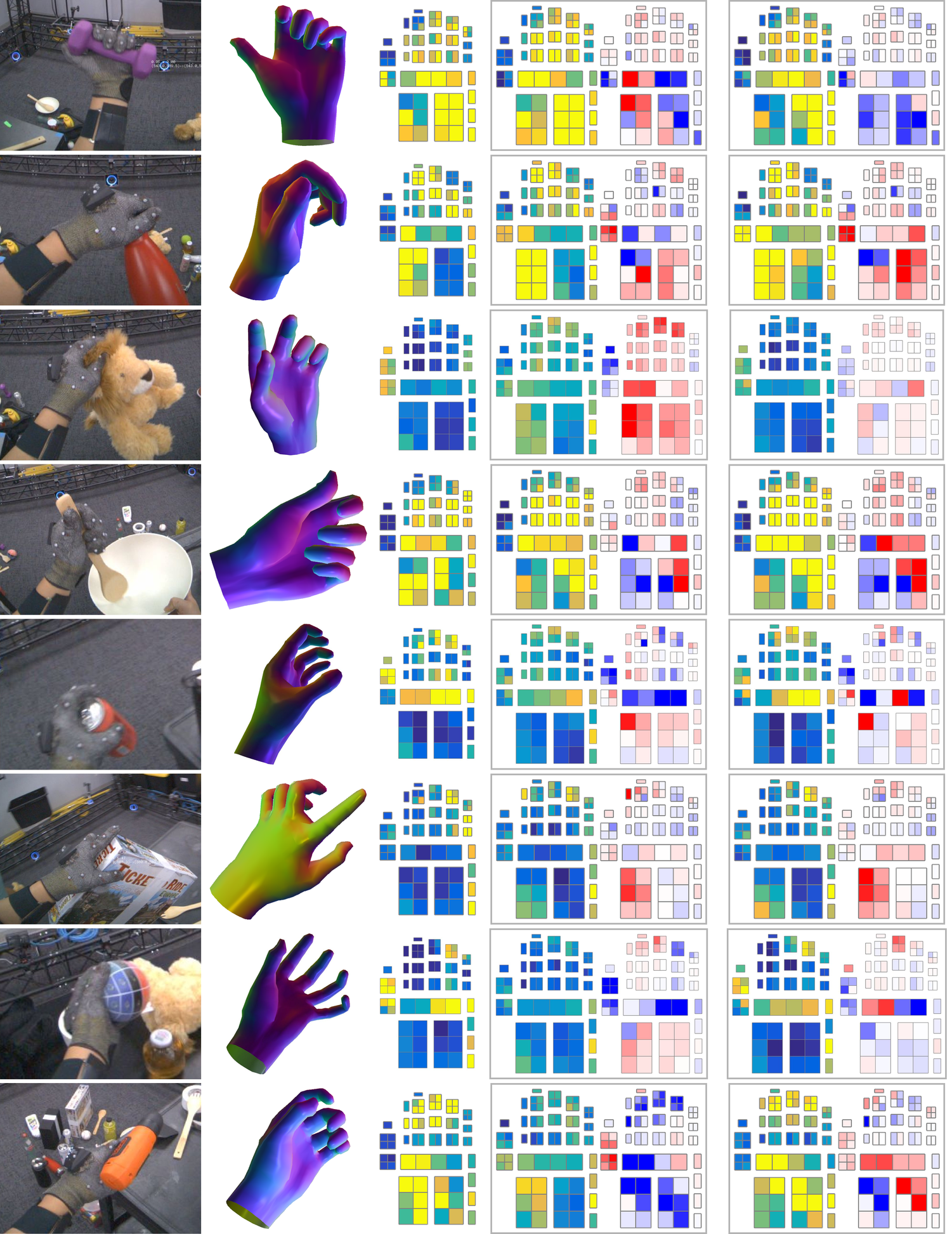}};
    \foreach \x/\lbl in {%
        0.11/{RGB Scene},%
        0.30/{Hand\\Pose},%
        0.45/{Ground\\Truth},%
        0.62/{Prediction --- Error\\Pressure Wristband},%
        0.87/{Prediction --- Error\\Wristband + Pose}%
    } {
      \node[font=\fontsize{6pt}{7pt}\selectfont\sffamily, anchor=north, align=center,
            inner sep=0pt, outer sep=1pt]
        at ([xshift=\x*\columnwidth, yshift=-2pt]img.south west) {\lbl};
    }
  \end{tikzpicture}
  \caption{Sample visualization of force predictions on the HOM dataset. Each row is one timestep; columns, from left to right: (1) RGB image of the scene; (2) MoCap-extracted 3D hand shape; (3) ground-truth tactile-glove pressure map; (4) predicted pressure map and the corresponding error map (pred~$-$~GT) for the \emph{pressure-wristband-only} model; and (5) the same prediction--error pair for the \emph{wristband + pose} model. In the error maps, maximum red and maximum blue correspond to $-1$~N and $+1$~N of error, respectively. Note the substantial improvement of the pose-conditioned model in several scenes.}
  \label{fig:hom-results}
\end{figure}

\begin{figure}[t]
  \centering
  \includegraphics[width=\columnwidth]{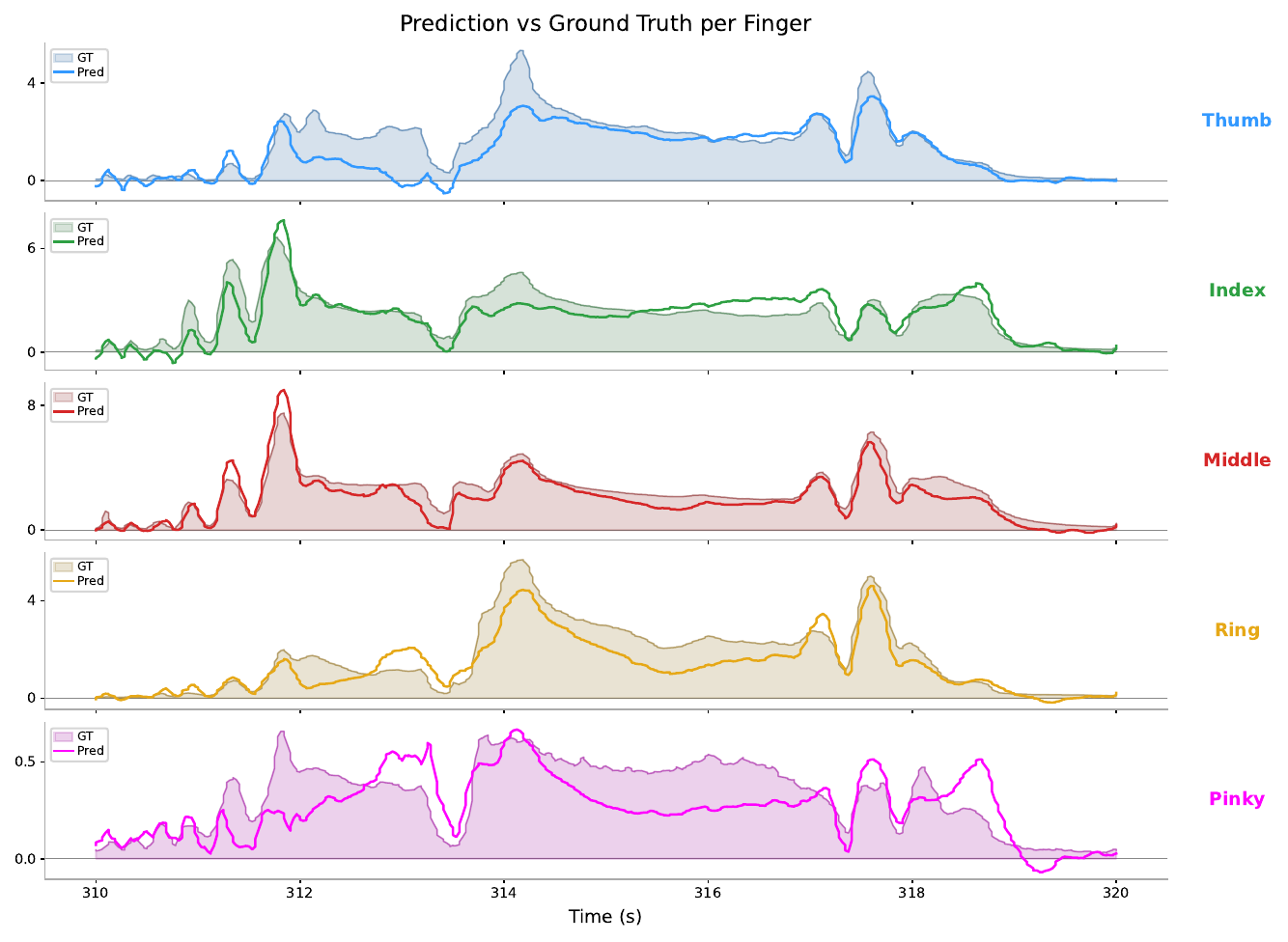}
  \caption{Sample time-series visualization of per-finger force prediction over a 10-second window of typical manipulation behavior. The predicted waveform tracks the overall shape of the ground-truth waveform, even when the absolute accuracy is imperfect.}
  \label{fig:finger-pred-vs-gt}
\end{figure}

\Cref{fig:hom-results} provides a qualitative view of these HOM force-estimation results by showing predicted glove-map projections alongside ground truth and per-map error, and \cref{fig:finger-pred-vs-gt} shows a sample graphical time-series visualization of a typical force prediction for each finger. Although the estimate does not match ground truth exactly, it consistently tracks its rises and falls. This is a result we observe throughout the datasets and across different tasks.

\subsubsection{Force error vs.\ ground-truth force magnitude}
\label{sect:results-force-binned}

To complement the aggregate HOM force MAE in \cref{tab:pilot_force} (and the auxiliary FF results in \cref{tab:force5}), we decompose the force error by ground-truth magnitude. For each (user, configuration), we bin samples by the magnitude of the ground-truth force and compute the mean $|pred - GT|$ within each bin.

\begin{figure}[t]
  \centering
  \includegraphics[width=\columnwidth]{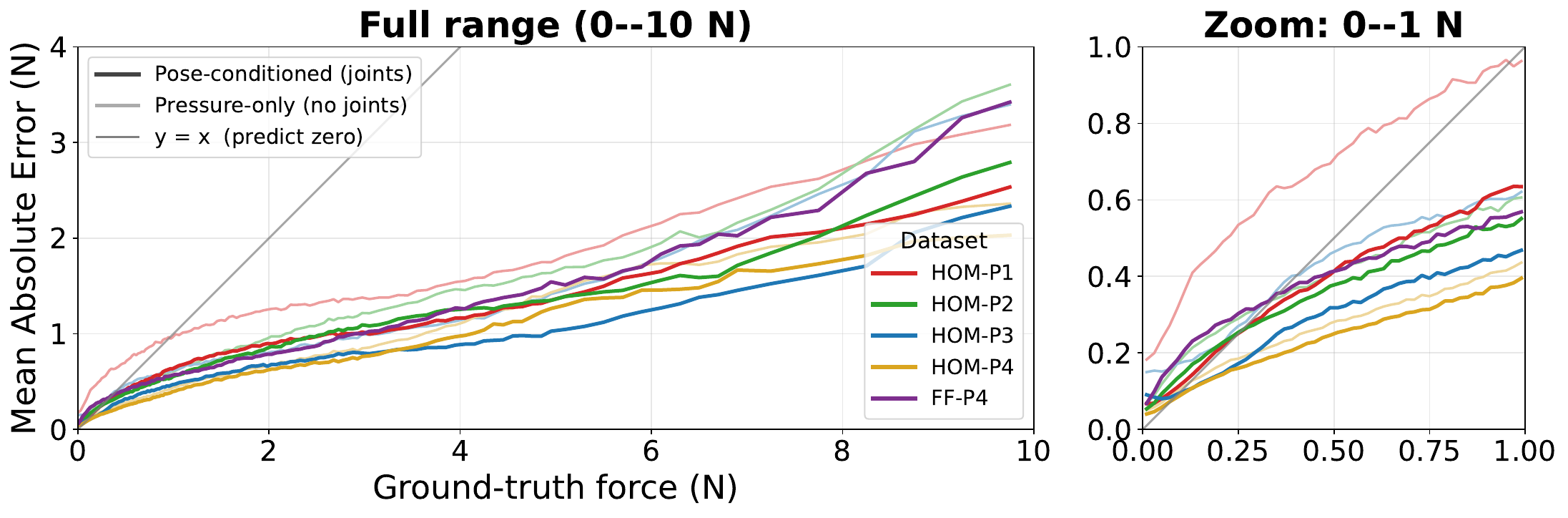}
  \caption{Per-bin force-prediction MAE versus ground-truth force, for the four HOM participants and the FF baseline. For HOM the pose-conditioned variant (pose input) is drawn thick and saturated, and the pressure-only variant (no pose input) thin and lightened in the same hue.}
  \label{fig:force-mae-vs-gt}
\end{figure}

\Cref{fig:force-mae-vs-gt} shows three consistent regimes. (i) Below ${\sim}0.3$~N, curves run close to the $y=x$ baseline: the wrist-pressure signal carries little information about very small forces, and the tactile glove itself cannot cleanly separate ``barely-touching'' from ``not touching,'' so this floor is a property of the modality--reference pair rather than a model deficiency. (ii) From ${\sim}0.3$ to ${\sim}5$~N, MAE grows sub-linearly, i.e.\ roughly constant \emph{relative} accuracy, and pose conditioning improves it throughout. (iii) Above ${\sim}5$~N curves steepen and inter-user spread widens as sample support in the tail thins. The per-user ordering is stable across all force levels (P4 best, then P3, P2, P1).

\subsection{Data Scaling}
\label{sect:results-data-scaling}
\label{sect:ablation-data}

We measure how prediction quality depends on the amount of training data by training on subsets of the available recordings. For each task (force on FF, kinematics on HP, per-user force on HOM), we use several held-out test sets (each ${\sim}20\%$ of recordings); for each, we repeatedly draw random $k$-recording training subsets from the remainder, using an equal number of draws at every subset size $k$. Each run is scored per \cref{sect:training-eval} and normalised by the corresponding full-training run; each plotted point aggregates all draws at a fixed $k$, placed on the $x$-axis at their mean total training duration.

\begin{figure}[t]
  \centering
  \begin{tikzpicture}[inner sep=0pt]
    \node (a) {\includegraphics[width=0.49\columnwidth]{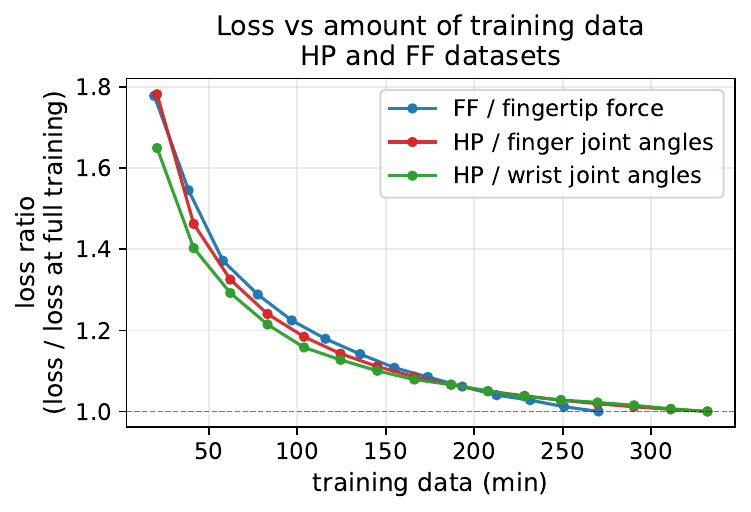}};
    \node (b) at (a.east) [anchor=west, xshift=2pt] {\includegraphics[width=0.49\columnwidth]{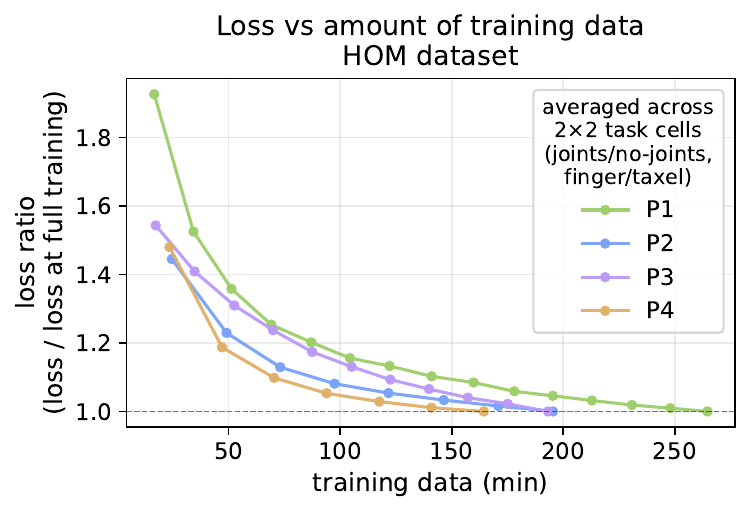}};
  \end{tikzpicture}
  \caption{Data-scaling curves. $y$ is validation $\ell_1$ loss normalised by the full-training run for the same held-out set; $x$ is training data in minutes. \emph{Left:} three tasks --- FF fingertip force, HP finger-joint kinematics, and HP wrist kinematics. \emph{Right:} the four HOM users (P1--P4), each averaged across the four HOM configurations. The dashed line at $y=1$ is the full-training baseline.}
  \label{fig:data-ablation}
\end{figure}

Both panels show a monotone, diminishing-returns trend: a single recording (${\sim}20$~min) yields ${\sim}1.5$--$1.8\times$ the full-training loss, two recordings recover most of the gap, and on HP the loss is within ${\sim}5\%$ of full-training once ${\sim}4$--$5$ recordings (${\sim}100$~min) are available. Crucially, even at the largest data budget the curves are still decreasing rather than flat --- performance has not converged, so more data would continue to improve accuracy. The shape holds across tasks and users and is essentially independent of the HOM configuration (pose/no-pose, finger/taxel). This behaviour is consistent with what large-scale sEMG studies report for participant-count scaling~\cite{salter2024emg2pose,kaifosh2025generic}, with the caveat that our sweep varies data \emph{within} a fixed user. Because single-user data does not saturate the model, pooling data across users and adapting to a new one would be a natural way to close the remaining gap --- though, as noted above, we find no such cross-user transfer in the present cohort.

\section{Discussion}
\label{sect:discussion}

We have shown that a wrist-worn capacitive pressure array recovers continuous full-hand pose and distributed contact force on a single wearable. The central finding is that the modality distal-forearm pressure supports both targets, separately or together, at a competitive accuracy (\cref{sect:results-hand-pose,sect:results-force}). The wristband works as a self-contained force sensor, and becomes a stronger force channel still when an external pose source --- e.g.\ a vision pipeline --- supplies kinematics that vision recovers but wrist pressure cannot directly. The ${\sim}25\%$ force-MAE reduction from pose conditioning is \emph{not} reproduced by predicting force alone, by predicting pose and force together, or by conditioning on the model's own pressure-derived pose estimate --- only an \emph{independent} pose signal helps. The gain therefore reflects informational access to pose rather than a multi-task or representation-learning effect.

The hardest target --- finger flexion and per-finger force --- is driven primarily by tendons that slide \emph{longitudinally} within the forearm, and longitudinal motion produces little signature at a circumferential pressure band (a sliding rod presents a nearly invariant cross-section along its length). What the wristband actually measures are the \emph{second-order} effects that accompany this motion: muscle bellies bulging, soft tissue stiffening under load, bones rotating at the wrist, and tensioned tendons pressing transversely against the retinacular pulleys. These transverse and bulk-deformation effects are what make wrist-side FMG work, and likely also bound it --- signals more distantly coupled to the underlying motion are more spatially diffuse and noise-prone. Wrist rotation is the strongest of them, and therefore also a confound: its signature can swamp the subtler patterns of individual finger flexion and contact force when the two co-occur. Finger error is in practice lowest near the centre of the wrist's range of motion and rises toward its sparsely sampled extremes (\cref{sec:supp_wrist_analysis}).

A distinguishing feature of our evaluation is that force is measured during \emph{natural object manipulation} with a tactile glove, rather than isometric exertions against a load cell or dynamometer in fixed postures, as in most wearable force work~\cite{sakr2019forcetorque,rahimi2024simulcontrol}. This is more representative of real use, but also harder and differently bounded: everyday hand--object contact occupies a low-dimensional, structured manifold, so part of our per-taxel accuracy reflects that structure rather than a fully general mechanical model. The reference is a bound as well: the tactile glove leaves the lateral, dorsal, and webbing surfaces uninstrumented, and we do not cross-check it against an independent reference (instrumented objects, force plates, or torque-sensorized fingers), which together cap per-taxel $R^2$ independently of model capacity. The pose-conditioned result, moreover, uses MoCap-derived hand pose as input --- a near-oracle that a real egocentric-vision pipeline would only approximate under the same occlusion --- so the ${\sim}25\%$ MAE reduction reported here is best read as an \emph{upper bound} a realistic-pose pipeline would target rather than immediately achieve, and closing that gap under vision-supplied pose is the natural next experiment. Pose is also not the only thing vision could supply. Coarse contextual cues --- whether the hand is holding an object, which object it is, how it is grasped --- are easier to recover reliably under occlusion than accurate 3D pose, and they constrain both the plausible contact geometry and the force range the model has to cover. Conditioning on such context rather than on pose alone is a plausible route to improving hand- and object-state estimation and force estimation together.

Our wristband is an FMG sensor: it measures the mechanical loading that accompanies hand action, not the muscle activity that produces it~\cite{zheng2022reviewemgfmgeit}. Surface EMG senses that activation directly and so registers static poses, isometric holds, and co-contraction --- activity with little external movement, exactly the regime where FMG's mechanical signal is weakest. FMG is not blind to these conditions --- they still deform the forearm, so some signal remains --- but a co-located sEMG channel captures them more directly, and EMG--FMG fusion outperforms either modality alone~\cite{rohr2025fmgemgfusion}.

Many downstream consumers do not need per-taxel reconstruction. Task-level abstractions --- contact onset, per-finger load, grasp phase, grip stiffness --- sit far closer to what a wrist band can observe than full per-taxel force does, and are naturally evaluated over the continuous temporal windows where our model is strongest. The most useful next datasets are therefore likely not ``larger and more natural'' in the abstract, but ones collected around a concrete consumer --- a contextual hand-state classifier, an imitation-learning policy, or a teleoperation pipeline~\cite{verma2025emg2tendon} --- so that the behaviours we capture match what the model will be asked to do. Scaling toward cross-user deployment, along the large-pretrain-plus-short-calibration trajectory of recent sEMG work~\cite{salter2024emg2pose,kaifosh2025generic}, is the natural path there.

\clearpage
\bibliographystyle{ACM-Reference-Format}
\bibliography{main}

\clearpage
\appendix
\section{Appendix}
\label{sect:appendix}


\subsection{Extended Background}
\label{app:background}

\subsubsection{Biomechanical Background}
\label{sec:supp_biomechanics}

At the anatomical level, finger and wrist motion is driven by both extrinsic muscles in the forearm and intrinsic muscles in the hand. Each degree of freedom is regulated by opposing muscle--tendon groups, whose differential activation produces a characteristic pressure pattern at the band--skin interface. The wrist contributes three rotational degrees of freedom, which recur throughout our results: \emph{flexion--extension} (bending the palm toward the inner forearm, or the back of the hand toward the outer forearm), \emph{radial--ulnar deviation} (tilting the hand within the plane of the palm toward the thumb or the little-finger side), and \emph{pronation--supination} (rotating the forearm so the palm turns to face downward or upward). Extrinsic muscle--tendon pathways often span multiple joints, and their loading transmits radially through the fascia to the skin, where it is measured by the wristband; forearm pronation--supination additionally changes wrist contour through relative motion of the radius and ulna, while wrist flexion--extension and radial--ulnar deviation are driven primarily by muscle--tendon actuation rather than skeletal reconfiguration. Bones and ligaments contribute little local pressure variation outside the pronation--supination case. Intrinsic muscle activity, in contrast, is largely invisible above the wrist, which sets a fundamental limit on how observable purely intrinsic actuation patterns --- including parts of thumb motion --- can be from any forearm-worn modality~\cite{salter2024emg2pose}. In practice, however, everyday hand behaviour rarely isolates intrinsic muscles, so the combined activation patterns still produce measurable mechanical signatures at the distal forearm.

\subsubsection{Why Pressure? The Surface-Deformation Regime of Wrist Modalities}
\label{sec:supp_why_pressure}

Beyond the biomechanical coupling that motivates a forearm-worn sensor (\cref{sect:overview}), a second observation motivates our specific choice of \emph{pressure} as the sensing modality --- and the EIT comparison of \cref{app:eit}. In early experiments with electric-based wristbands, such as EIT and sEMG~\cite{kyu2024eitpose, salter2024emg2pose, emg2qwerty}, we observed a consistent phenomenon: these devices rely on spring-based metal electrodes to maintain strong skin contact, and their measurements vary with the physical pressure exerted against the sensors. This motivates our hypothesis that, in tightly worn wrist-EIT used for hand-state inference, task-relevant measurements are often dominated by pressure-dependent electrode--skin contact impedance and near-surface deformation rather than only by conductivity changes deep within the forearm. Prior work supports the mechanism without claiming this dominance universally: Yao et al.~\cite{yao2020wearableeit} show that contact impedance is affected by pressure and contact area, changes with skin deformation, and improves gesture recognition; Lou et al.~\cite{lou2024advancing} describe measured impedance as a combination of contact-impedance change, internal tissue deformation, and outer-shape deformation. More generally, contact impedance is high, movement-dependent, and capable of producing EIT reconstruction errors~\cite{pennati2023eitsurvey}.

Under this hypothesis, tightly worn wrist-EIT and capacitive systems contain a mechanically mediated component that overlaps with the surface deformation measured directly by force myography (FMG); they are not physically equivalent, because EIT also responds to internal conductivity changes. Measuring pressure therefore isolates the mechanical component without the confound of an electrode--skin interface. We hypothesize that this component contains much of the information needed to decode hand state, and the matched-protocol EIT comparison in \cref{app:eit} provides direct empirical support: on the same participant, anatomy, and protocol, dense pressure sensing outperforms the EIT band on every pose axis.

\subsubsection{A History of Wrist- and Forearm-Worn Hand Sensing}
\label{sec:supp_related}

The compact taxonomy of \cref{sect:literature} groups prior work by sensing physics. Here we give the fuller, chronological account that space precludes in the main text, tracing how each modality matured and where our approach --- continuous full-hand pose and spatially resolved force from a dense capacitive pressure band --- fits the arc. We proceed modality by modality, beginning with force myography (FMG), the family our system belongs to.

\paragraph{Force myography: from prosthetic control to continuous sensing.}
FMG --- inferring hand state from the mechanical deformation of forearm tissue --- predates the term itself. Its origins lie in prosthetics: Abboudi et al.~\cite{abboudi1999tap} decoded individual finger intent in transradial amputees from foam pressure pads over residual tendons, and Phillips and Craelius~\cite{phillips2005rki} generalised this into \emph{residual kinetic imaging}, establishing that a ring of pressure sensors over the forearm carries per-finger information. In parallel, the wristband form factor for interaction emerged with GestureWrist~\cite{rekimoto2001gesturewrist}. Through the 2010s the modality was carried forward largely with force-sensitive resistors (FSRs): WristFlex~\cite{dementyev2014wristflex} demonstrated low-power always-on gesture input, and a substantial body of work from Menon and colleagues explored FMG for prosthetic and gesture control. The transducers then diversified --- barometric sensors and flexible capacitive arrays~\cite{truong2018capband, wang2023flexible} --- each producing a richer pressure ``image'' around the wrist. Two barometric results are especially relevant to our sparsity findings: Zhu et al.~\cite{zhu2018wrist} showed that ten sensors suffice to capture distinct tendon--muscle profiles, and Shull et al.~\cite{shull2019hand} reached $94\%$ gesture accuracy --- $90\%$ with only five optimally placed sensors --- with the flexor-tendon underside of the wrist identified as the most informative site. The overwhelming majority of this work, however, targets \emph{discrete gestures}. Continuous regression is rarer: Kadkhodayan et al.~\cite{kadkhodayan2016continuous} predicted fingertip displacement for three fingers from eight forearm FSRs (achieving correlations that exceeded contemporaneous sEMG), Shull et al.~\cite{shull2019hand} additionally regressed per-finger angles as single scalars ($R^2{=}0.79$), and more recent networks estimate finger joint angles from fused FMG signals~\cite{chen2024grufmg, kammarchedu2024realtime}. All of these recover only individual joint angles or per-finger scalars; the one system to regress a full articulated (MANO) hand from pressure, SleevePose~\cite{niu2026sleevepose}, does so from a 117-taxel full-forearm textile sleeve, but targets pose alone and degrades sharply across users. A recurring practical concern in this literature is session-to-session robustness --- FMG signals drift with donning and skin state~\cite{islam2019daytoday, wang2022interday} --- which motivates the strict cross-session, leave-one-donning-out protocol we adopt throughout. None of these systems recovers a full articulated hand model together with spatially distributed contact force, which is the gap our work addresses.

\paragraph{Electromyography: the best-benchmarked competitor.}
Surface electromyography (sEMG) senses the electrical correlates of muscle contraction and is the most mature wrist-worn modality. Interaction-oriented work began with Saponas et al.~\cite{saponas2008muscle}, who established forearm sEMG as a muscle--computer interface for finger gestures, and the field was subsequently anchored by public benchmarks --- most notably Ninapro~\cite{atzori2014ninapro} --- that enabled the deep-learning gesture-recognition literature. Continuous pose from sEMG followed: NeuroPose~\cite{liu2021neuropose} reconstructed 3D hand pose from an eight-channel forearm EMG armband, per user and stable across sensor remountings and days, and large-scale corpora have recently pushed the modality toward generalisation, with emg2pose~\cite{salter2024emg2pose} and emg2qwerty~\cite{emg2qwerty} providing hundreds of hours of data and generic neuromotor decoders reporting cross-user transfer~\cite{kaifosh2025generic}. sEMG is the natural channel for intrinsic and isometric activity that produces no external tissue displacement, and we regard it as complementary to --- not competing with --- surface-deformation sensing; its main practical costs are electrode drift and skin-condition sensitivity~\cite{salter2024emg2pose, kaifosh2025generic}.

\paragraph{Electrical impedance and bioimpedance.}
Electrical Impedance Tomography (EIT) reconstructs an internal conductivity map from boundary current injections. Tomo~\cite{zhang2015tomo} introduced wrist-worn EIT for gesture recognition, followed by higher-resolution and 3D variants, and the modality was extended to continuous pose by EITPose~\cite{kyu2024eitpose}; related bioimpedance and RF-reflection bands (EtherPose~\cite{kim2022etherpose}, EI-Lite~\cite{zhu2025eilite}) target pose and, more recently, pinch force. As we argue in \cref{sec:supp_why_pressure}, pressure-dependent contact impedance and tissue deformation can contribute substantially to wrist-EIT measurements~\cite{yao2020wearableeit,lou2024advancing,pennati2023eitsurvey}. We hypothesize that this mechanically mediated component often dominates the task-relevant signal in tightly worn hand-sensing bands, which makes EIT a useful internal control for our pressure results; the matched-protocol comparison in \cref{app:eit} tests that interpretation rather than assuming the two modalities are physically identical.

\paragraph{Ultrasound and acoustics.}
Ultrasound is the one wrist-morphology modality that genuinely penetrates tissue and resolves muscle and tendon structure. A-mode ultrasound classifies discrete finger gestures from four forearm transducers~\cite{yang2018amode} and sonomyographic control provides proportional, continuous outputs~\cite{dhawan2019sonomyography}, while contact ultrasound has been used for per-finger force~\cite{bimbraw2023ultrasoundforce}, joint-angle estimation~\cite{bimbraw2022config}, and fine finger-motion decoding~\cite{zadok2022ultrasound}, in each case from a forearm probe evaluated within-user. Its accuracy comes at the cost of bulk, power, and the need for acoustic coupling between transducer and skin. A distinct, non-imaging class of acoustic methods sidesteps that coupling entirely: instead of transmitting sound \emph{into} the tissue, a small speaker emits inaudible airborne signals (typically frequency-modulated continuous waves in the $20$--$29$\,kHz band) that travel across and around the wrist and reflect or diffract off the \emph{external} skin contour, with a nearby microphone capturing the resulting echo profile. These systems therefore measure the changing surface geometry of the hand and forearm --- not internal structure --- much as our pressure band does, but sensed at a distance through air rather than by mechanical contact. EchoWrist~\cite{lee2024echowrist} tracks $20$ finger joints this way (mean joint error $4.8$\,mm), and EchoForce~\cite{mahmoodi2025echoforce} regresses continuous grip force from flexor-driven surface deformation. Because a handful of air-coupled echoes carry far less spatial detail than a dense contact array, they recover coarser targets --- joint sets or aggregate force rather than per-taxel contact --- in exchange for a lighter, contact-free package.

\paragraph{Vision.}
Environment-mounted cameras remain the accuracy ceiling for hand tracking, with parametric models such as MANO~\cite{romero2017mano} as the representational backbone, but require instrumented spaces. Wrist-worn cameras bring this on-body --- Digits~\cite{kim2012digits} with an IR camera and, most recently, WristP$^2$~\cite{xi2026wristp2} with a fisheye camera achieving state-of-the-art wrist-worn pose --- but both depend on an unobstructed optical path to the fingers, exactly the line-of-sight constraint that surface sensing avoids.

\paragraph{Combined pose--force estimation and where we sit.}
The systems closest to ours estimate kinematics and kinetics together. Multi-task HD-EMG models predict wrist angle or full joint positions alongside grip force~\cite{li2024graphdriven, rahimi2024simulcontrol}, but rely on dense electrode grids far from a practical wearable; emg2tendon~\cite{verma2025emg2tendon} instead couples the two through a musculoskeletal model on the 16-channel emg2pose band. PiMForce~\cite{seo2024pimforce} showed that conditioning force estimation on 3D hand posture markedly improves prediction, using posture from an external camera and force from sEMG; we adopt the same pose-conditioning in our force model. Wrist2Finger~\cite{xiao2025wrist2finger} recovers per-finger pose and force from a watch plus an instrumented ring, and WristP$^2$~\cite{xi2026wristp2} adds per-vertex pressure but needs line-of-sight. Numerically (\cref{tab:comparison}), the wrist-worn full-hand pose landscape spans roughly an order of magnitude in MPJPE, from a fisheye camera with direct line-of-sight ($\sim$$3$\,mm) through eyes-free acoustic sensing ($\sim$$5$\,mm), our capacitive band ($7.2$\,mm), EIT ($\sim$$11$--$12$\,mm), and large-scale cross-user sEMG ($\sim$$16$\,mm). These figures are not a controlled comparison --- protocols and metrics differ in per-user versus cross-user evaluation, dataset scale, and joint set (for instance, NeuroPose is a per-user-calibrated median; emg2pose and EITPose are quoted at held-out-user and within-session operating points; Li et al.\ report a single wrist angle as Pearson $R$, shown as $R^2$; Wrist2Finger's force is a thumb-only best case; Rahimi's force is offline; Sakr is a combined 6-DoF $R^2$; and our own row pairs single-user HP pose with four-user HOM force) --- but they place dense capacitive pressure favourably among eyes-free modalities at a comparable within-user operating point, and, uniquely, our system recovers spatially resolved per-taxel contact force on the same wearable.

\subsection{Datasets}
\label{app:datasets}
\label{sec:supp_data}

\subsubsection{Recording Protocols}
\label{sec:supp_protocols}
This section documents the per-dataset recording protocols summarized in \cref{sect:data}. Across all datasets, the wristband is donned and doffed for every recording. During donning, we place the D-ring at the center of the dorsal distal forearm and set strap tension by visual and tactile inspection; participants are asked to keep tension consistent across sessions. For datasets that use the MoCap glove and forearm marker cluster, we place the forearm marker-cluster armband just below the elbow using the same visual procedure. Wristband placement and tension affect the measured signals, and we account for this variability during learning via data augmentation. Marker-cluster placement can also bias pronation--supination angle estimates; unlike wristband placement/tension effects, this error is not explicitly corrected.

\paragraph{Hand Pose (HP).}
\label{sec:supp_protocol_hp}
The participant wears a motion-capture glove, a forearm marker cluster (for 20 finger-joint angles plus 3 wrist angles), and the pressure wristband. The dataset emphasizes combined finger--wrist motion to disambiguate their overlapping effects on the wristband signal, where wrist motion is often dominant. The protocol includes full and partial finger flexion (distal to MCP), multi-finger combinations at different wrist orientations, thumb--finger tapping, micro-gestures (thumb swipes), and static hand poses with active wrist rotation. Motions are intentionally exaggerated and non-naturalistic to stress-test disambiguation. The dataset contains 20 recordings of 16--24 minutes each, totaling 6 hours 55 minutes. Sample frames are shown in \cref{fig:dataset-handpose}.

\paragraph{Fingertip Force (FF).}
\label{sec:supp_protocol_ff}
The participant wears both a PPS tactile glove and the pressure-sensing wristband. No MoCap is used, so this dataset contains no ground-truth kinematic information: only the wristband pressure input and the tactile-glove force target are recorded. This dataset is a stress test that maximally spans fingertip force, hand pose, and wrist pose, so that the model must learn fingertip force despite strong kinematic interference, especially from wrist motion, which strongly affects the wristband signal. Recorded behaviours include grabbing and holding everyday objects with different shapes and weights (roll of tape, earbuds case, paper cups, and a MacBook), clicking a computer mouse, and tapping/pressing on rigid and soft surfaces (desk and chair). During object interactions, the user applies squeezing forces with varying finger combinations (typically thumb plus one or more fingers) to promote finger-specific force disambiguation. The protocol also includes periods of constant-force grasping while rotating the wrist to encourage invariance to wrist motion. Most motions are intentionally exaggerated and non-naturalistic. The dataset contains 18 recordings, mostly around 20 minutes each, for a total of 5 hours 48 minutes. Sample frames are shown in \cref{fig:force5}.

\paragraph{Hand-Object Manipulation (HOM).}
\label{sec:supp_protocol_hom}
Participants wear all three sensing modalities: a PPS tactile glove instrumented with motion-capture markers, a forearm marker cluster, and the pressure wristband. This provides synchronized streams of hand pose, wrist pose, tactile force, and wristband pressure. Each recording contains two activity blocks: (1) pick-and-place of everyday objects between two tables, and (2) object use/manipulation while moving objects between tables, with the users alternating between the two activities throughout the recording. We use a diverse set of about 40 different objects, including household and office items such as aluminum cans, water bottles, plush toys, board-game boxes, a drill, a hammer, and a tape roll. Manipulation behaviours include opening/closing reusable bottle caps, in-hand reorientation while reading labels, simulated hammering, drinking gestures, and other in-hand rotation and repositioning tasks. Participants are instructed to maintain firm, stable grasps without using overly exaggerated or unnatural forces. The dataset includes 4 users and 55 recordings (total duration 19 hours 19 minutes). Most recordings are 21--28 minutes long; a small number of shorter recordings ($<$15 minutes) occur when tactile-glove connectivity issues require glove restart. Sample frames are shown in \cref{fig:dataset-manipulation}.

\subsubsection{Auxiliary Fingertip-Force (FF) Dataset and Results}
\label{sec:supp_ff}
The FF dataset is a single-participant, MoCap-free stress test of fingertip force under strong kinematic interference (full protocol above). It is not a headline result in the main paper, but serves as auxiliary training data and appears in several ablations (e.g.\ \cref{sect:ablation-data,sect:ablation-sensor}). Its standalone force-estimation results are reported in \cref{sec:supp_ff_results}.

\subsubsection{Dataset Samples}
\label{sec:supp_data_samples}
\begin{figure}[t]
  \centering
  \def\iw{0.195\columnwidth}
  \begin{tikzpicture}[inner sep=1pt]
    \node(p01)                            {\includegraphics[width=\iw]{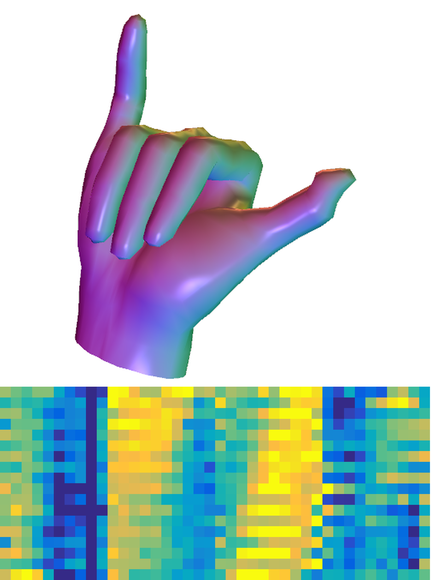}};
    \node(p02) at (p01.east) [anchor=west]{\includegraphics[width=\iw]{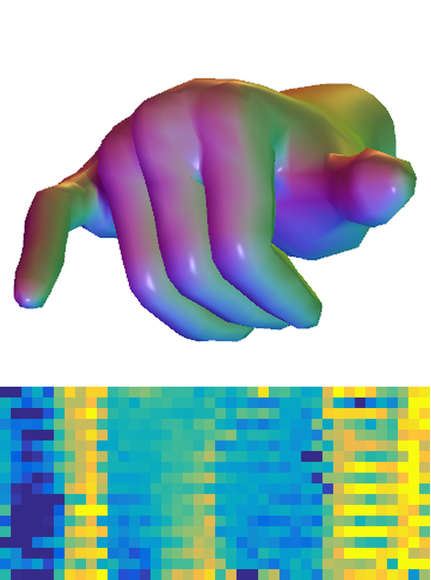}};
    \node(p03) at (p02.east) [anchor=west]{\includegraphics[width=\iw]{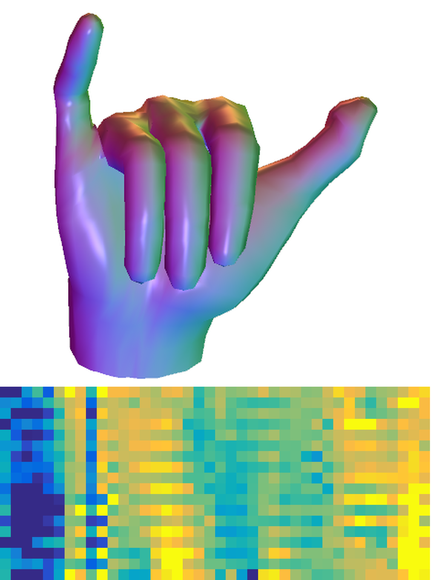}};
    \node(p04) at (p03.east) [anchor=west]{\includegraphics[width=\iw]{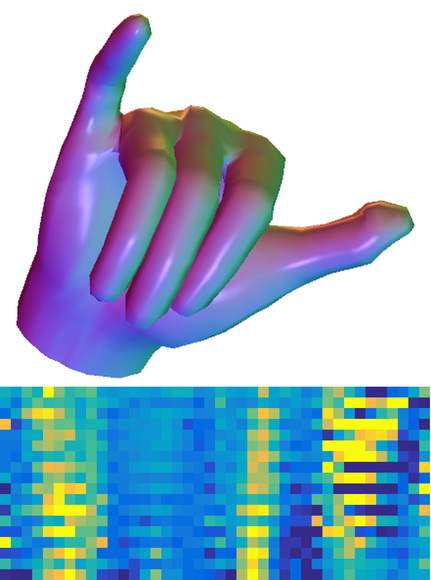}};
    \node(p05) at (p04.east) [anchor=west]{\includegraphics[width=\iw]{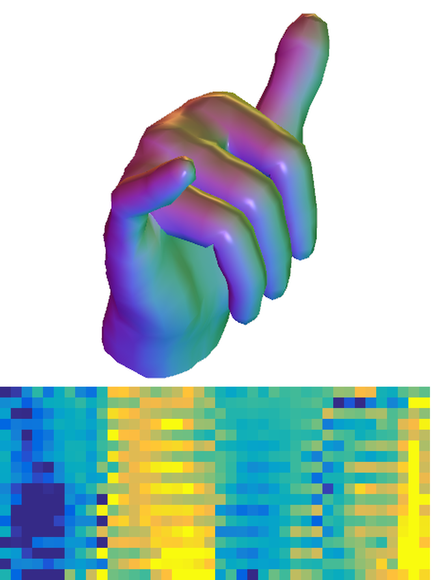}};
    \node(p06) at (p01.south west) [anchor=north west]{\includegraphics[width=\iw]{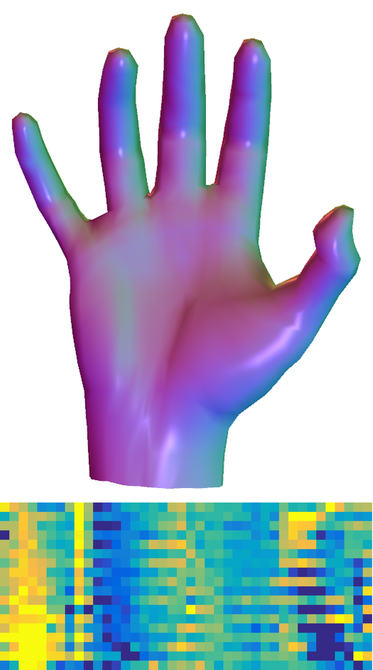}};
    \node(p07) at (p06.east) [anchor=west]{\includegraphics[width=\iw]{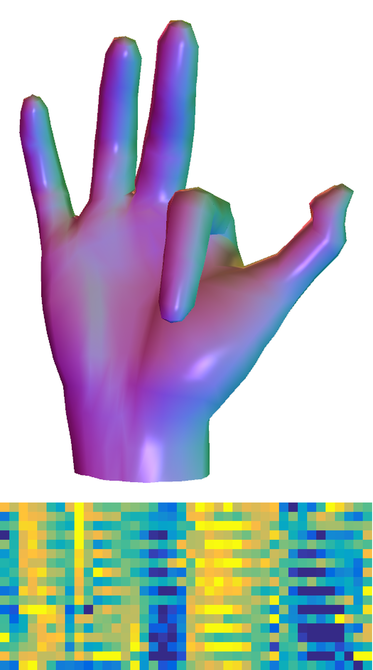}};
    \node(p08) at (p07.east) [anchor=west]{\includegraphics[width=\iw]{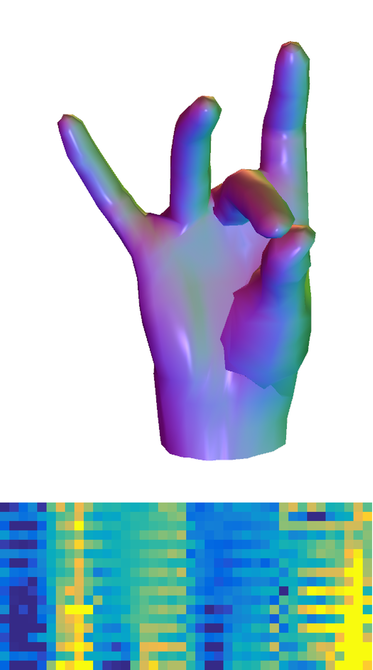}};
    \node(p09) at (p08.east) [anchor=west]{\includegraphics[width=\iw]{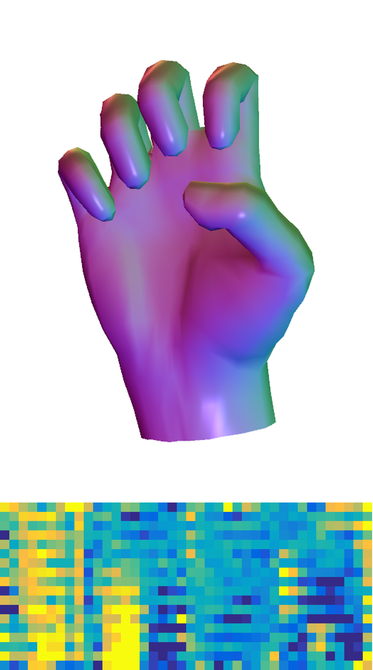}};
    \node(p10) at (p09.east) [anchor=west]{\includegraphics[width=\iw]{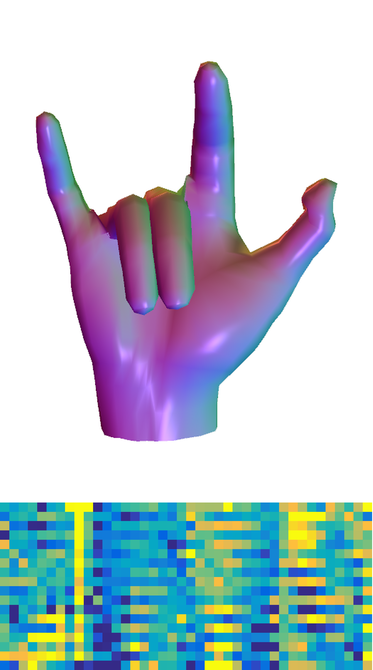}};
  \end{tikzpicture}
  \caption{Hand-pose dataset (HP). Top row: a fixed finger configuration under varying wrist orientations. Bottom row: diverse finger articulations at a neutral wrist orientation. Each panel shows the ground-truth hand pose alongside the corresponding wristband pressure map.}
  \label{fig:dataset-handpose}
\end{figure}

\begin{figure}[t]
  \centering
  \def\iw{0.243\columnwidth}
  \begin{tikzpicture}[inner sep=1pt]
    \node(p01)                            {\includegraphics[width=\iw]{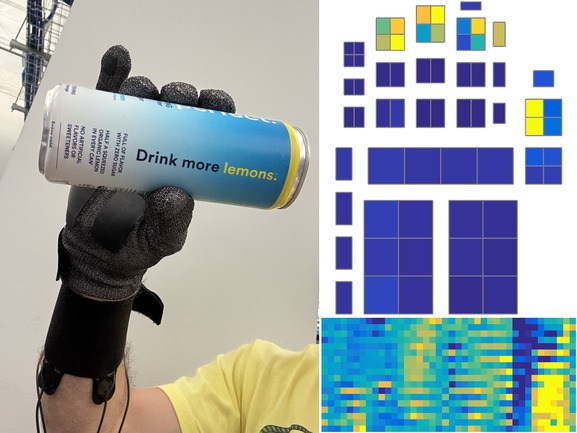}};
    \node(p02) at (p01.east) [anchor=west]{\includegraphics[width=\iw]{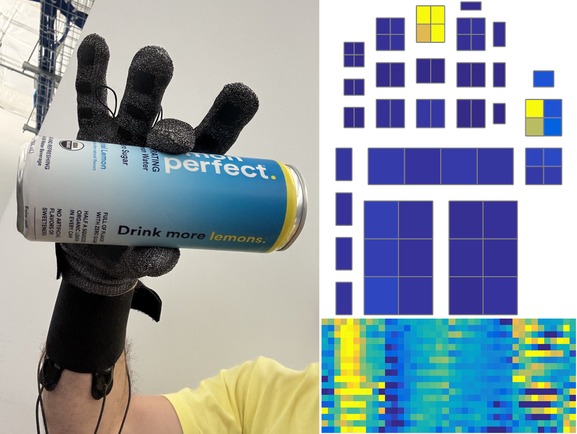}};
    \node(p04) at (p02.east) [anchor=west]{\includegraphics[width=\iw]{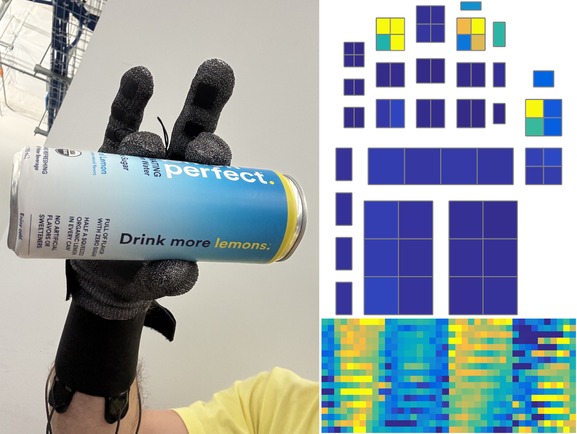}};
    \node(p05) at (p04.east) [anchor=west]{\includegraphics[width=\iw]{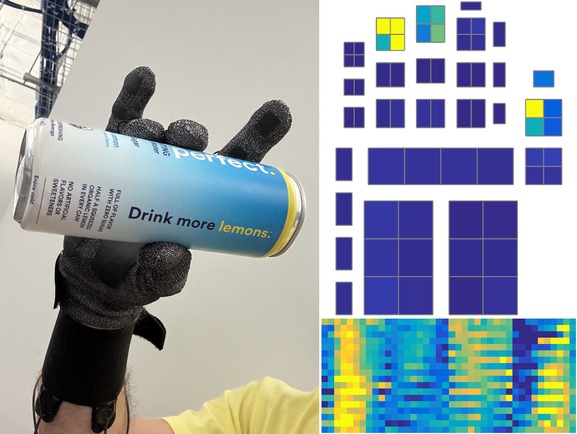}};
    \node(p06) at (p01.south west) [anchor=north west]{\includegraphics[width=\iw]{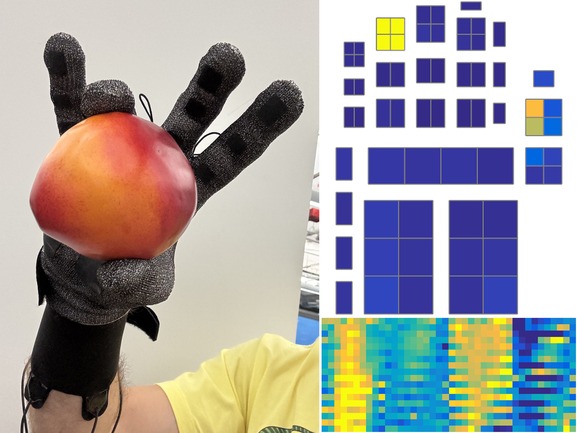}};
    \node(p07) at (p06.east) [anchor=west]{\includegraphics[width=\iw]{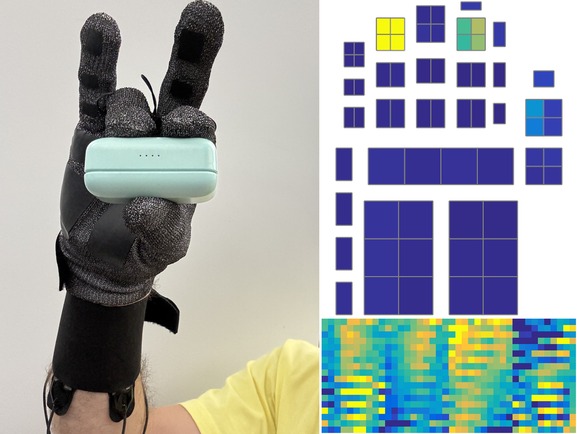}};
    \node(p09) at (p07.east) [anchor=west]{\includegraphics[width=\iw]{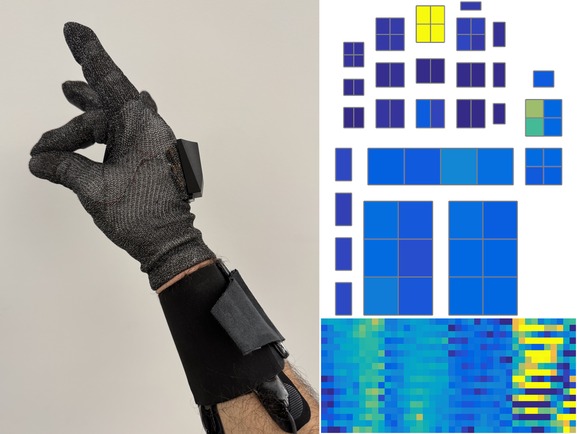}};
    \node(p10) at (p09.east) [anchor=west]{\includegraphics[width=\iw]{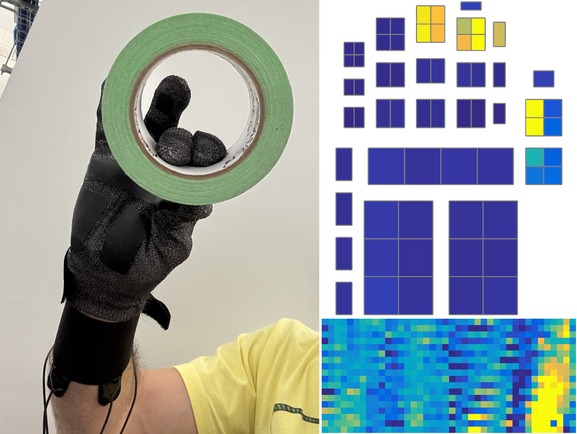}};
  \end{tikzpicture}
  \caption{Fingertip-force dataset (FF). Each panel shows the RGB scene frame (left), the ground-truth 2D taxel layout (top right), and the wristband pressure map (bottom right) captured at the same instant. Top row: grasping a can with various finger combinations and grip forces. Bottom row: diverse everyday object interactions.}
  \label{fig:force5}
\end{figure}

\begin{figure}[t]
  \centering
  \def\iw{0.243\columnwidth}
  \begin{tikzpicture}[inner sep=1pt]
    \node(p01)                            {\includegraphics[width=\iw]{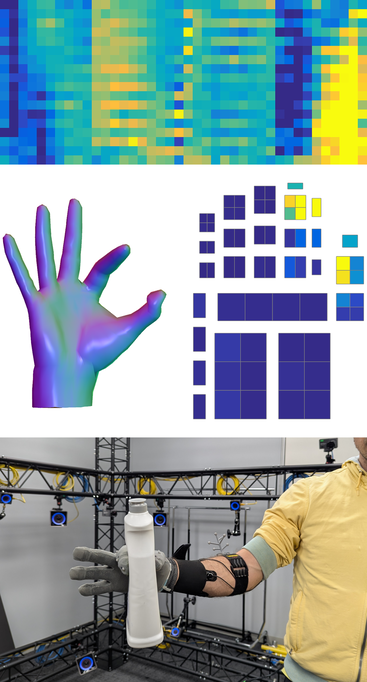}};
    \node(p02) at (p01.east) [anchor=west]{\includegraphics[width=\iw]{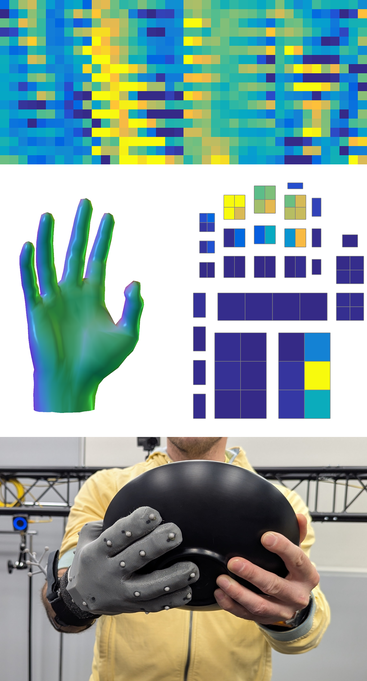}};
    \node(p03) at (p02.east) [anchor=west]{\includegraphics[width=\iw]{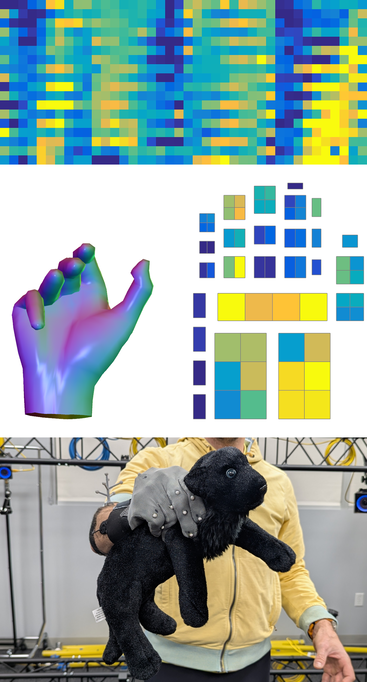}};
    \node(p04) at (p03.east) [anchor=west]{\includegraphics[width=\iw]{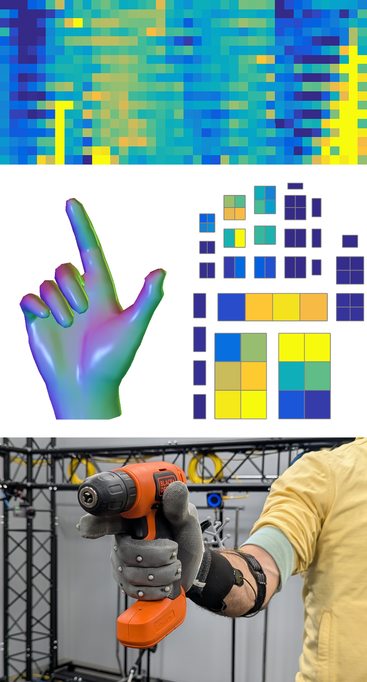}};
    \node(p05) at (p01.south west) [anchor=north west]{\includegraphics[width=\iw]{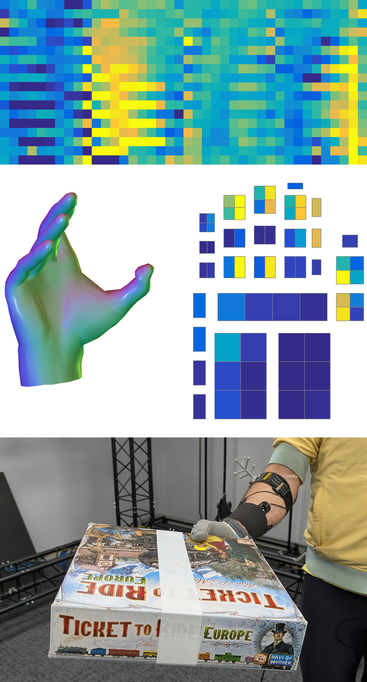}};
    \node(p06) at (p05.east) [anchor=west]{\includegraphics[width=\iw]{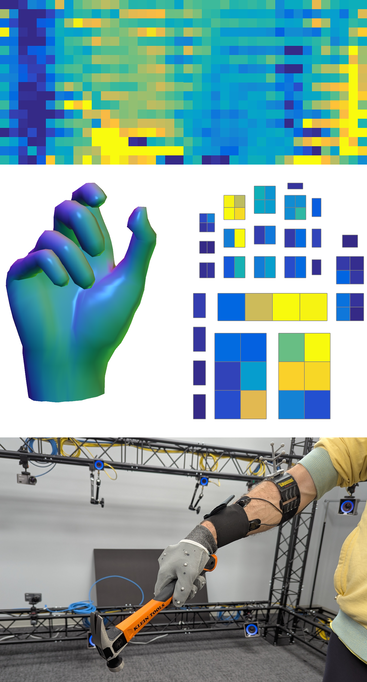}};
    \node(p07) at (p06.east) [anchor=west]{\includegraphics[width=\iw]{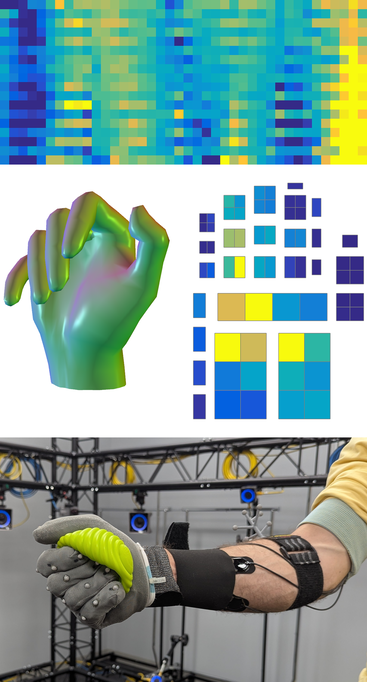}};
    \node(p08) at (p07.east) [anchor=west]{\includegraphics[width=\iw]{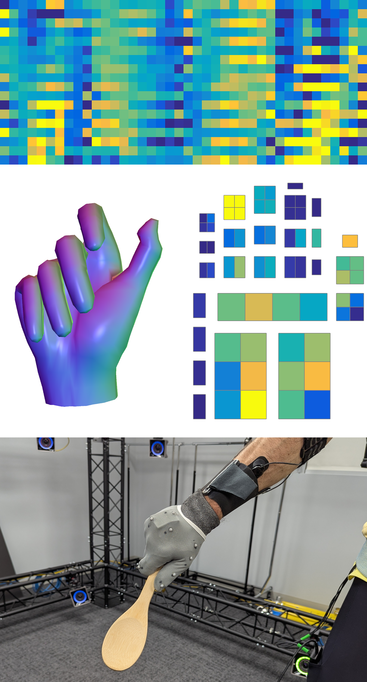}};
  \end{tikzpicture}
  \caption{Hand-object manipulation dataset (HOM). Representative scenarios showing the wristband pressure map, the ground-truth hand pose with glove taxel readings, and the corresponding RGB scene frame captured at the same instant.}
  \label{fig:dataset-manipulation}
\end{figure}

\Cref{fig:dataset-handpose,fig:force5,fig:dataset-manipulation} show representative samples from the three datasets.

\subsubsection{Data Processing}
\label{sect:data-processing}
We now briefly discuss the processing steps after data collection. For MoCap hand poses, we linearly interpolate missing data to fill data gaps shorter than 200~ms. These gaps are due to marker occlusion during casual object manipulation. Since MoCap runs at 60~Hz, we find linear interpolation effective. For larger gaps, and during sensor warm-up, we exclude the corresponding training windows. For tactile glove data, we first apply factory calibration to obtain per-taxel pressure in PSI, which we then convert to per-taxel normal force as described in \cref{sect:glove}. We observe that the baseline can drift during long recordings. To compensate, we estimate a time-varying baseline per taxel and subtract it from the measurements. Since taxels may fail after extended use, we also detect faulty recordings per taxel and mask the invalid ground truth out during training and evaluation. Finally, we time-align and linearly interpolate all sensor streams onto a common uniform timeline at 50~Hz sampling rate. To this end, we first apply zero-phase exponential smoothing with forward and backward filtering. Given a signal $\boldsymbol{x}(t)$ sampled at irregular times $t_1 < t_2 < \cdots$, the forward pass computes
\begin{equation}
  \bar{x}_k = (1-\alpha_k)\,\bar{x}_{k-1} + \alpha_k\,x_k, \quad \alpha_k = 1 - e^{-(t_k - t_{k-1})/\tau}
\end{equation}
where $\tau$ is a time constant that controls the smoothing bandwidth. The backward pass applies the same recurrence in reverse, yielding a zero-phase (non-causal) filter that preserves signal timing. We set $\tau = 1/(2 F_{\text{target}})$ with $F_{\text{target}} = 50$~Hz, which ensures the filter attenuates frequencies above the target sampling rate while retaining the slower physiological dynamics. After smoothing, we resample all streams onto the uniform 50~Hz timeline via linear interpolation.

\subsection{Model and Training Details}
\label{app:training}
\label{sec:supp_training}

We expand here on the training and inference mechanics summarized in \cref{sect:training-eval}. With a context length of $N=100$ samples (2~s at 50~Hz) and batch size 128, one optimization step processes 256~s of context, corresponding to roughly 80 iterations per epoch and about 600 effective epochs over 50k iterations.

The GRU produces one output per input step, but supervision is applied only on the last $N/2$ outputs so that the recurrent state has sufficient warm-up context; the hidden state is initialized to zero at both train and test time, and the context length is kept identical across training and testing. At inference, each length-$N$ window predicts its last $N/2$ steps, so with stride~1 each timestamp is typically covered by multiple overlapping windows (up to $N/2$ predictions); we average all valid predictions per timestamp before computing metrics.

\subsection{Metrics}
\label{app:metrics}
\label{sec:supp_metrics}

\subsubsection{Scalar Error Metrics}
\label{sec:supp_scalar_metrics}

We evaluate each estimated scalar quantity --- an individual joint angle, wrist angle, or force channel --- with the following metrics. For a predicted sequence $\hat{y}_i$ and ground truth $y_i$ over $N$ samples, the \emph{mean absolute error} is
\begin{equation}
  \mathrm{MAE} = \frac{1}{N}\sum_{i=1}^{N} |y_i - \hat{y}_i|,
\end{equation}
reported in degrees for joint angles, Newtons for force, and PSI for taxel pressure. For each degree of freedom (DOF), the \emph{coefficient of determination} measures the fraction of ground-truth variance explained by the predictions,
\begin{equation}
  R^2 = 1 - \frac{\sum_{i=1}^{N}(\hat{y}_i - y_i)^2}{\sum_{i=1}^{N}(y_i - \bar{y})^2},
\end{equation}
where $\bar{y}$ is the ground-truth mean of that DOF; $R^2 = 1$ indicates perfect prediction, $R^2 = 0$ matches predicting the mean, and negative values indicate worse-than-mean predictions. The \emph{ground-truth standard deviation} $\sigma(\mathrm{GT})$ is the standard deviation of $y_i$, reported as a reference scale for MAE.

\subsubsection{Pose Position Metrics}
\label{sec:supp_pose_metrics}

\begin{figure}[t]
  \centering
  \includegraphics[width=0.6\columnwidth]{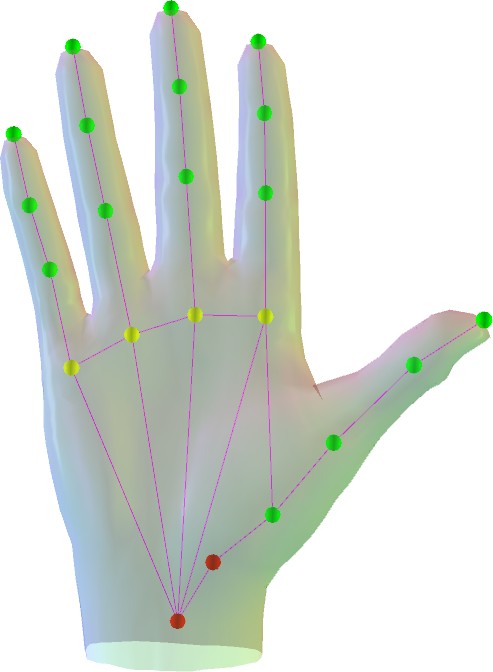}
  \caption{Hand skeleton joints used for the position metrics. Green joints are included in both variants (finger-only and with-wrist). Yellow joints (MCP base) are included only in the with-wrist variant, as they remain at fixed positions without wrist rotation. Red joints (wrist root and palm) are excluded from both.}
  \label{fig:skeleton_mpjpe}
\end{figure}

For hand pose we report fingertip position error and mean per-joint position error (MPJPE), both in millimetres, obtained via forward kinematics from the predicted joint angles. MPJPE is the mean Euclidean distance between predicted and ground-truth 3D skeleton joint positions. As illustrated in \cref{fig:skeleton_mpjpe}, the hand skeleton comprises 22 points: four per finger plus a wrist root and palm. We report two variants. \emph{Finger-only} (green joints in \cref{fig:skeleton_mpjpe}) uses the ground-truth wrist orientation as the root and evaluates 16 joints --- excluding the four MCP base joints (indices 6, 10, 14, 18), which remain at fixed positions without wrist rotation --- so it isolates finger-articulation error from wrist-prediction error. \emph{With-wrist} applies the predicted wrist orientation and includes the MCP base joints (yellow in \cref{fig:skeleton_mpjpe}), evaluating all 20 finger skeleton points and thus capturing the full end-to-end error. The wrist root and palm (red) are excluded from both. Per-user MPJPE is computed per-finger first, then averaged weighted by the number of skeleton points per finger.

\subsubsection{Aggregation of \texorpdfstring{$R^2$, $\sigma$(GT), and MAE}{R\^2, sigma(GT), and MAE}}
\label{sec:supp_aggregation}

When reporting aggregated metrics across multiple degrees of freedom (DOFs)---for example, a per-finger $R^2$ summarizing four joint types, or a per-joint-type $R^2$ summarizing five fingers---care must be taken to avoid inflation from between-DOF mean differences.

\paragraph{The pooled \texorpdfstring{$R^2$}{R\^2} problem.}
A na\"ive approach computes $R^2$ by pooling all DOFs into a single vector:
\begin{equation}
  R^2_{\text{pooled}} = 1 - \frac{\sum_{j}\sum_{i}(\hat{y}_{ij} - y_{ij})^2}{\sum_{j}\sum_{i}(y_{ij} - \bar{y})^2},
\end{equation}
where $\bar{y}$ is the \emph{global} mean across all samples $i$ and all DOFs $j$. The denominator $\mathrm{SS}_{\mathrm{tot}}$ then includes both within-DOF variance and between-DOF mean differences. For example, if one joint angle centers near $5^\circ$ and another near $40^\circ$, the $35^\circ$ gap inflates $\mathrm{SS}_{\mathrm{tot}}$, artificially raising $R^2$ even when per-DOF predictions are mediocre.

\paragraph{Variance-weighted \texorpdfstring{$R^2$}{R\^2}.}
We instead compute $R^2$ per DOF using each DOF's own mean, then combine with variance weighting:
\begin{equation}
  R^2_j = 1 - \frac{\mathrm{SS}_{\mathrm{res},j}}{\mathrm{SS}_{\mathrm{tot},j}}, \qquad
  R^2_{\text{agg}} = 1 - \frac{\sum_j \mathrm{SS}_{\mathrm{res},j}}{\sum_j \mathrm{SS}_{\mathrm{tot},j}} = \frac{\sum_j \mathrm{SS}_{\mathrm{tot},j} \cdot R^2_j}{\sum_j \mathrm{SS}_{\mathrm{tot},j}}.
\end{equation}
This is equivalent to a weighted average of per-DOF $R^2$ values, where each DOF is weighted by its ground-truth variance ($\mathrm{SS}_{\mathrm{tot},j} \propto \sigma_j^2$). High-variance DOFs contribute proportionally more. Crucially, between-DOF mean differences no longer inflate the denominator, since each $\mathrm{SS}_{\mathrm{tot},j}$ is computed relative to its own DOF mean $\bar{y}_j$.

\paragraph{Variance-weighted standard deviation \texorpdfstring{$\sigma$(GT)}{sigma(GT)}.}
For consistency, the aggregated ground-truth standard deviation is also variance-weighted:
\begin{equation}
  \sigma_{\text{agg}}(\mathrm{GT}) = \frac{\sum_j \sigma_j^2 \cdot \sigma_j}{\sum_j \sigma_j^2} = \frac{\sum_j \sigma_j^3}{\sum_j \sigma_j^2},
\end{equation}
giving higher-variance DOFs proportionally more influence on the reported spread.

\paragraph{Sample-count-weighted MAE.}
Unlike $R^2$ and $\sigma$(GT), MAE is not distorted by between-DOF mean differences, since it measures absolute prediction errors regardless of each DOF's range. We still need to choose how to combine it across DOFs that may have different valid sample counts (e.g., when one sensor channel has missing or masked-out frames). We aggregate using a sample-count-weighted mean:
\begin{equation}
  \mathrm{MAE}_{\text{agg}} = \frac{\sum_j n_j \cdot \mathrm{MAE}_j}{\sum_j n_j},
\end{equation}
where $n_j$ is the number of valid samples for DOF $j$ (i.e., after any per-DOF masking due to missing data). When all DOFs share the same $n_j$, this reduces to the unweighted arithmetic mean; when sample counts differ (e.g., a sensor with partial failure on one channel), DOFs with more data carry proportionally more weight in the aggregate.

\subsection{Detailed Results}
\label{app:results}

This section collects the detailed, per-joint, per-user, and per-finger results that expand the summary tables of \cref{sect:results}, together with full-resolution qualitative visualizations.

\subsubsection{Fingertip-Force (FF) Results}
\label{sec:supp_ff_results}
\Cref{tab:force5} reports per-finger force-estimation accuracy on the auxiliary FF dataset (\cref{sec:supp_ff}). Thumb performance is consistently worse than for the other fingers, reflecting both physiology (greater contribution of intrinsic hand muscles, which are less observable at the forearm) and reduced training data: thumb sensor failures during collection left only ${\sim}34\%$ of thumb samples valid.

\begin{table}[t]
  \centering
  \caption{Per-finger force prediction error (Newtons) on the Fingertip Force (FF) dataset.}
  \label{tab:force5}
  \small
  \setlength{\tabcolsep}{6pt}
  \begin{tabular}{l c cc}
    \toprule
    \textbf{Finger} & $\sigma$(GT) (N) & MAE (N) & $R^2$ \\
    \midrule
    Thumb    & 0.950 & 0.503 & 0.268 \\
    Index    & 1.696 & 0.489 & 0.661 \\
    Middle   & 1.518 & 0.430 & 0.718 \\
    Ring     & 0.985 & 0.245 & 0.730 \\
    Pinky    & 0.308 & 0.087 & 0.588 \\
    \midrule
    \textit{Mean} & \textit{1.277} & \textit{0.328} & \textit{0.696} \\
    \bottomrule
  \end{tabular}
\end{table}

\subsubsection{HOM Pose Estimation}
\label{sec:supp_hom_pose}
On the HOM dataset the bulkier tactile glove and cluttered object interaction make the MoCap ground truth noisier (marker occlusion, dropped frames), so pose quality is lower than on HP. We nonetheless run the same experiments; \cref{tab:pps_kinematics_summary} summarizes per-user kinematics for the multi-task Force+Pose model, with the single-user HP result for P4 as a reference. A full per-user, per-finger breakdown is given in \cref{tab:pps_kinematics}.

\begin{table}[t]
  \centering
  \caption{Per-user kinematics estimation on the HOM dataset for the multi-task \textit{Force+Pose} model, which simultaneously estimates per-taxel glove force ($\lambda_F{=}1.2$, see \cref{sec:supp_loss_weights}) and joint angles.}
  \label{tab:pps_kinematics_summary}
  \small
  \setlength{\tabcolsep}{4pt}
  \begin{tabular}{l ccc !{\hskip 4pt\vrule\hskip 4pt} ccc}
    \toprule
    &
      \multicolumn{3}{c}{\textbf{Finger}} &
      \multicolumn{3}{c}{\textbf{Wrist}} \\
    \cmidrule(lr){2-4} \cmidrule(lr){5-7}
    \textbf{User} &
      $\sigma$(GT)$^\circ$ & MAE$^\circ$ & $R^2$ &
      $\sigma$(GT)$^\circ$ & MAE$^\circ$ & $R^2$ \\
    \midrule
    P1 & 22.19 & 12.14 & 0.260 & 26.47 & 8.97 & 0.755 \\
    P2 & 21.01 &  9.10 & 0.431 & 25.12 & 9.16 & 0.716 \\
    P3 & 19.67 &  8.76 & 0.445 & 28.80 & 8.50 & 0.779 \\
    P4 & 17.88 &  7.19 & 0.582 & 19.66 & 6.75 & 0.763 \\
    \midrule
    \textit{Mean} & \textit{20.19} & \textit{9.30} & \textit{0.430} & \textit{25.01} & \textit{8.35} & \textit{0.753} \\
    \midrule
    P4 (HP) & 21.81 & 4.56 & 0.859 & 19.48 & 4.75 & 0.863 \\
    \bottomrule
  \end{tabular}
\end{table}

\subsubsection{Per-Joint Hand-Pose Errors (HP)}
\label{sec:supp_per_joint}
Extending the summary in \cref{tab:joint_angle_errors}, \cref{tab:per_joint_errors} breaks the HP error down to each individual joint of each finger.

\begin{table}[t]
  \centering
  \caption{\textbf{Per-finger per-joint estimation error ($^\circ$) for the combined wrist and finger model on the Hand Pose (HP) dataset.}}
  \label{tab:per_joint_errors}
  \small
  \setlength{\tabcolsep}{4pt}
  \begin{tabular}{ll c cc}
    \toprule
    \textbf{Finger} & \textbf{Joint} & $\sigma$(GT)$^\circ$ &
      MAE$^\circ$ & $R^2$ \\
    \midrule
    \multirow{4}{*}{Thumb} & CMC FE     & 16.15 & 3.63 & 0.897 \\
                           & CMC AA     & 10.11 & 3.61 & 0.758 \\
                           & MCP FE     & 17.06 & 5.49 & 0.790 \\
                           & IP FE      & 15.21 & 5.84 & 0.731 \\
    \cmidrule(lr){2-5}
                           & \textit{Mean} & \textit{15.51} & \textit{4.64} & \textit{0.802} \\
    \midrule
    \multirow{4}{*}{Index} & MCP AA     & 5.37 & 2.63 & 0.559 \\
                           & MCP FE     & 23.48 & 6.32 & 0.852 \\
                           & PIP FE     & 22.74 & 6.11 & 0.863 \\
                           & DIP FE     & 11.31 & 3.37 & 0.810 \\
    \cmidrule(lr){2-5}
                           & \textit{Mean} & \textit{21.47} & \textit{4.61} & \textit{0.846} \\
    \midrule
    \multirow{4}{*}{Middle} & MCP AA     & 5.09 & 2.43 & 0.577 \\
                           & MCP FE     & 26.09 & 6.37 & 0.886 \\
                           & PIP FE     & 23.68 & 5.61 & 0.886 \\
                           & DIP FE     & 19.45 & 3.89 & 0.904 \\
    \cmidrule(lr){2-5}
                           & \textit{Mean} & \textit{23.41} & \textit{4.58} & \textit{0.886} \\
    \midrule
    \multirow{4}{*}{Ring}  & MCP AA     & 6.95 & 2.62 & 0.736 \\
                           & MCP FE     & 26.00 & 6.61 & 0.869 \\
                           & PIP FE     & 24.60 & 5.49 & 0.897 \\
                           & DIP FE     & 14.91 & 3.38 & 0.889 \\
    \cmidrule(lr){2-5}
                           & \textit{Mean} & \textit{23.27} & \textit{4.53} & \textit{0.878} \\
    \midrule
    \multirow{4}{*}{Pinky} & MCP AA     & 10.34 & 3.05 & 0.795 \\
                           & MCP FE     & 27.38 & 6.57 & 0.865 \\
                           & PIP FE     & 15.83 & 4.44 & 0.832 \\
                           & DIP FE     & 16.49 & 3.76 & 0.865 \\
    \cmidrule(lr){2-5}
                           & \textit{Mean} & \textit{21.81} & \textit{4.45} & \textit{0.854} \\
    \midrule
    \multicolumn{2}{l}{\textit{Overall Mean}} & \textit{21.65} & \textit{4.56} & \textit{0.859} \\
    \bottomrule
  \end{tabular}
\end{table}

\subsubsection{Hand-Pose Method Comparison (HP)}
\label{sec:supp_hp_methods}
\Cref{tab:kinematics_comparison} compares model variants on HP --- estimating finger and wrist together, each alone, and conditioning finger estimation on wrist inputs --- under matched settings.

\begin{table}[t]
  \centering
  \caption{\textbf{Comparison of kinematics estimation approaches on the Hand Pose (HP) dataset. Mean MAE, $R^2$, and fingertip position error across all DOFs. Ground-truth standard deviations: Finger $\sigma$(GT)=21.81$^\circ$, Wrist $\sigma$(GT)=19.48$^\circ$.}}
  \label{tab:kinematics_comparison}
  \small
  \setlength{\tabcolsep}{3pt}
  \begin{tabular}{l @{\hspace{6pt}} c c !{\hskip 3pt\vrule\hskip 3pt} c c !{\hskip 3pt\vrule\hskip 3pt} c}
    \toprule
    & \multicolumn{2}{c}{\textbf{Finger}} & \multicolumn{2}{c}{\textbf{Wrist}} & \multicolumn{1}{c}{\textbf{Tip}} \\
    \cmidrule(lr){2-3} \cmidrule(lr){4-5} \cmidrule(lr){6-6}
    \textbf{Experiment} &
      MAE$^\circ$ & $R^2$ &
      MAE$^\circ$ & $R^2$ &
      (mm) \\
    \midrule
    Both wrist \& finger joints          & 4.56 & 0.860 & 4.75 & 0.863 & 10.7 \\
    Only finger joints                   & 4.54 & 0.862 & --- & --- & 10.6 \\
    Only wrist joints                    & --- & --- & 4.35 & 0.884 & --- \\
    Finger w/ pronation input            & 4.46 & 0.865 & --- & --- & 10.4 \\
    Finger w/ all wrist inputs           & 4.26 & 0.879 & --- & --- & 9.9 \\
    \bottomrule
  \end{tabular}
\end{table}

\subsubsection{Per-Finger HOM Kinematics}
\label{sec:supp_hom_kin_detail}
\Cref{tab:pps_kinematics} gives the per-user, per-DOF joint-angle errors on HOM underlying the summary in \cref{tab:pps_kinematics_summary}.

\begin{table}[t]
  \centering
  \caption{\textbf{Per-user joint angle estimation error ($^\circ$) on the HOM dataset for the combined wrist and finger model.}}
  \label{tab:pps_kinematics}
  \scriptsize
  \begin{minipage}[t]{0.49\columnwidth}
    \centering
    \textbf{P1}\\[2pt]
    \setlength{\tabcolsep}{3pt}
    \begin{tabular}{ll c cc}
      \toprule
      & \textbf{DOF} & $\sigma$(GT)$^\circ$ &
        MAE$^\circ$ & $R^2$ \\
      \midrule
      \multirow{3}{*}{\rotatebox[origin=c]{90}{Wrist}} & Pronation  & 31.5 & 9.2 & 0.860 \\
       & Flexion    & 20.9 & 8.3 & 0.728 \\
       & Deviation  & 14.9 & 7.8 & 0.543 \\
      \cmidrule(lr){2-5}
      & \textit{Mean} & \textit{26.5} & \textit{8.4} & \textit{0.783} \\
      \midrule
      \multirow{5}{*}{\rotatebox[origin=c]{90}{Finger}} & Thumb      & 17.4 & 9.5 & 0.266 \\
       & Index      & 20.9 & 12.4 & 0.160 \\
       & Middle     & 22.1 & 12.7 & 0.195 \\
       & Ring       & 24.7 & 14.2 & 0.210 \\
       & Pinky      & 23.1 & 13.3 & 0.212 \\
      \cmidrule(lr){2-5}
      & \textit{Mean} & \textit{22.2} & \textit{12.4} & \textit{0.205} \\
      \midrule
      \multirow{4}{*}{\rotatebox[origin=c]{90}{Joint}} & MCP Abd.   & 10.1 & 7.0 & 0.142 \\
       & MCP Flex.  & 22.6 & 12.8 & 0.287 \\
       & PIP Flex.  & 23.7 & 15.3 & 0.171 \\
       & DIP Flex.  & 22.8 & 14.6 & 0.180 \\
      \bottomrule
    \end{tabular}
  \end{minipage}\hfill
  \begin{minipage}[t]{0.49\columnwidth}
    \centering
    \textbf{P2}\\[2pt]
    \setlength{\tabcolsep}{3pt}
    \begin{tabular}{ll c cc}
      \toprule
      & \textbf{DOF} & $\sigma$(GT)$^\circ$ &
        MAE$^\circ$ & $R^2$ \\
      \midrule
      \multirow{3}{*}{\rotatebox[origin=c]{90}{Wrist}} & Pronation  & 29.8 & 11.2 & 0.771 \\
       & Flexion    & 19.9 & 8.7 & 0.709 \\
       & Deviation  & 13.7 & 6.6 & 0.606 \\
      \cmidrule(lr){2-5}
      & \textit{Mean} & \textit{25.1} & \textit{8.9} & \textit{0.734} \\
      \midrule
      \multirow{5}{*}{\rotatebox[origin=c]{90}{Finger}} & Thumb      & 14.5 & 7.2 & 0.458 \\
       & Index      & 24.9 & 8.9 & 0.453 \\
       & Middle     & 18.1 & 8.9 & 0.417 \\
       & Ring       & 20.1 & 9.0 & 0.488 \\
       & Pinky      & 21.7 & 10.8 & 0.397 \\
      \cmidrule(lr){2-5}
      & \textit{Mean} & \textit{20.6} & \textit{9.0} & \textit{0.440} \\
      \midrule
      \multirow{4}{*}{\rotatebox[origin=c]{90}{Joint}} & MCP Abd.   & 9.9 & 6.3 & 0.257 \\
       & MCP Flex.  & 21.5 & 10.2 & 0.538 \\
       & PIP Flex.  & 24.1 & 11.8 & 0.452 \\
       & DIP Flex.  & 15.0 & 7.6 & 0.286 \\
      \bottomrule
    \end{tabular}
  \end{minipage}
  \vspace{14pt}
  \begin{minipage}[t]{0.49\columnwidth}
    \centering
    \textbf{P3}\\[2pt]
    \setlength{\tabcolsep}{3pt}
    \begin{tabular}{ll c cc}
      \toprule
      & \textbf{DOF} & $\sigma$(GT)$^\circ$ &
        MAE$^\circ$ & $R^2$ \\
      \midrule
      \multirow{3}{*}{\rotatebox[origin=c]{90}{Wrist}} & Pronation  & 34.2 & 9.3 & 0.878 \\
       & Flexion    & 19.6 & 9.1 & 0.594 \\
       & Deviation  & 13.6 & 6.8 & 0.532 \\
      \cmidrule(lr){2-5}
      & \textit{Mean} & \textit{28.8} & \textit{8.4} & \textit{0.778} \\
      \midrule
      \multirow{5}{*}{\rotatebox[origin=c]{90}{Finger}} & Thumb      & 13.0 & 6.7 & 0.405 \\
       & Index      & 19.2 & 9.3 & 0.357 \\
       & Middle     & 18.1 & 8.3 & 0.456 \\
       & Ring       & 20.9 & 9.2 & 0.484 \\
       & Pinky      & 22.3 & 10.3 & 0.458 \\
      \cmidrule(lr){2-5}
      & \textit{Mean} & \textit{19.6} & \textit{8.8} & \textit{0.440} \\
      \midrule
      \multirow{4}{*}{\rotatebox[origin=c]{90}{Joint}} & MCP Abd.   & 7.8 & 4.7 & 0.327 \\
       & MCP Flex.  & 19.7 & 8.6 & 0.575 \\
       & PIP Flex.  & 21.0 & 11.2 & 0.388 \\
       & DIP Flex.  & 20.0 & 10.6 & 0.390 \\
      \bottomrule
    \end{tabular}
  \end{minipage}\hfill
  \begin{minipage}[t]{0.49\columnwidth}
    \centering
    \textbf{P4}\\[2pt]
    \setlength{\tabcolsep}{3pt}
    \begin{tabular}{ll c cc}
      \toprule
      & \textbf{DOF} & $\sigma$(GT)$^\circ$ &
        MAE$^\circ$ & $R^2$ \\
      \midrule
      \multirow{3}{*}{\rotatebox[origin=c]{90}{Wrist}} & Pronation  & 22.9 & 10.4 & 0.705 \\
       & Flexion    & 17.9 & 5.0 & 0.866 \\
       & Deviation  & 13.5 & 3.9 & 0.860 \\
      \cmidrule(lr){2-5}
      & \textit{Mean} & \textit{19.7} & \textit{6.4} & \textit{0.783} \\
      \midrule
      \multirow{5}{*}{\rotatebox[origin=c]{90}{Finger}} & Thumb      & 13.4 & 6.4 & 0.467 \\
       & Index      & 16.4 & 6.6 & 0.578 \\
       & Middle     & 17.4 & 6.6 & 0.626 \\
       & Ring       & 18.9 & 6.8 & 0.668 \\
       & Pinky      & 20.4 & 8.0 & 0.628 \\
      \cmidrule(lr){2-5}
      & \textit{Mean} & \textit{17.8} & \textit{6.9} & \textit{0.607} \\
      \midrule
      \multirow{4}{*}{\rotatebox[origin=c]{90}{Joint}} & MCP Abd.   & 9.9 & 4.9 & 0.493 \\
       & MCP Flex.  & 21.0 & 8.3 & 0.689 \\
       & PIP Flex.  & 17.7 & 7.9 & 0.581 \\
       & DIP Flex.  & 15.5 & 6.6 & 0.538 \\
      \bottomrule
    \end{tabular}
  \end{minipage}
\end{table}

\subsubsection{Per-Finger HOM Force}
\label{sec:supp_hom_force_detail}
\Cref{tab:pilot_force_per_finger} gives the per-user, per-finger force breakdown underlying \cref{tab:pilot_force}. The thumb and pinky carry the lowest $R^2$ across users.

\begin{table}[t]
  \centering
  \caption{\textbf{Per-user per-finger force prediction error (Newtons) on the HOM dataset at finger granularity, with data augmentation.} Across all users, the thumb and pinky carry the lowest $R^2$ relative to the middle three fingers.}
  \label{tab:pilot_force_per_finger}
  \small
  \setlength{\tabcolsep}{4pt}
  \begin{tabular}{ll c cc !{\hskip 4pt\vrule width 0.6pt\hskip 4pt} cc}
    \toprule
    & & &
      \multicolumn{2}{c}{\textbf{Pressure Input Only}} &
      \multicolumn{2}{c}{\textbf{Pose Input}} \\
    \cmidrule(lr){4-5} \cmidrule(lr){6-7}
    \textbf{User} & \textbf{Finger} & $\sigma$(GT) (N) &
      MAE (N) & $R^2$ &
      MAE (N) & $R^2$ \\
    \midrule
    \multirow{6}{*}{P1}           & Thumb  & 3.607 & 1.805 & 0.458 & 1.427 & 0.613 \\
                                   & Index  & 2.207 & 0.981 & 0.511 & 0.715 & 0.717 \\
                                   & Middle & 2.321 & 1.216 & 0.250 & 0.723 & 0.737 \\
                                   & Ring   & 1.748 & 0.928 & 0.190 & 0.549 & 0.719 \\
                                   & Pinky  & 0.297 & 0.166 & 0.177 & 0.125 & 0.442 \\
    \cmidrule(lr){2-7}
                                   & \textit{Mean} & \textit{2.860} & \textit{0.875} & \textit{0.393} & \textit{0.572} & \textit{0.669} \\
    \midrule
    \multirow{6}{*}{P2}          & Thumb  & 1.992 & 0.686 & 0.529 & 0.559 & 0.683 \\
                                   & Index  & 2.086 & 0.641 & 0.708 & 0.537 & 0.794 \\
                                   & Middle & 2.555 & 0.756 & 0.752 & 0.669 & 0.808 \\
                                   & Ring   & 1.270 & 0.381 & 0.659 & 0.330 & 0.758 \\
                                   & Pinky  & 0.284 & 0.113 & 0.431 & 0.106 & 0.469 \\
    \cmidrule(lr){2-7}
                                   & \textit{Mean} & \textit{2.160} & \textit{0.527} & \textit{0.676} & \textit{0.448} & \textit{0.768} \\
    \midrule
    \multirow{6}{*}{P3}         & Thumb  & 1.948 & 0.952 & 0.401 & 0.812 & 0.541 \\
                                   & Index  & 1.913 & 0.743 & 0.582 & 0.509 & 0.796 \\
                                   & Middle & 1.890 & 0.693 & 0.682 & 0.506 & 0.821 \\
                                   & Ring   & 1.713 & 0.570 & 0.726 & 0.425 & 0.840 \\
                                   & Pinky  & 0.710 & 0.282 & 0.627 & 0.220 & 0.743 \\
    \cmidrule(lr){2-7}
                                   & \textit{Mean} & \textit{1.834} & \textit{0.594} & \textit{0.590} & \textit{0.438} & \textit{0.742} \\
    \midrule
    \multirow{6}{*}{P4}          & Thumb  & 0.968 & 0.303 & 0.656 & 0.268 & 0.721 \\
                                   & Index  & 2.030 & 0.562 & 0.781 & 0.496 & 0.825 \\
                                   & Middle & 1.404 & 0.421 & 0.736 & 0.373 & 0.787 \\
                                   & Ring   & 1.433 & 0.309 & 0.836 & 0.280 & 0.866 \\
                                   & Pinky  & 0.442 & 0.117 & 0.737 & 0.102 & 0.794 \\
    \cmidrule(lr){2-7}
                                   & \textit{Mean} & \textit{1.624} & \textit{0.312} & \textit{0.770} & \textit{0.277} & \textit{0.815} \\
    \bottomrule
  \end{tabular}
\end{table}

\subsubsection{Wrist-Angle Distribution and Finger Error}
\label{sec:supp_wrist_analysis}
\begin{figure}[t]
  \centering
  \includegraphics[width=0.49\columnwidth]{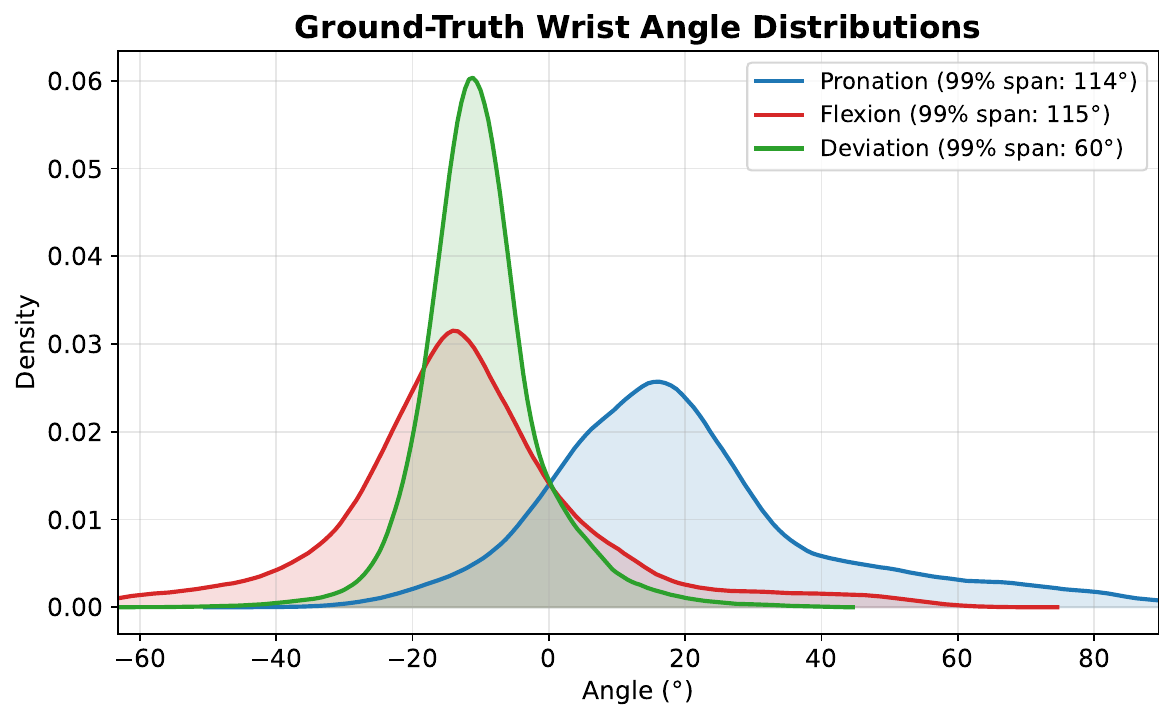}\hfill
  \includegraphics[width=0.49\columnwidth]{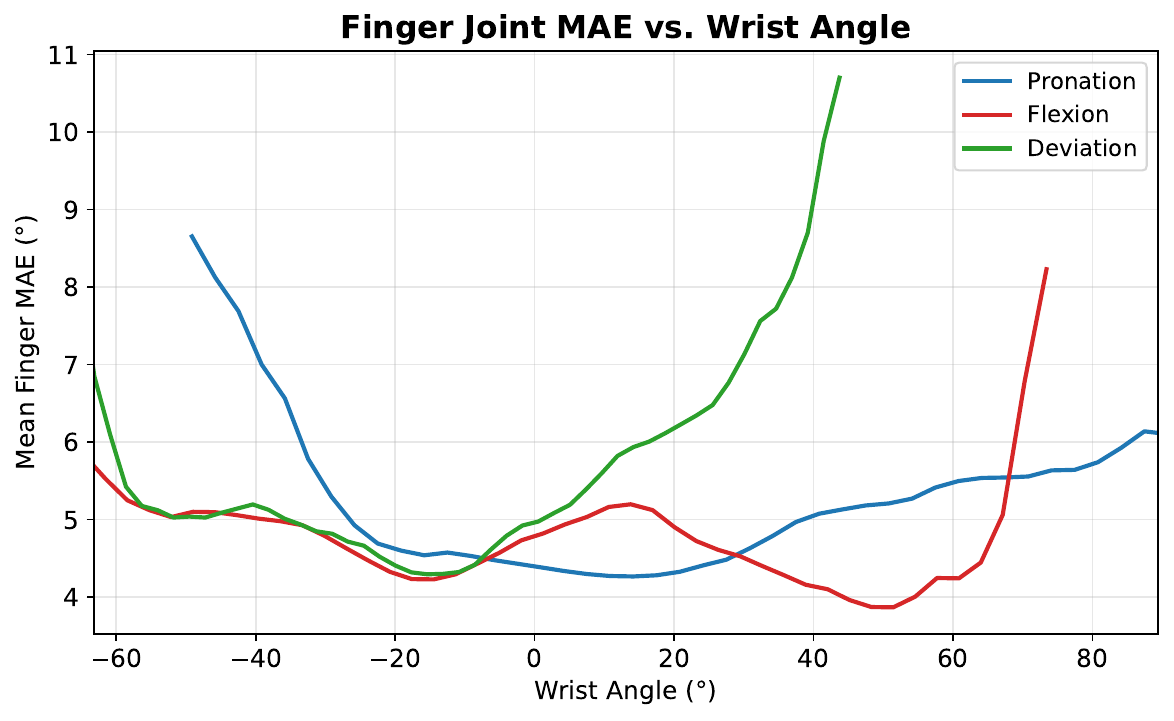}
  \caption{\emph{Left:} distribution of the three wrist angles in the HP dataset. \emph{Right:} mean finger-joint MAE as a function of wrist angle; error rises toward the sparsely-sampled extremes of the range of motion.}
  \label{fig:wrist-analysis}
\end{figure}

Because wrist rotation dominates the pressure signal, we examine how finger error varies across the wrist's range of motion. \Cref{fig:wrist-analysis} (left) shows the distribution of wrist angles in the HP dataset, and (right) the mean finger-joint error as a function of wrist angle: error is lowest near the center of each wrist-angle distribution and rises toward the extremes, where training data is sparser --- most pronounced for radial--ulnar deviation.

\subsubsection{Qualitative Visualizations}
\label{sec:supp_qualitative}
\begin{figure*}[p]
  \centering
  \begin{tikzpicture}
    \node[inner sep=0] (img) {\includegraphics[width=\linewidth]{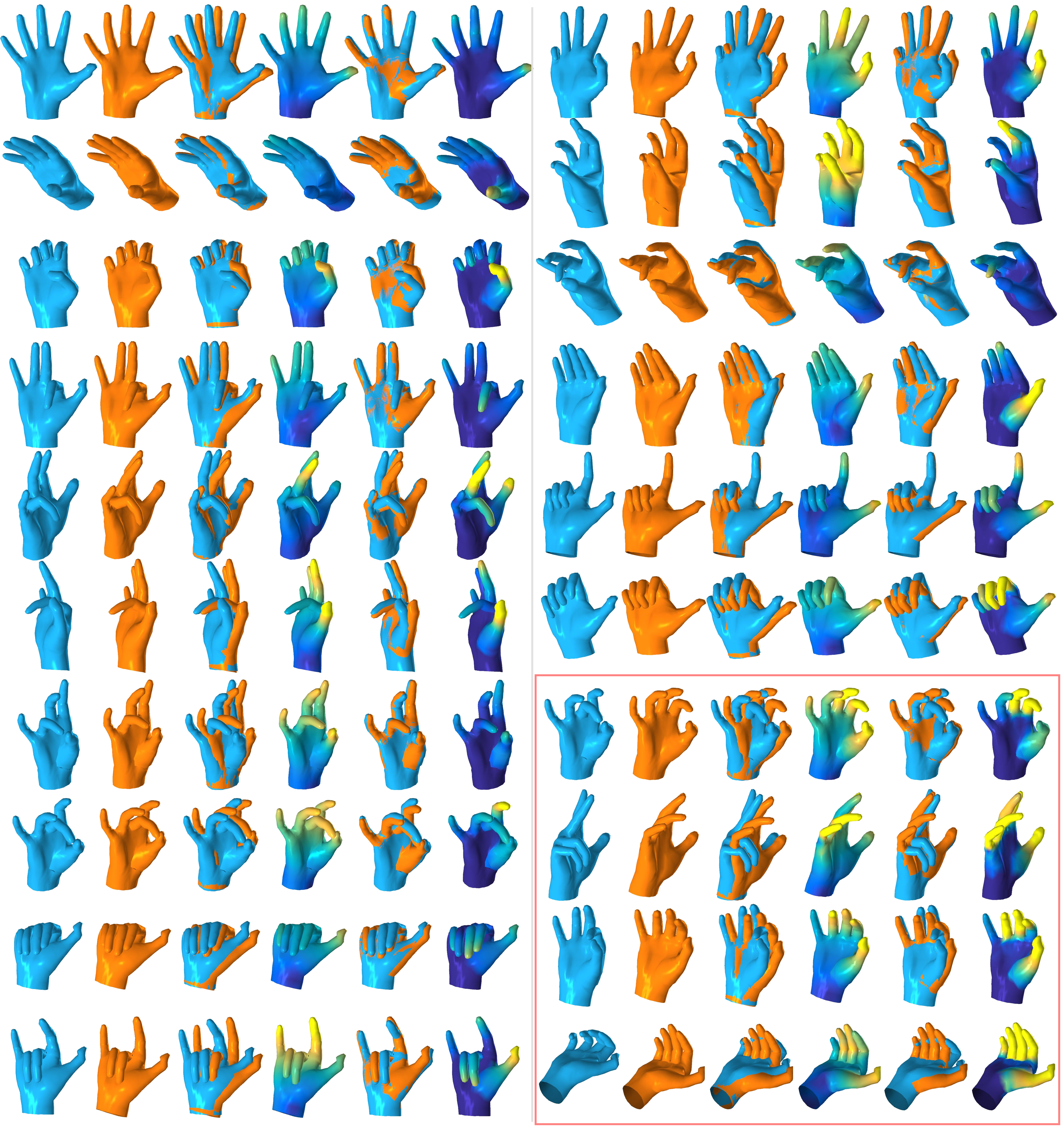}};
    \foreach \x/\lbl in {%
        0.040/{Ground\\Truth},
        0.115/Prediction,
        0.200/{GT/Prediction\\Overlay},
        0.280/{Vertex Dist\\Error},
        0.365/{GT Wrist\\Overlay},
        0.450/{GT Wrist\\Vertex Dist},
        0.535/{Ground\\Truth},
        0.610/Prediction,
        0.695/{GT/Prediction\\Overlay},
        0.775/{Vertex Dist\\Error},
        0.860/{GT Wrist\\Overlay},
        0.945/{GT Wrist\\Vertex Dist}%
    } {
      \node[font=\fontsize{6pt}{7pt}\selectfont\sffamily, anchor=south, align=center,
            inner sep=0pt, outer sep=1pt]
        at ([xshift=\x*\linewidth, yshift=2pt]img.north west) {\lbl};
      \node[font=\fontsize{6pt}{7pt}\selectfont\sffamily, anchor=north, align=center,
            inner sep=0pt, outer sep=1pt]
        at ([xshift=\x*\linewidth, yshift=-2pt]img.south west) {\lbl};
    }
  \end{tikzpicture}
  \caption{Sample visualization on 20 hand poses, blue = ground truth, orange = prediction. \emph{Col.\ 1--2:} GT and prediction shown side-by-side, both rendered with the predicted wrist orientation. \emph{Col.\ 3:} GT and prediction overlaid in the predicted wrist orientation --- the full end-to-end result. \emph{Col.\ 4:} predicted hand mesh color-coded by per-vertex distance to GT (yellow = larger). \emph{Col.\ 5:} GT and prediction overlaid in the \emph{ground-truth} wrist orientation, isolating finger-estimation errors from wrist errors. \emph{Col.\ 6:} per-vertex distance heat map under the ground-truth wrist orientation. The two heat-map columns use different saturation scales chosen per column to make the dynamic range visible. The last 4 rows are highlighted as examples where the estimate diverges significantly from the ground truth.}
  \label{fig:hand-pose-selected-full}
\end{figure*}

\begin{figure*}[p]
  \centering
  \begin{tikzpicture}[inner sep=0]
    \node (left)                                           {\includegraphics[width=0.495\linewidth]{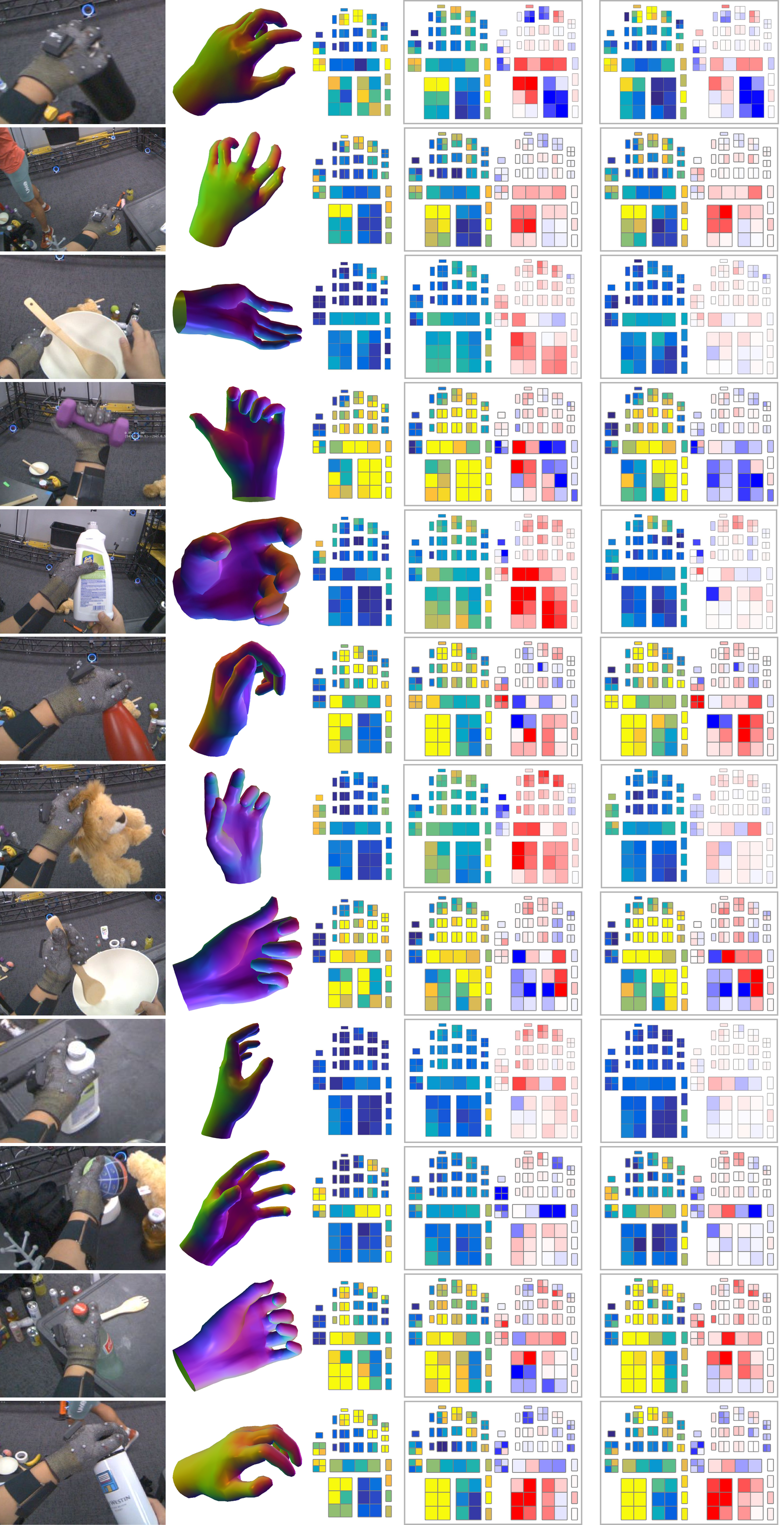}};
    \node (right) at (left.east) [anchor=west, xshift=2pt] {\includegraphics[width=0.495\linewidth]{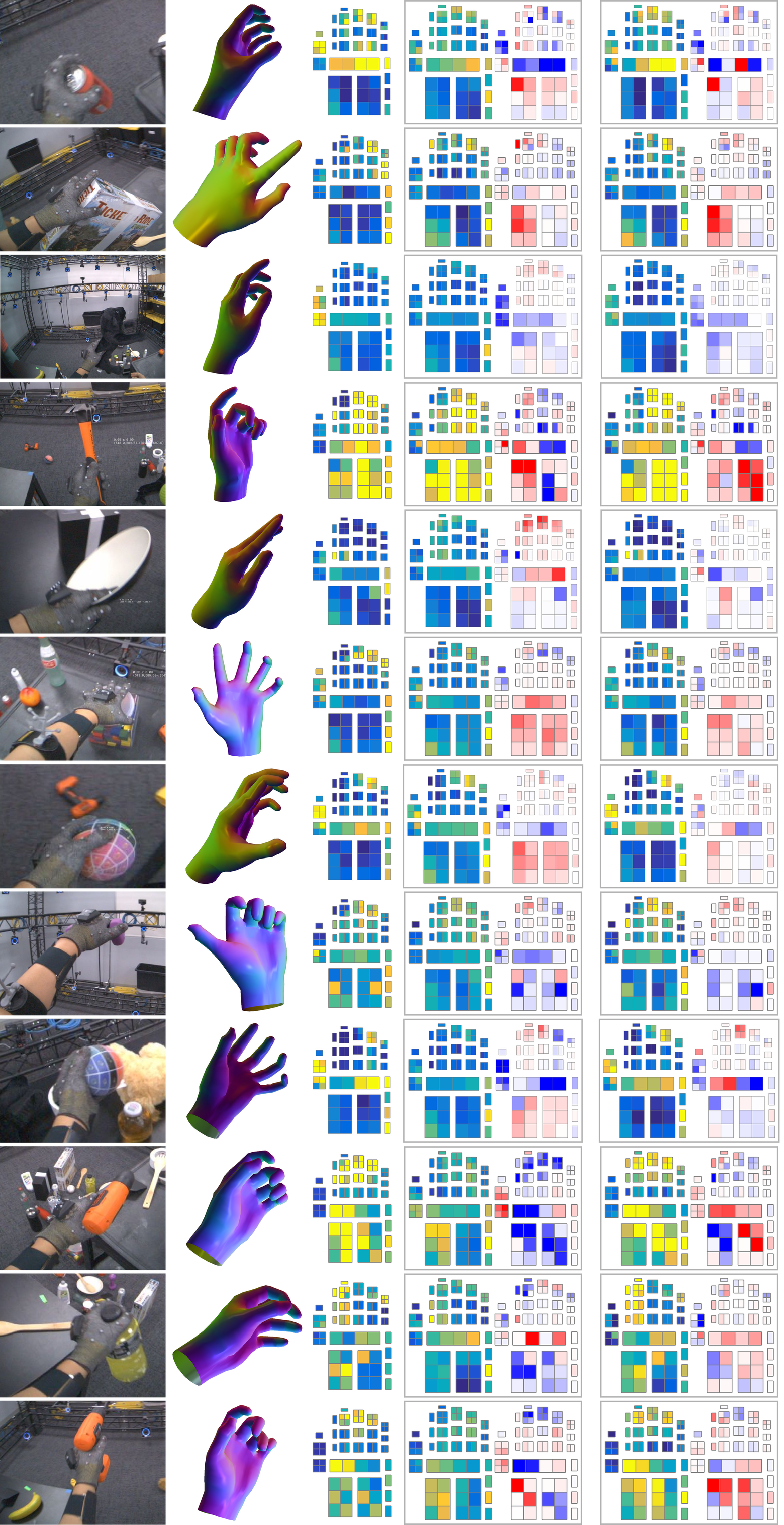}};
    \foreach \x/\lbl in {%
        0.050/{RGB Scene},%
        0.150/{Hand\\Pose},%
        0.225/{Ground\\Truth},%
        0.315/{Prediction --- Error\\Pressure Wristband},%
        0.440/{Prediction --- Error\\Wristband + Pose},%
        0.550/{RGB Scene},%
        0.650/{Hand\\Pose},%
        0.725/{Ground\\Truth},%
        0.815/{Prediction --- Error\\Pressure Wristband},%
        0.940/{Prediction --- Error\\Wristband + Pose}%
    } {
      \node[font=\fontsize{6pt}{7pt}\selectfont\sffamily, anchor=south, align=center, inner sep=0pt, outer sep=1pt]
        at ([xshift=\x*\linewidth, yshift=2pt]left.north west) {\lbl};
      \node[font=\fontsize{6pt}{7pt}\selectfont\sffamily, anchor=north, align=center, inner sep=0pt, outer sep=1pt]
        at ([xshift=\x*\linewidth, yshift=-2pt]left.south west) {\lbl};
    }
  \end{tikzpicture}
  \caption{Sample visualization of force predictions, showing two example timesteps side by side with identical column layout. Left to right within each half: (1) RGB image of the scene; (2) MoCap-extracted 3D hand shape; (3) ground-truth tactile-glove pressure map; (4) predicted pressure map and the corresponding error map (pred~$-$~GT) for the \emph{pressure-wristband-only} model; (5) the same prediction--error pair for the \emph{wristband + pose} model. In the error maps, maximum red and maximum blue correspond to $-1$~N and $+1$~N of error, respectively. Focus on the substantial improvement of the pose-conditioned models in some scenes.}
  \label{fig:hom-results-full}
\end{figure*}

\Cref{fig:hand-pose-selected-full} shows a larger set of hand-pose predictions (the full companion to \cref{fig:hand-pose-results}), and \cref{fig:hom-results-full} the full HOM force visualization (companion to \cref{fig:hom-results}).

\subsection{Model Capacity Ablation}
\label{app:ablation-capacity}
\label{sect:ablation-model-capacity}

We sweep a $10\times10$ grid of network widths, varying the \emph{Input Reduce} layer (the first dense layer that compresses the raw pressure features) and the \emph{RNN Hidden} layer (the recurrent layer that follows). Both axes take values $\{32, 40, 48, 64, 80, 96, 128, 160, 192, 256\}$, giving 100 configurations per ablation. All other hyperparameters follow the defaults of \cref{sect:training-eval}, with a fixed random seed. Each configuration's mean L1 validation error is averaged across all available recordings for that dataset, then expressed as a percentage change relative to the \textbf{nominal} configuration with both layers at width 128 (cell outlined in black; positive values are worse than baseline, negative values are better). The color scale saturates at $\pm 10\%$ on each side.

\begin{figure}[t]
  \centering
  \begin{tikzpicture}[every node/.style={inner sep=0pt, outer sep=0pt}]
    \def\panelw{0.49\columnwidth}
    \node (a) at (0, 0)
      {\includegraphics[width=\panelw]{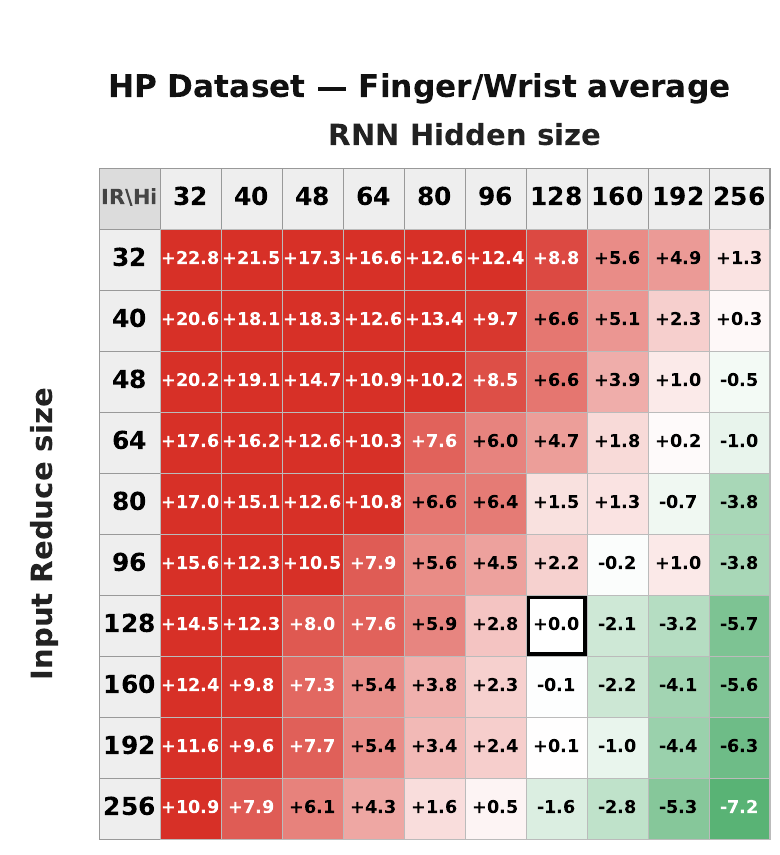}};
    \node (b) at ($(a.east) + (1mm, 0)$) [anchor=west]
      {\includegraphics[width=\panelw]{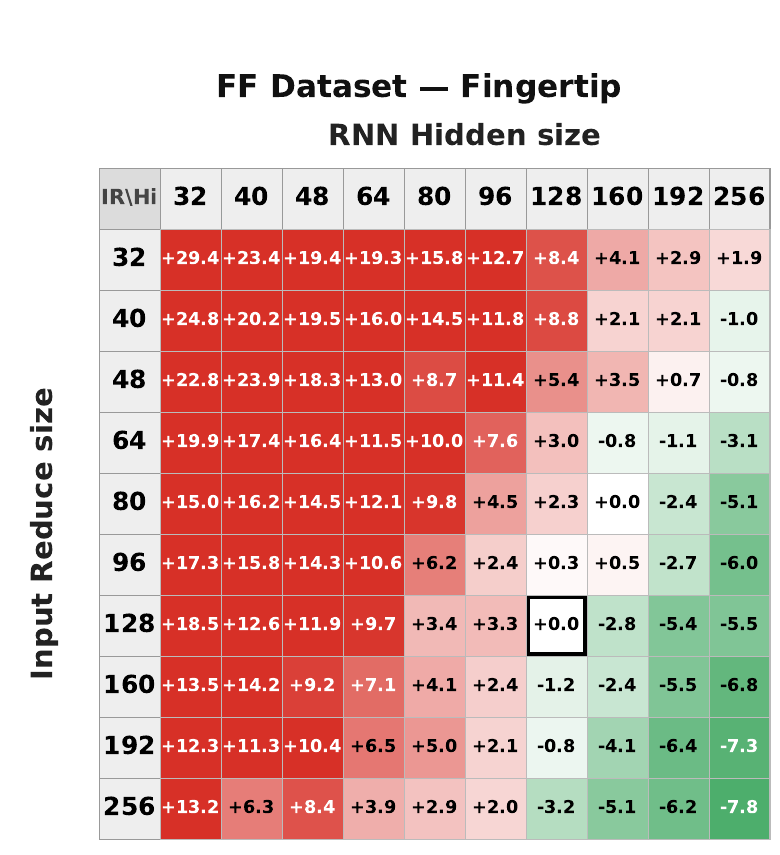}};
    \node (c) at ($(a.south) + (0, -1mm)$) [anchor=north]
      {\includegraphics[width=\panelw]{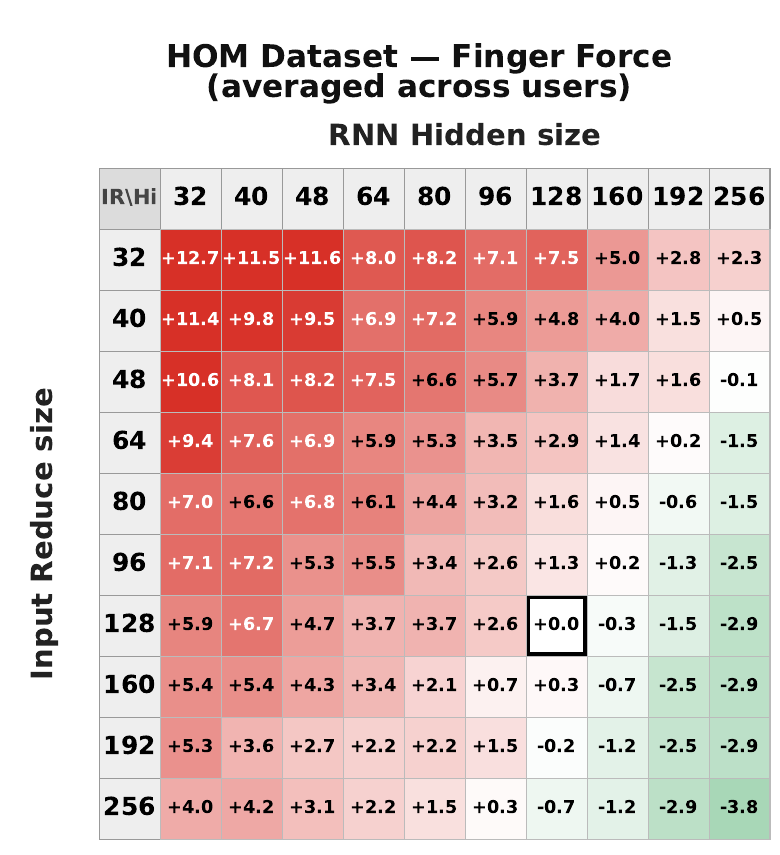}};
    \node (d) at ($(b.south) + (0, -1mm)$) [anchor=north]
      {\includegraphics[width=\panelw]{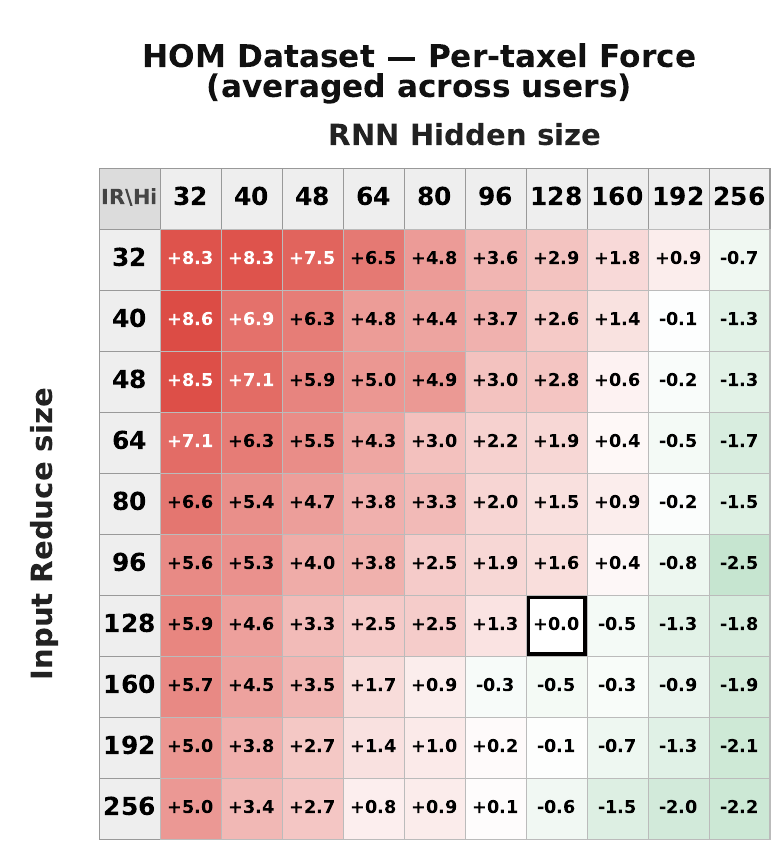}};
  \end{tikzpicture}
  \caption{Validation-error sensitivity to network width across our four datasets/tasks. Each cell is the per-configuration L1 error expressed as a percent change from the $(128,128)$ baseline (outlined in black); green cells are better than baseline, red cells are worse, with color saturating at $\pm 10\%$. (Top-left) \textbf{HP dataset}: per-channel percentages for the finger and wrist joint targets (degrees) are computed against their own baselines and then averaged. (Top-right) \textbf{FF dataset}: single-task fingertip force regression (Newtons), averaged over 18 recordings. (Bottom-left, bottom-right) \textbf{HOM dataset}: per-task models for finger force and per-taxel force (Newtons), averaged equally across four users.}
  \label{fig:capacity}
\end{figure}

The four panels in \cref{fig:capacity} tell a consistent story across otherwise quite different prediction problems.

\paragraph{The RNN hidden size dominates.}
Within every panel, the error gradient is markedly steeper along the horizontal axis (RNN hidden size) than along the vertical axis (input reduction). On HP, growing the RNN hidden layer from 32 to 256 removes roughly twice as much error as growing the input-reduce layer over the same range; on FF, the ratio is similar. This is the most actionable finding: under a fixed parameter budget, capacity should be spent on the recurrent layer first. A typical example is the $(64,256)$ configuration, which sits within $1$--$2$ percentage points of the $(256,256)$ corner on every dataset, while using a fraction of the parameters.

\paragraph{Returns diminish quickly past the baseline.}
On every panel, the $(128,128)$ baseline already sits close to the green-shaded plateau in the bottom-right quadrant. The best single-cell improvement is bounded by $\approx 7$--$8\%$ on HP and FF, and $3$--$4\%$ on the pooled HOM tables. Conversely, dropping back to $(64,64)$ or smaller increases loss by $6$--$15\%$, so the nominal configuration is a reasonable default in absolute terms.

\paragraph{Task difficulty does not change the shape of the surface.}
The HOM/glove panel has a much smaller dynamic range than HOM/finger (absolute errors are roughly 5--10$\times$ lower for glove targets), yet the relative response surface --- where the best cells sit, where the plateau begins, and the stronger sensitivity to RNN hidden size than to input-reduce width --- is essentially the same. This suggests that the asymmetry is structural to the network rather than a feature of any one task.

\paragraph{Default configuration.}
We adopt $(128,128)$ as our default. It sits in the plateau region on all four tasks, and given the limited amount of training data we have, the marginal gains from a larger model are not worth the additional overfitting risk.

\subsection{Context Length Ablation}
\label{app:ablation-context}
\label{sect:ablation-context-length}

\begin{figure*}[t]
  \centering
  \includegraphics[width=0.32\textwidth]{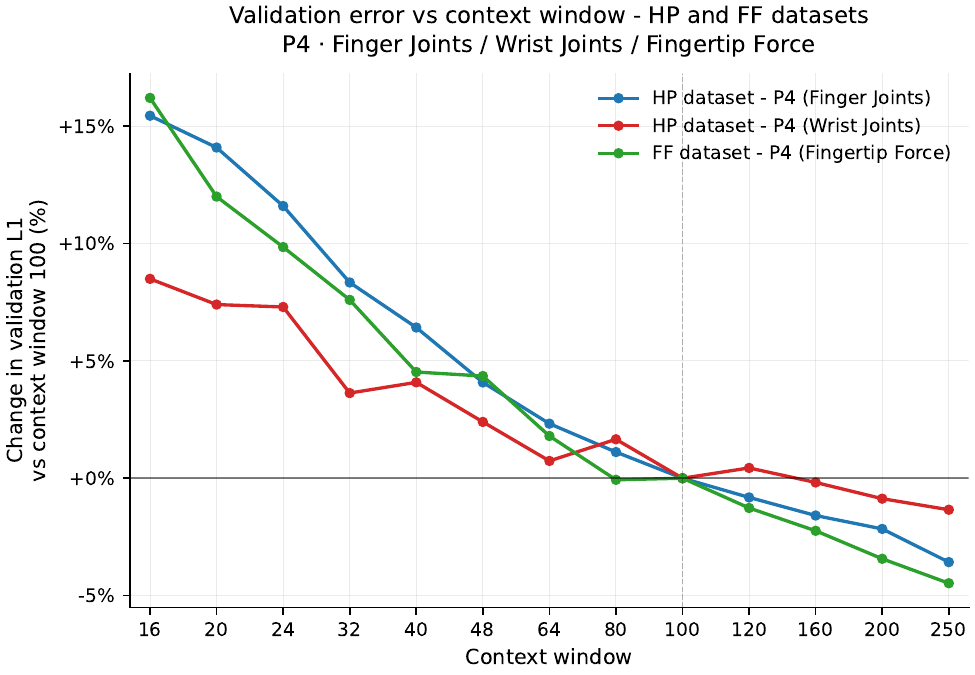}\hfill
  \includegraphics[width=0.32\textwidth]{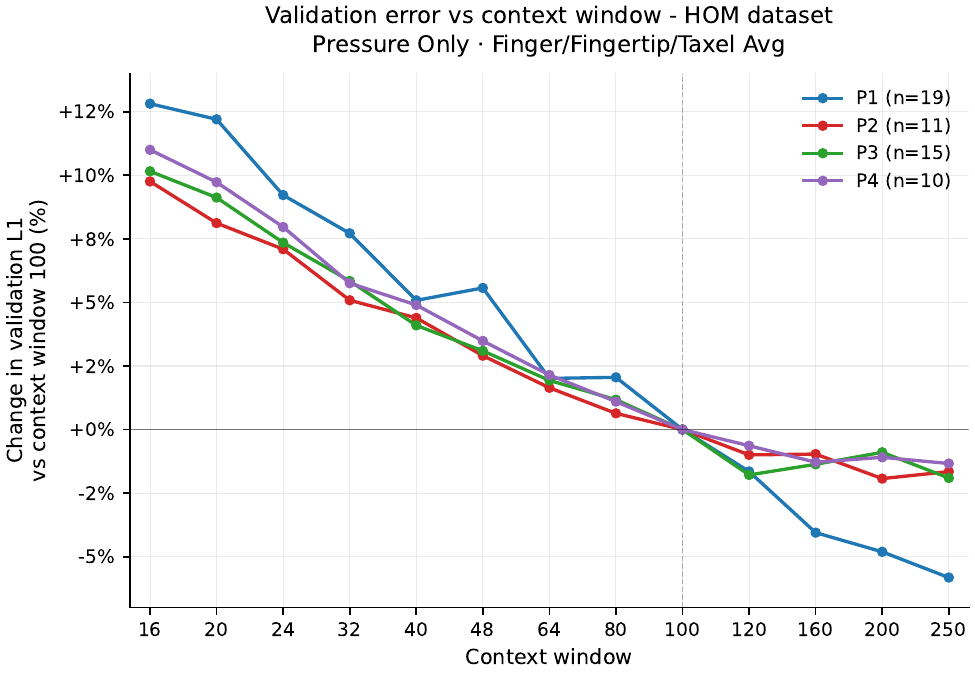}\hfill
  \includegraphics[width=0.32\textwidth]{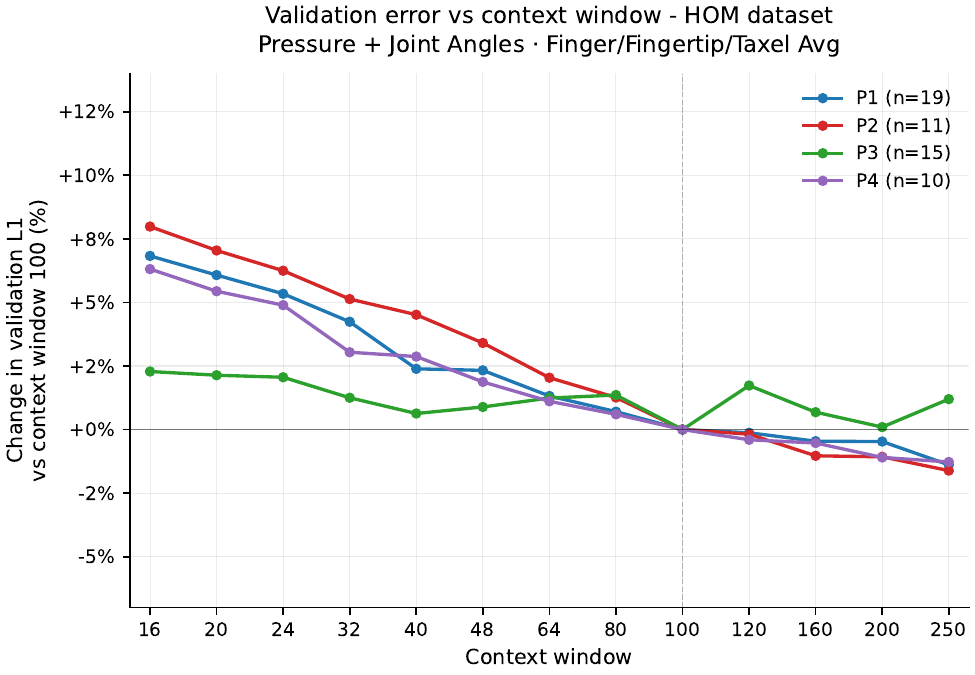}
  \caption{Effect of training context window on validation $L_1$ error, expressed as percent change relative to the model trained with a context window of $100$ samples. \emph{Left:} HP and FF datasets, single subject (P4) --- one curve per regression target. \emph{Middle:} HOM dataset with pressure-only input, four subjects; each curve is the per-subject average percent change across the Finger, Fingertip, and Taxel output tasks. \emph{Right:} same as middle, but with pose features concatenated to the pressure input. Shrinking the window to $16$ samples increases error by $8$--$16\%$ on HP/FF and $10$--$13\%$ on HOM pressure-only; extending it past $\approx100$ buys back only $1$--$6\%$, so all curves enter a diminishing-returns regime well before the longest setting. Adding the pose channel (right) more than halves the gain from longer context (e.g.\ P1 drops from $\sim13\%$ to $\sim7\%$ head-room), consistent with longer history partly compensating for missing kinematic input.}
  \label{fig:ctxwin-all}
\end{figure*}

Across all three datasets, validation error decreases monotonically with context window up to roughly $100$ samples, then enters a regime of strongly diminishing returns (\cref{fig:ctxwin-all}). Doubling the window from $100$ to $200$ reduces error by less than $\sim2\%$ in every condition; quadrupling it to $400$ would, by extrapolation, fall well inside the run-to-run noise observed across recordings. We therefore adopt a context window of $100$ as a reasonable operating point that captures essentially all of the benefit of temporal context, while keeping training and inference cost bounded.

The cross-dataset comparison (\cref{fig:ctxwin-all}, left panel) shows that the two HP outputs and the FF output respond similarly to context length, despite measuring very different quantities (angles vs.\ forces); this suggests the effect is driven by the temporal dynamics of the pressure input rather than by properties of the regression target.

Within the HOM dataset, the middle and right panels of \cref{fig:ctxwin-all} expose a clear interaction between input richness and context-window benefit. Removing the pose channel roughly \emph{doubles} the gain from longer context: e.g.\ P1 improves by $\sim13\%$ from blk-16 to blk-100 without pose, but only $\sim7\%$ with pose. In other words, longer context partially compensates for missing kinematic input, and once that input is available the model needs much less history to reach the same operating point. The per-subject spread is also larger in the pressure-only setting, consistent with longer history mattering more when the network has fewer cues for disambiguation.

\subsection{Sensor-Subset Ablation}
\label{app:ablation-sensor}
\label{sect:ablation-sensor}

The PPS \#6863 pressure-sensor array used in our wristband (\cref{sect:wristband}) provides an $18 \times 10$ taxel grid. We analyze whether the full footprint is necessary by ablating along three axes: (i) the number of contiguous rows used as input, (ii) the position of the row window along the forearm, and (iii) the combination of row length with column subsampling.

\subsubsection{Effect of window length on regression error.}
\label{sect:ablation-sensor-length}
\begin{figure*}[t]
  \centering
  \begin{tikzpicture}
    \node[anchor=south west, inner sep=0pt] (a) at (0,0)
      {\includegraphics[width=0.32\linewidth]{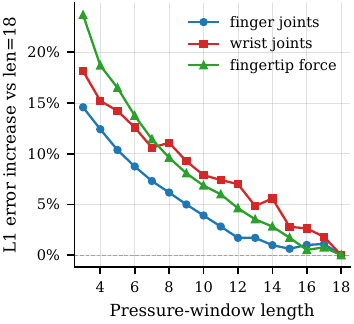}};
    \node[anchor=south west, inner sep=0pt] (b) at ($(a.south east) + (4pt,0)$)
      {\includegraphics[width=0.32\linewidth]{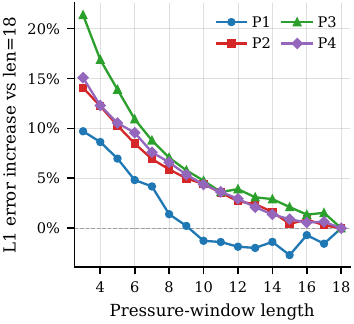}};
    \node[anchor=south west, inner sep=0pt] (c) at ($(b.south east) + (4pt,0)$)
      {\includegraphics[width=0.32\linewidth]{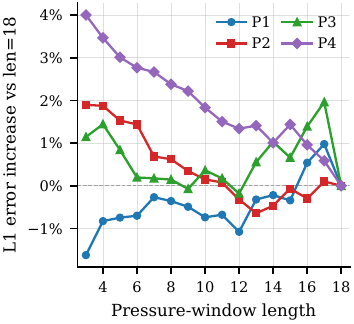}};
  \end{tikzpicture}
  \caption{Pressure-window length ablation across the three datasets. Each panel plots one or more estimation tasks as a function of the contiguous-row window length $L$ used as model input. Y-axis is per-curve MAE expressed as percent change relative to the same model trained on the full $L = 18$ strip, so every curve is anchored at $0\,\%$ on the right edge. X-axis is the window length $L \in \{3, 4, \dots, 18\}$. For each $L$ we sweep every valid start offset $s \in \{0, 1, \dots, 18 - L\}$ and average the resulting MAE across $s$ and across all recordings of the dataset before computing the percent change. All runs use the full 10-column resolution. \emph{Left:} HP and FF datasets --- three estimation tasks on a single axis (HP finger and wrist joint angles in degrees; FF fingertip force in Newtons). \emph{Middle:} HOM, \emph{no-pose} input. Each curve is one HOM user (P1--P4), averaged over the finger and glove estimation tasks. \emph{Right:} HOM, \emph{pose} input (sensor strip plus ground-truth finger and wrist angles). Same four users, averaged over the same two estimation tasks.}
  \label{fig:length-ablation}
\end{figure*}

We sweep the length of the contiguous pressure window used as model input, $L \in \{3, 4, \dots, 18\}$, over all valid start offsets $s \in \{0, 1, \dots, 18 - L\}$ within the 18-row strip, while keeping the full 10-column resolution. Row indices are 0-based throughout the paper, and the strip is worn along the forearm with a fixed anatomical convention: \emph{row 0 corresponds to the elbow side of the strip, and row 17 to the wrist side}. Each $(L, s)$ configuration is trained independently on every recording of the three datasets, scored per the protocol of \cref{sect:training-eval}, and the resulting MAE is averaged across $s$ and recordings to produce a single MAE-per-$L$ value. Each curve in \cref{fig:length-ablation} is normalised by its own MAE at $L = 18$, so that the comparison is dimensionless and every curve is anchored at $0\,\%$ on the right edge.

All three datasets exhibit the same qualitative behaviour: MAE decreases monotonically with window length, with most of the recoverable accuracy captured by $L \approx 10$--$12$ and only marginal additional gains thereafter. This common saturation pattern across heterogeneous output modalities --- joint angles in degrees, fingertip force in Newtons, and HOM per-finger and per-taxel force --- suggests it reflects a property of the input signal itself: a roughly constant ``information per added row'' profile that flattens once the spatial bandwidth of the pressure strip is saturated, rather than anything specific to the regressed quantity.

The magnitude of the loss differs across estimation tasks. The FF fingertip task shows the largest relative sensitivity to window length (\cref{fig:length-ablation} \emph{left}, $\approx 15\,\%$ degradation at $L = 4$). Two non-exclusive explanations are plausible: fingertip-force pressure signatures may be localized to a small subset of forearm muscles that a short window can miss, and force estimation may rely more heavily on wrist-pose disambiguation, which a wider strip provides. The HP estimation tasks degrade more gently (10--13\,\%), with the wrist channel being the more demanding of the two in absolute units.

The HOM panels make the most useful design point. When the network has only pressure as input (\cref{fig:length-ablation} \emph{middle}, \emph{no-pose}), reducing $L$ increases loss by 5--15\,\% across all four users. Adding the ground-truth finger and wrist angles to the input (\cref{fig:length-ablation} \emph{right}, \emph{pose}) reduces the ablation loss to within a few percent across the entire sweep --- the curves are essentially flat. Once exact joint angles are available, the pressure strip is largely redundant for force prediction, and progressively removing rows from it adds little loss. This dichotomy is the basis of our minimum-footprint sensor argument: a small pressure strip is sufficient \emph{provided} that some independent pose signal is available to the inference pipeline.

\subsubsection{Where on the forearm does the window matter?}
\label{sect:ablation-sensor-position}
The length sweep above establishes \emph{how many} rows are needed, but is agnostic to \emph{where} the window sits on the forearm. We re-examine the same runs by plotting each $(L, s)$ result as a single point at the spatial centre of its window --- $x = s + (L-1)/2$ on the row axis (0-based) --- with curves connecting all starts of a fixed length. Recall the anatomical convention: row 0 lies on the elbow side of the strip and row 17 on the wrist side, so the x-axis below reads forearm-proximal\,$\rightarrow$\,distal.

\begin{figure*}[t]
  \centering
  \begin{tikzpicture}
    \node[anchor=south west, inner sep=0pt] (a) at (0,0)
      {\includegraphics[width=0.325\linewidth]{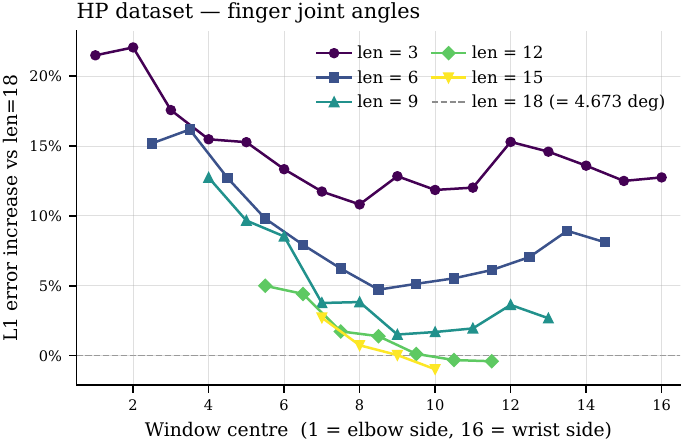}};
    \node[anchor=south west, inner sep=0pt] (b) at ($(a.south east) + (4pt,0)$)
      {\includegraphics[width=0.325\linewidth]{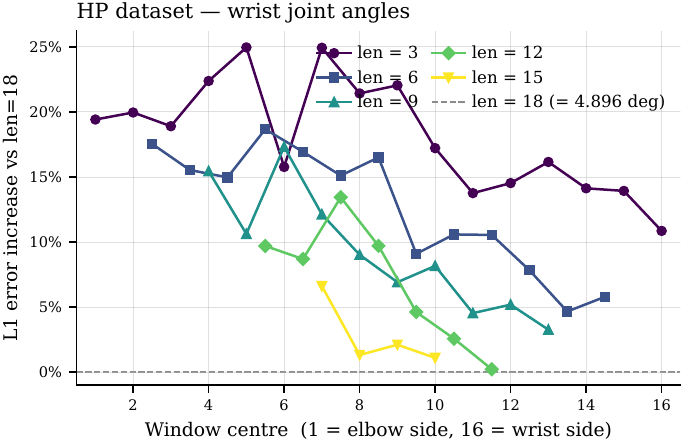}};
    \node[anchor=south west, inner sep=0pt] (c) at ($(b.south east) + (4pt,0)$)
      {\includegraphics[width=0.325\linewidth]{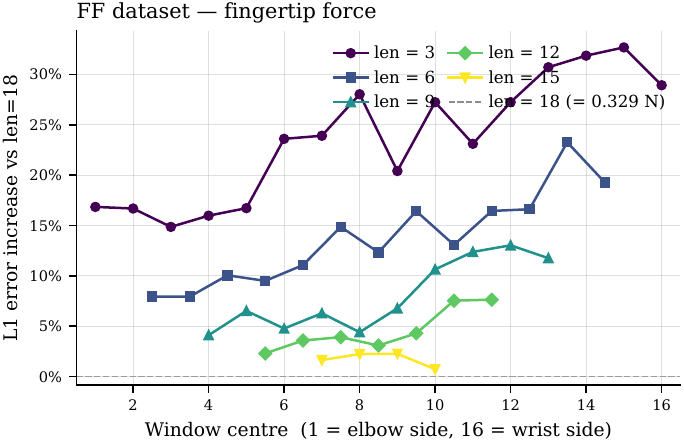}};
  \end{tikzpicture}
  \caption{Window-position ablation for the two higher-grade datasets: \emph{Left:} HP --- finger joint angles, \emph{Middle:} HP --- wrist joint angles, \emph{Right:} FF --- fingertip force. Each colored curve is one window length $L \in \{3, 6, 9, 12, 15\}$; each marker is one start offset $s$ positioned at the window's spatial centre. X-axis is the 0-based row index of the window centre (0 = elbow side, 17 = wrist side). Y-axis is MAE as percent change vs.\ the full $L = 18$ strip (dashed line at $0\,\%$). All runs use the full 10-column resolution.}
  \label{fig:window-position-main}
\end{figure*}

\begin{figure*}[t]
  \centering
  \begin{tikzpicture}
    \node[anchor=south west, inner sep=0pt] (a) at (0,0)
      {\includegraphics[width=0.243\linewidth]{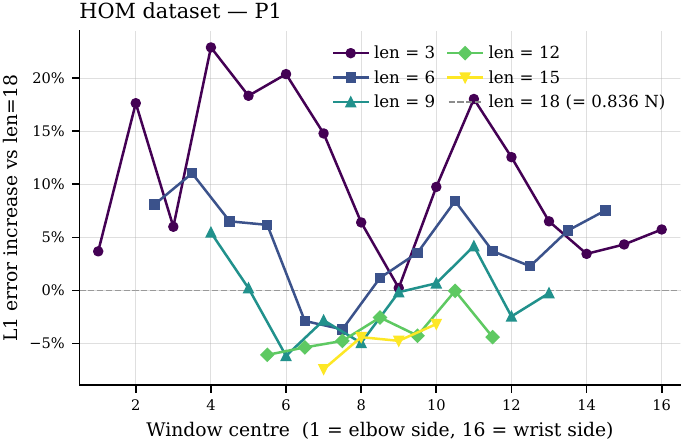}};
    \node[anchor=south west, inner sep=0pt] (b) at ($(a.south east) + (4pt,0)$)
      {\includegraphics[width=0.243\linewidth]{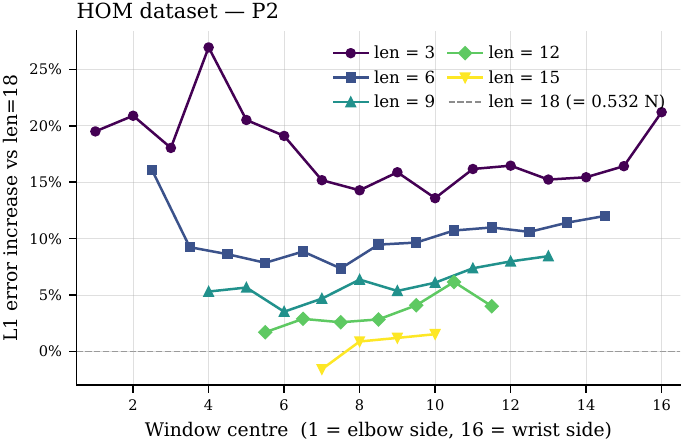}};
    \node[anchor=south west, inner sep=0pt] (c) at ($(b.south east) + (4pt,0)$)
      {\includegraphics[width=0.243\linewidth]{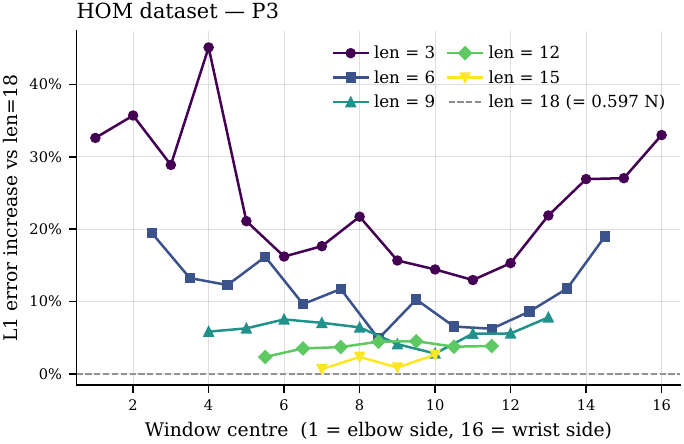}};
    \node[anchor=south west, inner sep=0pt] (d) at ($(c.south east) + (4pt,0)$)
      {\includegraphics[width=0.243\linewidth]{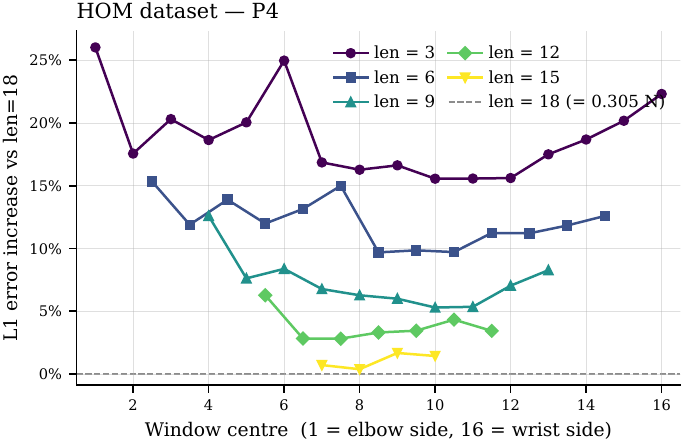}};
  \end{tikzpicture}
  \caption{Window-position ablation for the four HOM users P1--P4 (no-pose / finger configuration). Each colored curve is one window length $L \in \{3, 6, 9, 12, 15\}$; each marker is one start offset $s$ positioned at the window's spatial centre ($x = s + (L-1)/2$, 0-based), and starts of the same length are connected by a line. X-axis is the 0-based row index of the window centre (0 = elbow side, 17 = wrist side); the strip is worn so that this axis reads forearm-proximal\,$\rightarrow$\,distal. Y-axis is MAE as percent change vs.\ the full $L = 18$ strip (dashed line at $0\,\%$); the legend in each panel reports the corresponding $L = 18$ MAE in Newtons. All runs use the full 10-column resolution. One panel per HOM user in the order P1, P2, P3, P4.}
  \label{fig:window-position-hom}
\end{figure*}

For every estimation task, very short windows ($L = 3$) leave large gaps and are markedly worse than longer windows --- typically $10$--$30\,\%$ above the full-strip baseline depending on dataset --- but \emph{which} short window is least bad depends on the estimation task. Finger-joint and fingertip-force estimation tasks both prefer windows whose centre falls on the wrist side of the strip (high row index), which is anatomically consistent: the muscles whose tendons actuate the fingers are distributed along the forearm, but the most discriminative pressure signal for fingertip activity tends to come from the distal half of the strip closer to the wrist. A comparable wrist-side advantage has been reported in an adjacent modality: EI-Lite~\cite{zhu2025eilite} found that placing a 4-electrode bioimpedance band at the distal radioulnar joint approximately halved cross-user keypoint error relative to the proximal carpal row ($13.09$~mm vs $25.35$~mm MPJPE), which we read as cross-modality evidence that the distal half of the forearm is the more information-dense region for hand-state decoding. Wrist-joint regression (HP --- wrist joints in \cref{fig:window-position-main}) is, unsurprisingly, the most strongly biased of the three: the curves for $L \in \{3, 6\}$ drop steeply toward the wrist side and rise sharply when the window is centred on the elbow half of the strip. As $L$ grows, the curves flatten --- once the window is long enough to span the informative region regardless of start, position stops mattering. The four HOM users (\cref{fig:window-position-hom}) show the same overall shape, with larger absolute swings reflecting the higher per-recording variance of the at-home data; P4 in particular has the noisiest baseline. The curves themselves are visibly jagged. Two non-exclusive effects contribute. First, each marker is a single training run aggregated over relatively few valid $(L, s)$ combinations, so individual point-to-point fluctuations should be read as run-to-run noise around the smooth trend rather than as substantive structure. Second, the \#6863 array itself has a mild row-to-row baseline difference (the row-banding artefact described in \cref{para:row-banding}); since the x-axis here \emph{is} the row index, any residual per-row bias survives averaging and shows up as small repeatable bumps superimposed on the smoother shape.

\subsubsection{Combined length \texorpdfstring{$\times$}{x} column-subsampling tables.}
\label{sect:ablation-sensor-subsampling}
To complete the picture, we sweep \emph{both} spatial ablation axes together. In addition to the contiguous row window $L \in \{3, 6, 9, 12, 15, 18\}$ used as model input, we also subsample the 10 pressure columns by keeping only every $k$-th column with $k \in \{1, 2, 3, 4, 5\}$; the kept-column counts are $10, 5, 4, 3, 2$. Cells are coloured by percent change relative to the same dataset's $(L = 18, \text{cols}=10)$ baseline (per-table viridis scale, $0\,\%$ $\rightarrow$ table maximum).

\begin{figure}[t]
  \centering
  \begin{tikzpicture}[inner sep=0pt, outer sep=0pt, node distance=2pt]
    \node (a)              {\includegraphics[width=0.32\columnwidth]{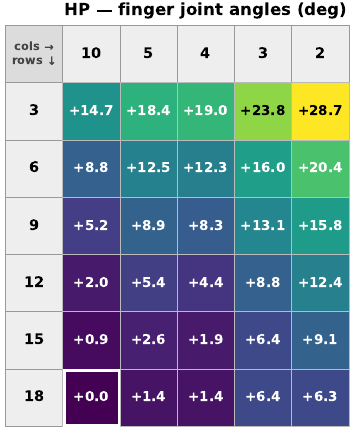}};
    \node (b) [right=of a] {\includegraphics[width=0.32\columnwidth]{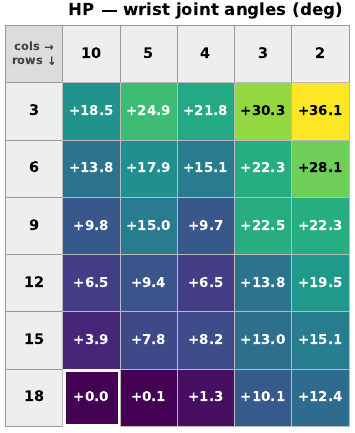}};
    \node (c) [right=of b] {\includegraphics[width=0.32\columnwidth]{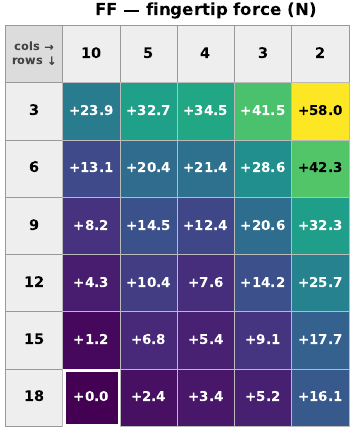}};
  \end{tikzpicture}
  \caption{Combined length $\times$ column-subsampling tables for the two higher-grade datasets: HP --- finger joint angles, HP --- wrist joint angles, FF --- fingertip force. Each cell shows the percent change in MAE relative to that dataset's own $(L = 18, \text{cols}=10)$ full-strip baseline. Rows correspond to the pressure-window length $L$; columns correspond to the number of pressure columns retained after subsampling (out of 10). Cells are averaged over all valid start offsets and over all recordings of the dataset. Colour is viridis on a per-table scale from $0\,\%$ to the table's own maximum increase; the absolute baseline MAE and the table-specific $v_{\max}$ are annotated below each table.}
  \label{fig:tables-main}
\end{figure}

\begin{figure}[t]
  \centering
  \begin{tikzpicture}[inner sep=0pt, outer sep=0pt, node distance=2pt]
    \node (a)                            {\includegraphics[width=0.49\columnwidth]{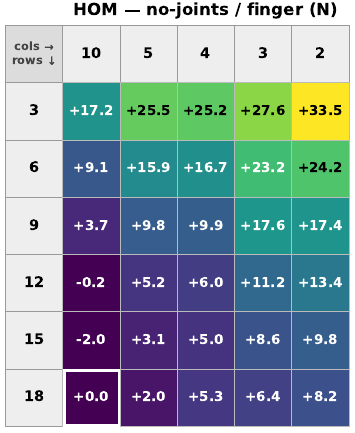}};
    \node (b) [right=of a]               {\includegraphics[width=0.49\columnwidth]{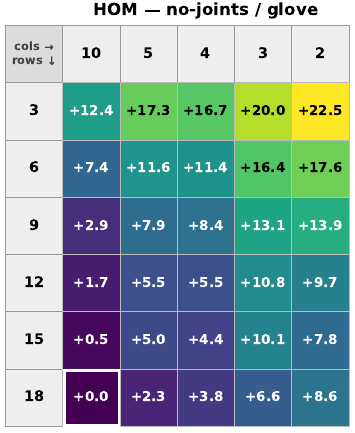}};
    \node (c) [below=of a.south west, anchor=north west] {\includegraphics[width=0.49\columnwidth]{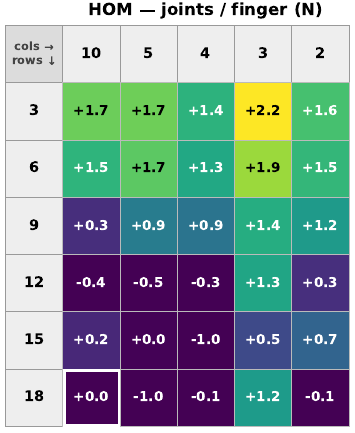}};
    \node (d) [right=of c]               {\includegraphics[width=0.49\columnwidth]{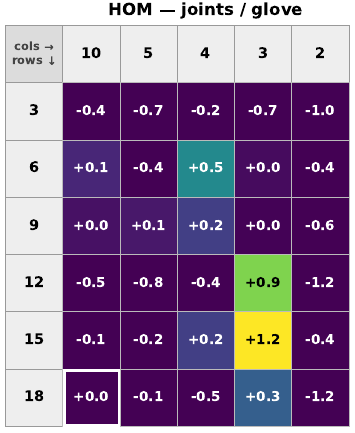}};
  \end{tikzpicture}
  \caption{Combined length $\times$ column-subsampling tables for the four HOM configurations (averaged across the four users P1--P4 and across all recordings and start offsets). Top row: \emph{no-pose} input (pressure strip alone), for the finger and glove estimation tasks respectively. Bottom row: \emph{pose} input (pressure strip concatenated with ground-truth finger and wrist joint angles), again for finger and glove. Axes, baseline-cell convention, and colour scale match \cref{fig:tables-main}: rows are window length $L$, columns are pressure-columns kept after subsampling, cells show percent change vs.\ the per-table $(L=18, \text{cols}=10)$ baseline, and each table uses its own viridis scale from $0\,\%$ to its table-specific maximum.}
  \label{fig:tables-hom}
\end{figure}

Three patterns emerge across all seven tables. First, length and column-count are roughly additive penalties: cells along the diagonal (short window \emph{and} few columns) are the worst, while either-or ablations are noticeably gentler. Second, the column direction is more forgiving than the row direction at moderate subsampling --- halving from 10 to 5 columns adds only $\sim$1--2\,\% loss across HP and FF (versus $5$--$10\,\%$ for the analogous row-halving from 18 to 9), and dropping to 4 columns is still essentially free on the kinematics tasks. The asymmetry breaks down at extreme column subsampling, where dropping to 3 or 2 columns adds $6$--$16\,\%$ on HP/wrist and FF (\cref{fig:tables-hom}, left two panels). Part of this is structural: row removal is contiguous and excises a whole anatomical band, whereas column removal is strided and only thins spatially-adjacent samples, which carry strongly redundant signal because soft tissue acts as a spatial low-pass filter between muscles and skin. Moderate column subsampling discards mostly that redundancy; only once the spacing approaches the natural correlation length does each removed column take real information with it. Sahnoun et al.~\cite{sahnoun2025eitreduction} report a comparable graceful-degradation pattern in EIT (halving 40 to 20 boundary voltages costs $1.4$--$6.4\%$ classification accuracy), suggesting that spatial redundancy at the band--skin interface is a structural property of wrist-worn sensing rather than a quirk of either modality. Third, and most useful for the minimum-footprint argument, \cref{fig:tables-hom} makes the pose-vs-no-pose contrast crisp: the two right-hand panels (\emph{pose} input) are essentially monochromatic --- every cell within a few percent of the baseline, including the extreme corner at $(L=3, \text{cols}=2)$ --- whereas without the side channel (left two panels, especially \emph{no-pose / finger}) the same aggressive subsampling adds $30\,\%$ or more loss on average.

We also tried a spatial pre-smoothing pass before column subsampling, intended to mitigate information loss from aliasing when columns are dropped. In practice, the smoothed variants differed from the no-smoothing runs by at most a few percent across all cells, so we report the unsmoothed configuration throughout.

\subsection{Loss-Weight Ablations}
\label{app:ablation-loss-weights}
\label{sec:supp_loss_weights}

Our multi-task losses combine separate MSE terms for finger angles, wrist angles, and (when applicable) force. The relative weights between these terms control how the model allocates capacity across tasks. We sweep two of them: the wrist/finger ratio inside the pose loss, and the force/pose ratio when force and pose are estimated together.

\subsubsection{Wrist vs.\ finger (pose only).}
\label{sec:supp_loss_weights_wrist}
Since we compute mean L2 losses separately for finger and wrist angles, and there are 20 finger angles versus 3 wrist angles, the relative wrist-loss weight controls how much emphasis the model places on wrist accuracy. Setting the wrist term to 0.4 assigns each wrist angle approximately $0.4/(3/20)=2.66$ times the relative weight of each finger angle. We sweep the wrist-loss weight $\lambda_W$ across 17 values spanning $0.01$ to $2.0$ on HP. \Cref{fig:loss_weight_ablations} (top-left) shows the expected trade-off: too little wrist weighting under-fits wrist angles, while too much weighting degrades finger-angle performance. A wrist-loss weight of $0.4$ provides a practical balance and is used as the default across all experiments.

\subsubsection{Force vs.\ pose (\textit{Force+Pose} model).}
\label{sec:supp_loss_weights_force}
\begin{figure}[htbp]
  \centering
  \begin{tikzpicture}[inner sep=0pt, outer sep=0pt]
    \node (a)                                {\includegraphics[width=0.49\columnwidth]{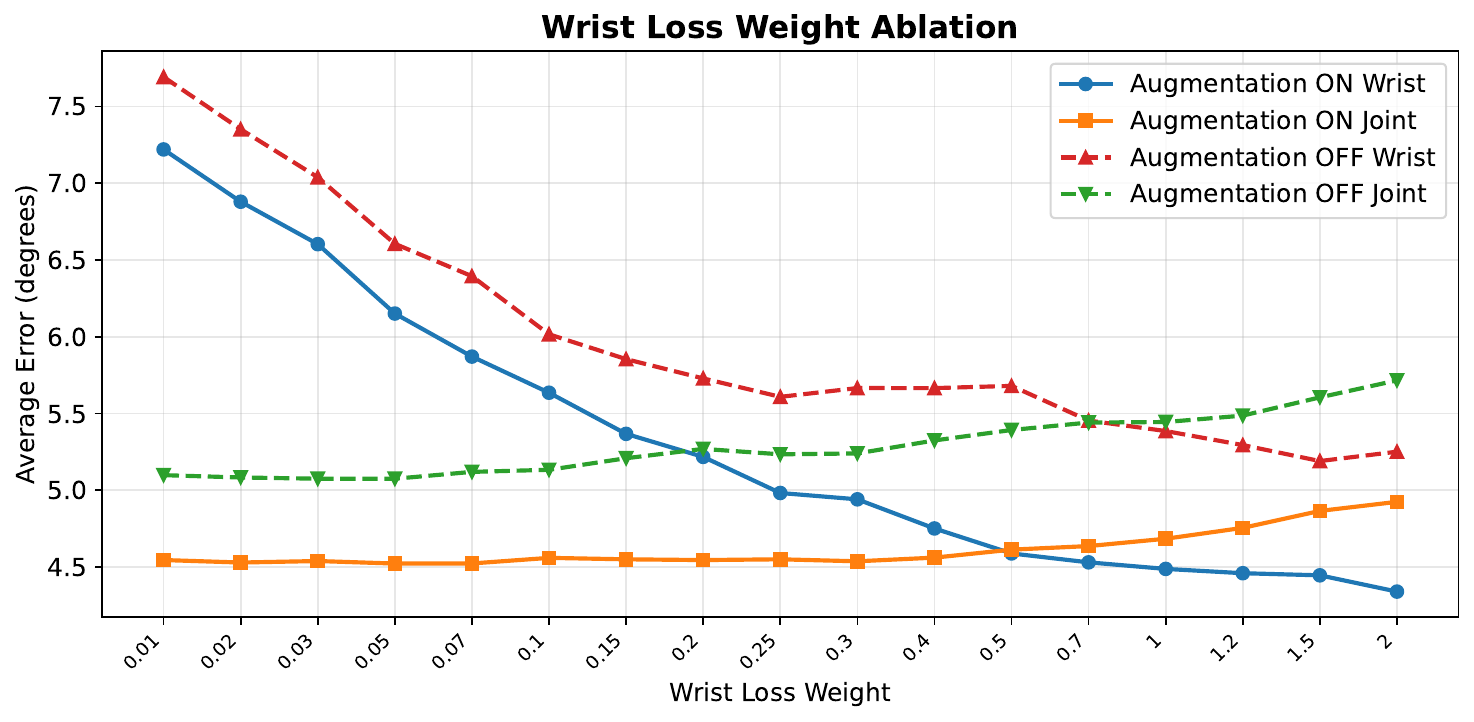}};
    \node (b) at (a.east)   [anchor=west, xshift=2pt]
                                             {\includegraphics[width=0.49\columnwidth]{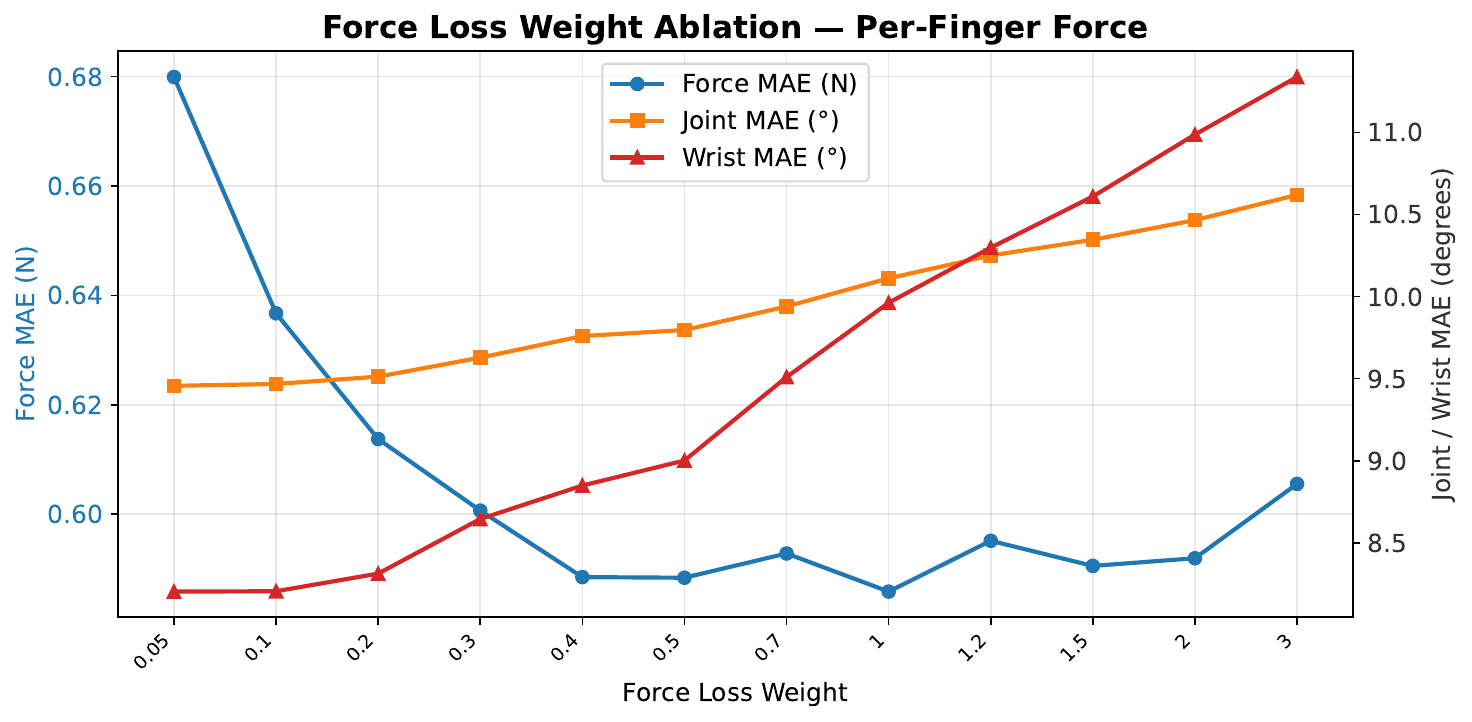}};
    \node (c) at (a.south)  [anchor=north, yshift=-2pt]
                                             {\includegraphics[width=0.49\columnwidth]{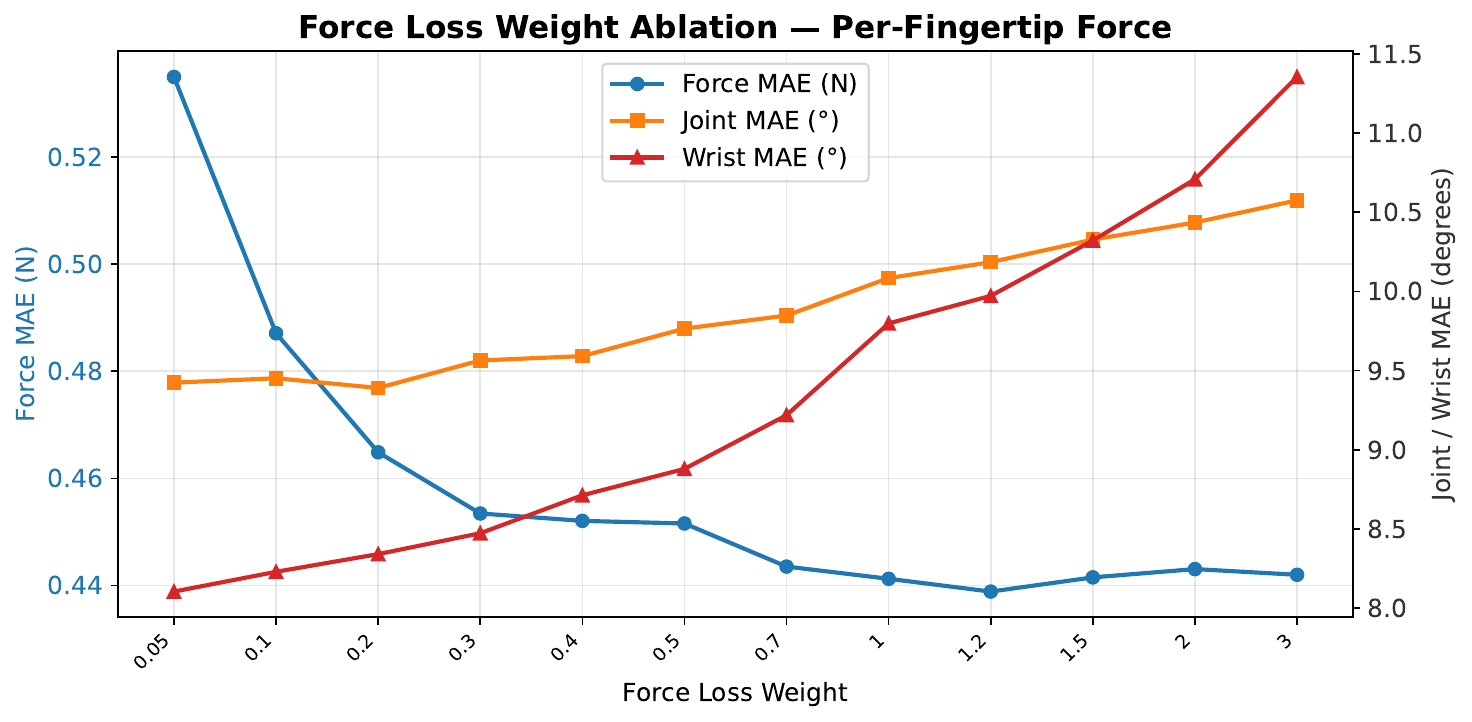}};
    \node (d) at (b.south)  [anchor=north, yshift=-2pt]
                                             {\includegraphics[width=0.49\columnwidth]{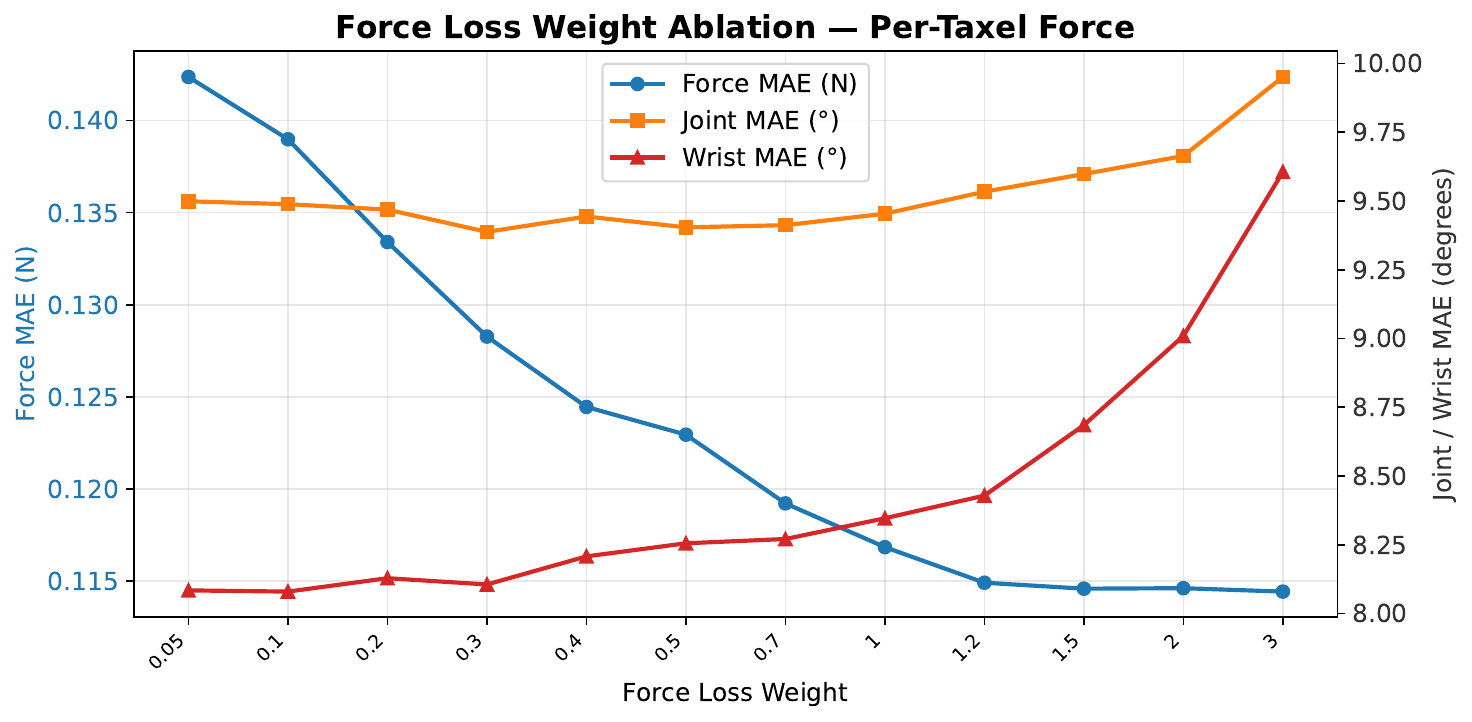}};
  \end{tikzpicture}
  \caption{Loss-weight ablations. \emph{Top-left:} wrist vs.\ finger trade-off as the relative wrist-loss weight is varied (pose-only model on HP); operating point $0.4$. \emph{Other three panels:} force vs.\ pose trade-off for the \textit{Force+Pose} model on HOM (averaged across four users) at three force granularities --- top-right per-finger force, bottom-left per-fingertip force, bottom-right per-taxel (glove) force. Each force panel plots force MAE on the left axis and finger/wrist joint MAE on the right axis as $\lambda_F$ varies. Operating points $\lambda_F = 0.4$, $0.4$, $1.2$ are picked at the elbow of each trade-off.}
  \label{fig:loss_weight_ablations}
\end{figure}

When the network is trained to predict force and joint angles together, the total loss combines a force MSE term (with weight $\lambda_F$) and the pose MSE terms above (relative wrist weight $0.4$). The relative force weight $\lambda_F$ controls how much the model trades off pose accuracy for force accuracy. We sweep $\lambda_F \in \{0.05, 0.1, 0.2, 0.3, 0.4, 0.5, 0.7, 1.0, 1.2, 1.5, 2.0, 3.0\}$ at three force granularities (per-finger, per-fingertip, and per-taxel), separately for each HOM user, and report the average across users. The three force-weight panels of \cref{fig:loss_weight_ablations} (top-right, bottom-left, bottom-right) are consistent across granularities: force MAE drops sharply at very small $\lambda_F$ then flattens, while joint and wrist MAE rise roughly monotonically as $\lambda_F$ grows --- the model is reallocating capacity from pose to force. We pick the operating point at the elbow of each curve. For per-finger and per-fingertip force, this lands at $\lambda_F{=}0.4$, where the force curve has reached within $\sim$2\,\% of its asymptote and wrist MAE has only degraded by about $7\,\%$ relative to the $\lambda_F{=}0.05$ extreme. Per-taxel (glove) force is a $65$-channel target whose per-channel absolute magnitudes are roughly $4$--$5\times$ lower than per-finger force, so each force-MSE term contributes much less when summed at a comparable $\lambda_F$; correspondingly, a higher weight ($\lambda_F{=}1.2$) is needed to put the force loss on the same effective footing and capture most of the achievable force improvement. The corresponding pose penalty is larger but still acceptable. These three operating points ($\lambda_F = 0.4$, $0.4$, $1.2$) are the defaults used wherever the main paper reports \textit{Force+Pose} results.

\subsection{Augmentation Details}
\label{app:augmentation}
\label{sec:supp_augmentation}

\subsubsection{Transforms and Ranges}
\label{sec:supp_aug_ranges}

We apply five augmentations to the wristband input during training, each modelling a session-to-session nuisance factor (cf.\ \cref{sect:data-augmentation}): vertical shifts (up/down, up to 0.5 pixels) for proximal--distal placement differences; horizontal shifts (left/right, up to 0.5 pixels) for circumferential rotation around the wrist; in-plane rotation (up to $\pm 3^{\circ}$) for slight sensor misalignment; isotropic zoom (up to $\pm 3\%$) for day-to-day wrist-contour variation; and global intensity scaling (up to $\pm 15\%$) for overall pressure-level shifts under tighter or looser fit. These ranges are determined empirically. The same perturbation set is applied independently to each pressure array using bilinear resampling with clamping; augmentation parameters are sampled once per mini-sequence to preserve temporal consistency within each RNN window, and no augmentation is applied at inference time.

\subsubsection{Augmentation Effect}
\label{sec:supp_aug_effect}

\Cref{tab:aug_effect} compares matched training runs with and without augmentation across datasets and targets. MAE improves on every target: the gains are largest on the single-user HP pose, where placement and strap-tension variation matter most, and more modest on the harder, higher-variance HOM targets. Because every recording is a separate donning of the band, we attribute these gains chiefly to reduced sensitivity to session-to-session placement and interaction variability rather than to a generic regularisation effect. We therefore apply augmentation in all reported results, and the per-target search below examines whether the nominal configuration can be improved on.

\begin{table}[t]
  \centering
  \caption{Effect of data augmentation on prediction accuracy (MAE), summarised across datasets and targets. Each row compares matched training runs without and with augmentation; the last column is the relative MAE change (green = improvement). HOM pose is averaged over the four users.}
  \label{tab:aug_effect}
  \small
  \setlength{\tabcolsep}{5pt}
  \begin{tabular}{ll cc r}
    \toprule
    \textbf{Dataset} & \textbf{Target} & \textbf{No Aug} & \textbf{With Aug} & \textbf{$\Delta$} \\
    \midrule
    FF & Fingertip force (N)      & 0.361 & 0.328 & \improv{$-$9.3\%} \\
    \midrule
    \multirow{2}{*}{HP} & Finger angle ($^\circ$) & 5.32 & 4.56 & \improv{$-$14.3\%} \\
                        & Wrist angle ($^\circ$)  & 5.67 & 4.75 & \improv{$-$16.1\%} \\
    \midrule
    \multirow{2}{*}{\shortstack[l]{HOM\\pose}} & Finger angle ($^\circ$) & 9.85 & 9.28 & \improv{$-$5.8\%} \\
                        & Wrist angle ($^\circ$)  & 8.65 & 8.03 & \improv{$-$7.2\%} \\
    \midrule
    \multirow{3}{*}{\shortstack[l]{HOM force\\(Force+Pose)}}
       & Per-finger (N)    & 0.63 & 0.59 & \improv{$-$6\%} \\
       & Per-fingertip (N) & 0.47 & 0.45 & \improv{$-$4\%} \\
       & Per-taxel (PSI)   & 0.46 & 0.43 & \improv{$-$5\%} \\
    \midrule
    \multirow{3}{*}{\shortstack[l]{HOM force\\(Pose Input)}}
       & Per-finger (N)    & 0.47 & 0.45 & \improv{$-$4\%} \\
       & Per-fingertip (N) & 0.37 & 0.35 & \improv{$-$5\%} \\
       & Per-taxel (PSI)   & 0.36 & 0.35 & \improv{$-$3\%} \\
    \bottomrule
  \end{tabular}
\end{table}

\subsubsection{Per-Target Augmentation Search}
\label{sec:supp_augsearch}

The nominal augmentation configuration $[0.5, 0.5, 3, 3, 0]$ used in the main paper was selected empirically from a small number of trial runs, biased toward conservative magnitudes (cf.\ \cref{sect:data-augmentation}). To check whether per-task refinement could meaningfully improve on this default, we run a Bayesian-optimisation loop (Gaussian-process surrogate, Thompson-sampling acquisition) over the five-parameter configuration $[v, h, r, z, i]$ --- vertical shift, horizontal shift, rotation, zoom, intensity --- across ten targets: the FF and HP datasets, plus the per-user force and kinematics targets for the four HOM users P1--P4. Each candidate is scored as the percent change in test L1 error against the zero-augment baseline. \Cref{tab:best-configs} reports the best per-target configuration recovered, together with its improvement against the nominal default.

\begin{table}[t]
\centering
\small
\setlength{\tabcolsep}{4pt}
\begin{tabular}{l @{\quad} c@{}r@{,\,}r@{,\,}r@{,\,}r@{,\,}r@{}c @{\quad} r @{\quad} r}
\toprule
Target & \multicolumn{7}{c}{Best config $[v, h, r, z, i]$} & Score & vs.\ nominal \\
\midrule
FF dataset            & [ & 1.5 & 0   & 4   & 3  & 40 & ] & $-7.48\%$  & $-2.16$\,pp \\
HP dataset            & [ & 0.5 & 0.5 & 1.5 & 3  & 5  & ] & $-11.42\%$ & $-0.46$\,pp \\
\midrule
HOM-P1-force          & [ & 3   & 5   & 3   & 22 & 0  & ] & $-6.95\%$  & $-4.95$\,pp \\
HOM-P2-force          & [ & 0.5 & 0.5 & 3   & 7  & 5  & ] & $-5.63\%$  & $-0.21$\,pp \\
HOM-P3-force          & [ & 0   & 0.5 & 1   & 3  & 15 & ] & $-2.75\%$  & $-1.23$\,pp \\
HOM-P4-force          & [ & 0.5 & 0.5 & 3   & 3  & 40 & ] & $-2.50\%$  & $-0.04$\,pp \\
\midrule
HOM-P1-kinematics     & [ & 1.5 & 0   & 7   & 4  & 10 & ] & $-6.81\%$  & $-1.57$\,pp \\
HOM-P2-kinematics     & [ & 1   & 0.5 & 2   & 0  & 5  & ] & $-7.56\%$  & $-1.64$\,pp \\
HOM-P3-kinematics     & [ & 0.5 & 0.5 & 0   & 3  & 40 & ] & $-5.00\%$  & $-0.15$\,pp \\
HOM-P4-kinematics     & [ & 1.5 & 0   & 5   & 8  & 20 & ] & $-8.46\%$  & $-1.52$\,pp \\
\midrule
\textbf{Average}      & \multicolumn{7}{c}{}                                 & $\mathbf{-6.46\%}$ & $\mathbf{-1.39}$\,\textbf{pp} \\
\bottomrule
\end{tabular}
\caption{Best per-target augmentation configuration (vertical shift, horizontal shift, rotation, zoom, intensity) from Bayesian-optimisation search, compared against the nominal config $[0.5, 0.5, 3, 3, 0]$. Score is the percent change in L1 error relative to the zero-augment baseline; more negative is better. The ``vs.\ nominal'' column reports the additional improvement (in percentage points) the per-target optimum provides over simply using the nominal config. P1--P4 are the four HOM participants.}
\label{tab:best-configs}
\end{table}

Per-target tuning yields an average of only $-1.39$\,pp on top of the nominal configuration. Two targets account for most of that gain --- the FF dataset ($-2.16$\,pp) and HOM-P1's force model ($-4.95$\,pp); for the remaining eight targets the per-target optimum sits within roughly $1.5$\,pp of the nominal score, and four of the ten gain less than $0.5$\,pp. The objective landscape is also flat near the nominal: rounding the GP-found optima to a $0.5$-shift / integer-rotation grid loses at most $\sim$$1.5$\,pp, and reverting insensitive parameters typically loses less than $1$\,pp. Given the small absolute deltas, the flatness around nominal, and the desire not to overfit to a particular scenario, we keep the single nominal configuration as the default.

The more interesting pattern is how widely the per-user optima differ from one another, particularly on the force targets: P1 prefers very aggressive augmentation ($v{=}3$, $h{=}5$, $z{=}22$), P2 sits near nominal except for elevated zoom, P3 prefers \emph{no} vertical shift and minimal rotation, and P4 lands at the nominal within noise. The kinematics targets show a similarly broad spread (zoom of $0$ for P2, rotation of $0$ for P3, elevated rotation/zoom for P1 and P4). Each user's gain from picking their own optimum is small ($\leq$2\,pp for most), yet the optima sit in different corners of parameter space --- the search is not converging on any single ``right'' setting. The augmentation parameters are therefore likely compensating for the session-to-session nuisance factors of \cref{sect:data-augmentation} rather than tuning the model itself. Per-user augmentation tuning for per-user models, or per-user refinement on top of a generic model, are natural directions for future work.

\subsection{EIT: A Comparison Modality}
\label{app:eit}

EIT (electrical impedance tomography) is an internal control: it is collected on the same participant, the same anatomy, and the same recording protocol as the pressure Hand Pose (HP) dataset, but with a different sensing modality. This lets us train and evaluate the same model architecture under matched conditions, isolating the contribution of dense pressure sensing from confounds of participant, protocol, and setup. This section presents the comparison end to end: the EIT hardware, the dataset and protocol, the results against the pressure HP baseline, and the input preprocessing sweep that fixes the reported operating point.

\subsubsection{EIT Hardware}
\label{sect:eit-hardware}
\begin{figure}[t]
  \centering
  \def\iw{10em}
  \resizebox{\columnwidth}{!}{%
  \begin{tikzpicture}[inner sep=1pt]
    \node(p01)                             {\includegraphics[height=\iw]{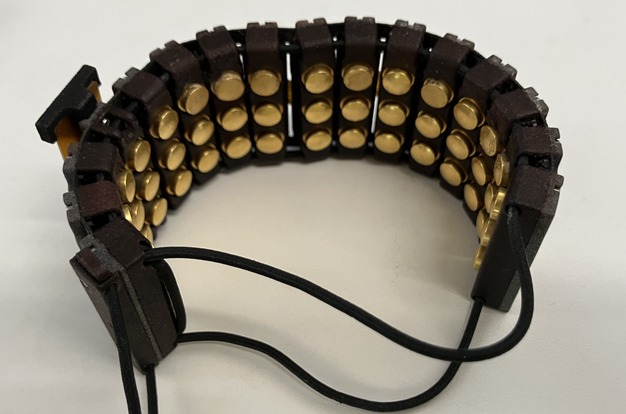}};
    \node(p02) at (p01.east) [anchor=west] {\includegraphics[height=\iw]{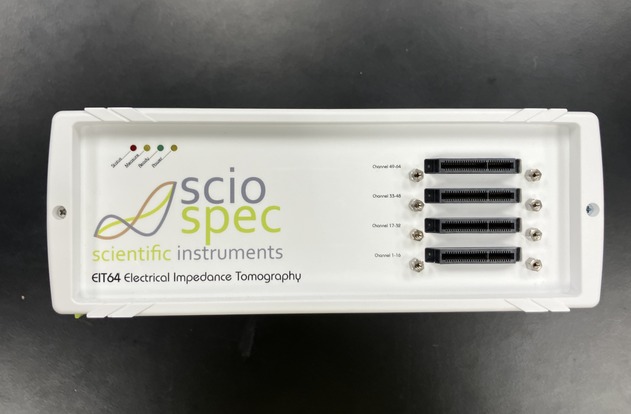}};
    \node(p03) at (p02.east) [anchor=west] {\includegraphics[height=\iw]{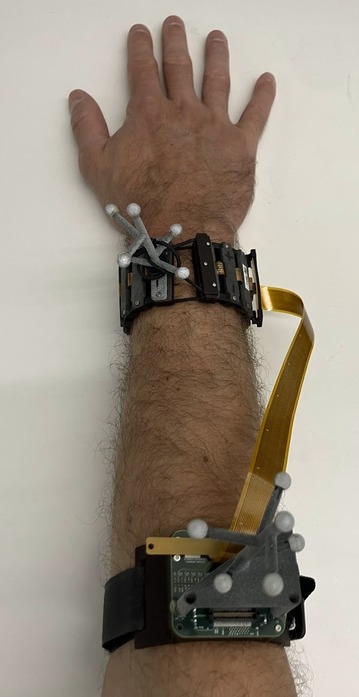}};

    \node[anchor=north, yshift=-2pt] at (p01.south) {(a)};
    \node[anchor=north, yshift=-2pt] at (p02.south) {(b)};
    \node[anchor=north, yshift=-2pt] at (p03.south) {(c)};
  \end{tikzpicture}%
  }
  \caption{EIT hardware. (a) the 16-electrode band; (b) the Sciospec EIT64 unit; (c) the band worn on the distal forearm, with a second band holding the proximal-forearm reference marker tree.}
  \label{fig:eit-hardware}
\end{figure}
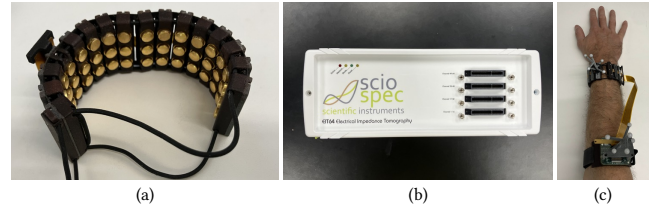

For comparison with our pressure-sensing wristband, we build an Electrical Impedance Tomography (EIT) setup. The acquisition unit is a Sciospec EIT64, a 64-channel impedance tomograph, of which we use 16 channels. The electrode array is a repurposed sEMG wristband with three rows of 16 spring-loaded brass electrodes, of which we drive only one row (the proximal row, closest to the elbow) (\cref{fig:eit-hardware}~(a); the Sciospec unit is shown in \cref{fig:eit-hardware}~(b)). \Cref{fig:eit-hardware}~(c) shows the wristband worn on the distal forearm, with a second wristband on the proximal forearm holding the reference marker tree.

We configure the EIT64 with 16 adjacent-pair current injections per frame; for each injection we read all 16 electrodes, yielding $16 \times 16 = 256$ measurements per frame. The injection current is 10~mA at 50~kHz, and the unit runs at 30~Hz frame rate in this configuration. A custom Sciospec module reads the standard OptiTrack timecode signal, so that EIT frames are timestamped against the MoCap clock for downstream synchronization.

\subsubsection{EIT Dataset and Protocol}
\label{sec:supp_data_eit}
This smaller dataset is collected with the EIT comparison wristband (\cref{sect:eit-hardware}) rather than the pressure wristband, and is included purely for cross-modality comparison: it lets us train and evaluate the same model architecture on a matched protocol and anatomy under a different sensing modality (\cref{sec:supp_results_eit}). The collection was carried out before we switched to the pressure wristband as our primary modality, and we keep it to anchor the pressure results against a directly comparable EIT measurement. The recording protocol is the same finger and wrist movement protocol as HP, and the rest of the setup (MoCap and tactile glove) is identical. In total, we collect 25 recordings totalling 398~min (6~h~38~min), with most sessions between 15 and 21 minutes.

The EIT dataset reuses the HP movement protocol verbatim (see \cref{sec:supp_protocol_hp,fig:dataset-handpose}), differing only in that the pressure wristband is replaced by our EIT hardware while keeping the MoCap glove and forearm marker cluster identical.

\subsubsection{EIT Results}
\label{sec:supp_results_eit}
For comparison with our primary pressure-based wristband, we evaluate the same combined wrist and finger model on the EIT dataset, collected with our EIT hardware on the same finger and wrist movement protocol as HP. Results are pooled over 25 recordings across 20 random seeds; \cref{tab:eit_angle_errors} reports the same metrics in the same format as the HP results.

\begin{table}[t]
  \centering
  \caption{Joint-angle and position errors for the combined wrist-and-finger model on the EIT dataset (best causal-filtering setting $\alpha{=}10^{-2}$, causal high-pass; pooled across 25 recordings $\times$ 20 random seeds). Same metrics and format as the HP results (\cref{tab:joint_angle_errors}, \cref{sect:metrics}).}
  \label{tab:eit_angle_errors}
  \small
  \setlength{\tabcolsep}{3pt}
  \begin{tabular}{ll c cc !{\hskip 3pt\vrule\hskip 3pt} cc}
    \toprule
    \textbf{Category} & \textbf{DOF} & $\sigma$(GT)$^\circ$ &
      MAE$^\circ$ & $R^2$ &
      \multicolumn{2}{c}{\textbf{Tip\,/\,MPJPE (mm)}} \\
    \midrule
    \multirow{4}{*}{Wrist}
      & Pronation     & 26.14 & 14.25 & 0.512 & & \\
      & Flexion       & 26.43 & 7.51  & 0.826 & & \\
      & Deviation     & 10.73 & 5.42  & 0.512 & & \\
    \cmidrule(lr){2-7}
      & \textit{Mean} & \textit{25.09} & \textit{9.06} & \textit{0.659} & {\footnotesize No Wrist} & {\footnotesize With Wrist} \\
    \midrule
    \multirow{6}{*}{Finger}
      & Thumb         & 14.13 & 5.69 & 0.633 & 14.7\,/\,8.1 & 35.1\,/\,24.1 \\
      & Index         & 18.48 & 5.25 & 0.725 & 12.3\,/\,8.5 & 35.8\,/\,28.7 \\
      & Middle        & 20.14 & 5.46 & 0.766 & 13.1\,/\,9.2 & 36.9\,/\,28.8 \\
      & Ring          & 19.50 & 5.54 & 0.732 & 14.3\,/\,9.9 & 36.9\,/\,28.3 \\
      & Pinky         & 17.34 & 5.41 & 0.696 & 12.4\,/\,8.5 & 35.6\,/\,27.5 \\
    \cmidrule(lr){2-7}
      & \textit{Mean} & \textit{18.32} & \textit{5.47} & \textit{0.719} & \textit{13.4\,/\,8.8} & \textit{36.1\,/\,27.5} \\
    \bottomrule
  \end{tabular}
\end{table}

Compared to the pressure-based HP results (\cref{tab:joint_angle_errors}), the EIT model is consistently worse on every axis. Wrist errors increase substantially --- pronation MAE rises from $6.4^\circ$ to $14.3^\circ$ (with $R^2$ dropping from $0.85$ to $0.51$) --- and finger MAE rises by roughly 20\% ($4.6^\circ \to 5.5^\circ$) at $R^2 = 0.72$ rather than $0.86$. The fingertip and MJPE position errors follow the same pattern: the With-Wrist MJPE jumps from $15.3$~mm to $27.5$~mm, dominated by the residual pronation error that rotates the entire finger chain about the forearm axis. The gap is consistent with our motivating observation that contact-pressure variation is the dominant signal at the band--skin interface; an EIT measurement on the same anatomy recovers only a fraction of the information that the dense pressure array captures directly.

Part of this gap is also attributable to sensor resolution. Our EIT wristband uses 16 electrodes, each roughly 1~cm wide, which spatially corresponds to about 3 rows and 4 columns of a single pressure sensor array. \Cref{app:ablation-sensor} shows that aggressively subsampling the pressure strip to a comparable footprint already loses a meaningful fraction of the achievable accuracy, with wrist regression degrading more sharply than finger regression under spatial subsampling. The remaining EIT-vs-pressure gap on top of that reflects the additional degradation intrinsic to the EIT modality itself, beyond the resolution mismatch alone.

The reported EIT numbers use a causal high-pass pre-filter ($\alpha{=}10^{-2}$) on the EIT input. EIT-style electrode systems share a well-known artefact: a $15$--$20$\,min settling transient after donning plus continued slow baseline drift due to changing skin conditions (sweat, moisture content, etc.), which high-pass filtering removes; \cref{sec:supp_eit_filter} sweeps the filter strength and the causal vs.\ non-causal variant.

\subsubsection{EIT Preprocessing Sweep}
\label{sec:supp_eit_filter}
The EIT results in \cref{sec:supp_results_eit} use a causal high-pass filter with $\alpha{=}10^{-2}$. This section unpacks that choice by sweeping the filter parameter $\alpha$ across both causal (real-time-safe) and non-causal (offline forward--backward) variants.

\paragraph{Why filter, causal vs.\ non-causal.}
EIT-style electrode systems exhibit a $15$--$20$\,min settling transient after donning, together with continued slow baseline drift due to changing skin conditions (sweat, moisture content, etc.), neither of which carries hand-pose information. A high-pass on the EIT input removes this nuisance component. Concretely, we estimate a running mean of the input as a first-order exponentially-decaying low-pass with smoothing parameter $\alpha$ (per-sample update $\bar{x}_t \leftarrow (1-\alpha)\bar{x}_{t-1} + \alpha\,x_t$ at our $50$\,Hz sampling rate, giving an effective time constant of $\tau \approx 1/(\alpha\,f_s)$ seconds) and subtract it from the signal. We compare a one-sided causal pass (past-only, real-time-safe but phase-distorting) against a two-sided non-causal forward--backward pass (the same exponential decay applied once forward and once in reverse, which has zero phase distortion but requires the entire recording up front and is therefore offline-only). At matched $\alpha$, the non-causal variant is uniformly slightly better (e.g.\ wrist mean MAE $8.58^\circ$ vs.\ $9.06^\circ$ at the best operating point), but the gap is small enough that the causal filter is a reasonable choice when real-time decoding is required.

\paragraph{Choice of \texorpdfstring{$\alpha$}{alpha}.}
\Cref{tab:eit_sweep} shows the full sweep. Several patterns emerge. (i) The unfiltered \textit{raw-data} baseline tracks fingers reasonably well ($5.85^\circ$ mean MAE) but is essentially blind to wrist pronation ($R^2{=}0.07$): the long-tailed settling/drift component dominates the input and swamps the much slower wrist signal. (ii) Light filtering ($\alpha{=}10^{-4}$, $3{\times}10^{-4}$) is not enough --- it leaves too much slow drift in place and finger accuracy actually degrades slightly relative to \textit{raw-data}. (iii) Filtering in the range $\alpha \in [10^{-3}, 10^{-2}]$ is the sweet spot: causal $\alpha{=}10^{-2}$ achieves the best finger mean ($5.47^\circ$) and the best wrist mean ($9.06^\circ$), and non-causal $\alpha{=}3{\times}10^{-3}$ is the best operating point offline. (iv) Filtering too aggressively ($\alpha{=}3{\times}10^{-2}$) starts to suppress real signal in addition to drift, with both wrist and finger accuracy regressing slightly. The trade-off arises because, under our protocol, the slow component of wrist motion overlaps spectrally with the settling-drift artefact, so the same filter that removes the artefact also discards some of the wrist signal that could have been used.

We pick causal $\alpha{=}10^{-2}$ for the main paper because it has the best causal MAE on both wrist and finger averages while remaining real-time-safe.

\begin{table}[t]
  \centering
  \caption{\textbf{EIT sweep: preprocessing settings on the EIT dataset.} \textit{raw-data} disables preprocessing; the remaining 12 settings apply a high-pass filter to the EIT input parameterized by $\alpha$ (smaller = lighter filtering), in two blocks: \textbf{causal} (real-time) and \textbf{non-causal} (offline only). Per-DOF entries are MAE ($^\circ$); the \textbf{Mean} pair in each section reports (MAE$^\circ$, $R^2$) averaged over the wrist and finger DOFs separately. Best value per column in \textbf{bold}.}
  \label{tab:eit_sweep}
  \scriptsize
  \setlength{\tabcolsep}{2.5pt}
  \newcommand{\vh}[1]{\rotatebox[origin=lB]{90}{#1}}
  \begin{tabular}{l !{\color{black!25}\vrule width 0.4pt} ccc !{\color{black!25}\vrule width 0.4pt} cc !{\color{black!55}\vrule width 0.7pt} ccccc !{\color{black!25}\vrule width 0.4pt} cc}
    \toprule
    & \multicolumn{3}{c}{\textbf{Wrist}} & \multicolumn{2}{c}{\textbf{W.\ Mean}}
    & \multicolumn{5}{c}{\textbf{Finger}} & \multicolumn{2}{c}{\textbf{F.\ Mean}} \\
    \cmidrule(lr){2-4} \cmidrule(lr){5-6} \cmidrule(lr){7-11} \cmidrule(lr){12-13}
    \textbf{Setting}
      & \vh{Pron.} & \vh{Flex.} & \vh{Dev.} & MAE & $R^2$
      & \vh{Thumb} & \vh{Index} & \vh{Middle} & \vh{Ring} & \vh{Pinky} & MAE & $R^2$ \\
    \midrule
    \textit{raw-data} & 20.62 & 9.42 & 5.90 & 11.98 & 0.41 & 6.11 & 5.61 & 5.89 & 5.87 & 5.74 & 5.85 & 0.68 \\
    \midrule
    \multicolumn{13}{l}{\textit{Causal (real-time)}} \\
    \midrule
    $10^{-4}$          & 19.82 & 13.24 & 7.65 & 13.57 & 0.20 & 7.17 & 6.31 & 6.71 & 6.75 & 6.48 & 6.68 & 0.57 \\
    $3{\times}10^{-4}$ & 16.86 & 9.99  & 6.41 & 11.09 & 0.45 & 6.78 & 6.12 & 6.58 & 6.64 & 6.45 & 6.52 & 0.57 \\
    $10^{-3}$          & 15.19 & 7.97  & 5.57 & 9.58  & 0.58 & 6.13 & 5.62 & 5.91 & 6.00 & 5.90 & 5.91 & 0.65 \\
    $3{\times}10^{-3}$ & 14.46 & \textbf{7.51} & 5.43 & 9.13 & 0.61 & 5.81 & 5.33 & 5.55 & 5.65 & 5.52 & 5.57 & 0.70 \\
    $10^{-2}$          & \textbf{14.25} & 7.51 & \textbf{5.42} & \textbf{9.06} & \textbf{0.62} & \textbf{5.69} & \textbf{5.25} & \textbf{5.46} & \textbf{5.54} & \textbf{5.41} & \textbf{5.47} & \textbf{0.71} \\
    $3{\times}10^{-2}$ & 14.37 & 7.69 & 5.45 & 9.17 & 0.61 & 5.70 & 5.27 & 5.48 & 5.54 & 5.44 & 5.48 & 0.71 \\
    \midrule
    \multicolumn{13}{l}{\textit{Non-causal (offline)}} \\
    \midrule
    $10^{-4}$          & 16.76 & 10.32 & 6.64 & 11.24 & 0.44 & 6.97 & 6.48 & 6.93 & 6.94 & 6.86 & 6.84 & 0.53 \\
    $3{\times}10^{-4}$ & 14.85 & 7.46  & 5.33 & 9.21  & 0.61 & 5.93 & 5.49 & 5.74 & 5.91 & 5.83 & 5.78 & 0.67 \\
    $10^{-3}$          & 13.85 & 6.96  & \textbf{5.14} & 8.65 & 0.65 & 5.61 & 5.21 & 5.40 & 5.55 & 5.43 & 5.44 & 0.71 \\
    $3{\times}10^{-3}$ & \textbf{13.64} & \textbf{6.93} & 5.18 & \textbf{8.58} & \textbf{0.65} & \textbf{5.54} & \textbf{5.13} & 5.34 & 5.44 & 5.33 & 5.36 & 0.72 \\
    $10^{-2}$          & 13.84 & 7.14 & 5.22 & 8.73 & 0.64 & 5.55 & 5.13 & \textbf{5.34} & \textbf{5.40} & \textbf{5.33} & \textbf{5.35} & \textbf{0.73} \\
    $3{\times}10^{-2}$ & 14.15 & 7.62 & 5.36 & 9.04 & 0.62 & 5.61 & 5.17 & 5.37 & 5.42 & 5.37 & 5.39 & 0.72 \\
    \bottomrule
  \end{tabular}
\end{table}

\end{document}